# Human Brain Mapping with Multi-Thousand Channel PtNRGrids Resolves Novel Spatiotemporal Dynamics

Youngbin Tchoe[1†], Andrew M. Bourhis[1†], Daniel R. Cleary[1,2†], Brittany Stedelin[3], Jihwan Lee[1], Karen J. Tonsfeldt[1,4], Erik C. Brown[3], Dominic Siler[3], Angelique C. Paulk[5], Jimmy C. Yang[5,6], Hongseok Oh[1], Yun Goo Ro[1], Keundong Lee[1], Samantha Russman[1], Mehran Ganji[1], Ian Galton[1], Sharona Ben-Haim[1,2], Ahmed M. Raslan[3], and Shadi A. Dayeh[1,2,7*]

[1]Integrated Electronics and Biointerfaces Laboratory, Department of Electrical and Computer Engineering, University of California San Diego, La Jolla, California 92093, United States
[2]Department of Neurological Surgery, University of California San Diego, La Jolla, California 92093, United States
[3]Department of Neurological Surgery, Oregon Health & Science University, Mail code CH8N, 3303 SW Bond Avenue, Portland, Oregon 97239- 3098, United States
[4]Department of Obstetrics, Gynecology, and Reproductive Sciences, Center for Reproductive Science and Medicine, University of California San Diego, La Jolla, California 92093, United States
[5]Department of Neurology, Massachusetts General Hospital, Boston, Massachusetts 02114, United States
[6]Department of Neurosurgery, Massachusetts General Hospital, Boston, Massachusetts 02114, United States
[7]Graduate Program of Materials Science and Engineering, University of California San Diego, La Jolla, California 92093, United States

* Corresponding author. Email: sdayeh@eng.ucsd.edu
[†] These authors contributed equally to this work

**One Sentence Summary:** PtNRGrids offer high spatial resolution and cortical coverage and reveal functional organization and neuronal dynamics not previously resolved.

**Abstract:** Electrophysiological devices are critical for mapping eloquent and diseased brain regions and for therapeutic neuromodulation in clinical settings and are extensively utilized for research in brain-machine interfaces. However, the existing devices are often limited in either spatial resolution or cortical coverage, even including those with thousands of channels used in animal experiments. Here, we developed scalable manufacturing processes and dense connectorization to achieve reconfigurable thin-film, multi-thousand channel neurophysiological recording grids using platinum-nanorods (PtNRGrids). With PtNRGrids, we have achieved a multi-thousand channel array of small (30 µm) contacts with low impedance, providing unparalleled spatial and temporal resolution over a large cortical area. We demonstrate that PtNRGrids can resolve sub-millimeter functional organization of the barrel cortex in anesthetized rats that captured the histochemically-demonstrated structure. In the clinical setting, PtNRGrids resolved fine, complex temporal dynamics from the cortical surface in an awake human patient performing grasping tasks. Additionally, the PtNRGrids identified the spatial spread and dynamics of epileptic discharges in a patient undergoing epilepsy surgery at 1 mm spatial resolution, including activity induced by direct electrical stimulation. Collectively, these findings demonstrate the power of the PtNRGrids to transform clinical mapping and research with brain-machine interfaces and highlights a path toward novel therapeutics.

## INTRODUCTION

Functional mapping with direct electrical stimulation paired with neurophysiological recording is the gold standard for mapping the human brain and delineating the margins between functional and pathological tissue.(*1-4*) Neurophysiological recording with non-penetrating surface electrocorticography (ECoG) grids have been used for over six decades to attain reliable clinical information and improve patient outcomes during surgical interventions.(*3, 4*) ECoG grids can have cortical coverage of up to 8 × 8 $cm^2$ and interelectrode pitch as small as 4 mm.(*2, 5-10*) Higher resolution grid such as the penetrating Utah arrays have less coverage (4 × 4 $mm^2$) and better pitch (0.4 mm) than ECoG grids but they penetrate into the brain parenchymal.(*1, 11-14*) These are the de facto standard for research on chronic neural prostheses for motor control and decoding language, as well as for providing sensory feedback in paraplegic individuals via closed loop devices.(*5-10, 12, 15-23*) While great progress has been made using these devices, the next steps in neuroprostheses and neural decoding requires higher spatial resolution(*24-26*) and expanded coverage of the cortex.

We used advanced thin-film microfabrication techniques and a newly developed, biocompatible platinum nanorod (PtNR)(*27*) microelectrode material to develop large surface-area ECoG grids with both high resolution and broad spatial coverage (PtNRGrids). Our PtNRGrids are built on thin, conformal parylene C substrates, and the distribution of contacts is reconfigurable for different pitches and area coverage. We utilized compact one-touch connectors to enable a simple and reliable interface with thousands of channels that is amenable to the constraints of the operating room. Here, we demonstrate the use of these PtNRGrids to isolate submillimeter functional boundaries of individual cortical columns in controlled animal experiments, and neural mapping from both awake and anesthetized patients undergoing tumor or epileptogenic tissue resection.

## RESULTS

### Fabrication of Multi-Thousand Channel PtNRGrids and Connectorization

PtNRGrids were composed contacts embedded in flexible sheets of 6.6 μm thick parylene C, specially designed for recording neural activity on the cortical surface (Fig. 1). The layout, shape, and size of the PtNRGrids were generated with customizable designs by leveraging established microelectromechanical systems (MEMS) techniques on large 18 × 18 $cm^2$ glass wafers and a newly developed, biocompatible platinum nanorod (PtNR)(*27*) microelectrode material. This process produced multiple 17 cm long and up to 8 × 8 $cm^2$ large area coverage electrodes ranging from 1024 to 2048 electrode contacts or channels with high uniformity and yield (Fig. 1). The 30 μm wide PtNR contacts were recessed by ~2 μm below the surface of parylene C to prevent shear forces on PtNRs during implant (Fig. 1B;(*27*)). Between the PtNR contacts and the bond pads, we routed gold traces which were 500 nm thick, 4 μm wide, 6 μm apart, and over 10 cm long, and fully encapsulated between two layers of parylene C (3.5 μm bottom and 3.1 μm top, Fig. S1, S2). We patterned perforation holes in the parylene C throughout the thin grid to perfuse saline and cerebrospinal fluid away from the electrode contacts. Therefore, an intimate interface between the PtNRGrids and the surface of the brain was maintained and electrochemical shunting between nearby recording contacts was avoided (Fig. S3). Additionally, the large-perfusion holes distributed across the grid provided access for probes of a handheld clinical stimulator to directly stimulate any point of the cortex through the grid. We varied the pitch of the 30 μm PtNR contacts from 150 μm for rodent brain mapping to up to 1.8 mm for human brain mapping with a coverage area ranging from 5 × 5 $mm^2$ (1024 PtNRGrids) to 8 × 8 $cm^2$ (2048 PtNRGrids) (Fig. 1D). The detailed fabrication process of the PtNRGrids can be found in the supplementary materials.

A major bottleneck for scaling microelectrode arrays towards hundreds or thousands of channels was the connectorization of electrodes to acquisition circuits. Inspired by solutions used in the microelectronics industry which can reliably route high bandwidth connections to thousands of channels(*28, 29*), we utilized an off-the-shelf land grid array (LGA) – LGA1155 CPU socket – that was originally designed for the Intel's Sandy Bridge computer processors. Manufacturing the grids on large-area substrates ensures sufficient space to bond the PtNRGrids to custom LGA- printed circuit boards that mate with the LGA1155 socket without compromising the large area coverage of PtNRGrids or their long thin-film metal leads (Fig. 1D). An additional extender board was used to further increase the separation between the surgical field and a custom-acquisition board (Fig. S4), and for improved intraoperative handling procedures. The acquisition board connects to a 1024-channel electrophysiology control system, provided by Intan Technologies LLC (Fig. S5). The entire PtNRGrid and connector (Fig. 1C) were compatible with the conventional processes(*30*) used to sterilize surgical equipment, maintaining contact yields up to 99.4% with a narrow 1 kHz impedance distribution centered at 11 kΩ with a standard deviation of 2 kΩ (Figs. 1E-F), achieved with manual bonding of the electrodes to the printed circuit board (PCB) (Fig. S3). The 1 kHz impedance magnitude, averaged over seven different PtNRGrids used in successful human recording experiments, was 10 ± 2 kΩ. The scalable process allowed us to obtain up to 95.2% contact yield with impedances ≤ 100 kΩ even when the total channel count increased to 2048 (Fig. 1F). The simple, one-touch connector methodology enabled our neurosurgical and research team to swiftly and reliably connect thousands of channels to the acquisition board across the boundary between the sterile and non-sterile zones. As a result, sterilization was limited to the disposable grid and its connector eliminating the need to sterilize the acquisition electronics. The setup allowed us to record simultaneously from 1024 channels with a sampling rate of 20,000 samples/second, thereby capturing full broadband neurophysiological activity.

A commonly voiced concern over increasing the density of microelectrode arrays is the potential for electrical crosstalk to introduce artifacts into the neurophysiological recordings. This electrical crosstalk is primarily a result of parasitic capacitance between neighboring leads and thus will scale directly with increasing trace length and inversely with their trace pitch. Thus, traces should be kept short to reduce these parasitic paths. The termination impedance of neighboring channels to tissue (i.e. the electrochemical interfacial impedance) also needs to be accounted for, especially for conventional high-impedance electrochemical interfaces which can affect crosstalk through parasitic capacitance paths. However, this is not a concern for the low impedance PtNR contacts which maintain 1 kHz impedances that are at least 10 million times lower than the impedance of the parasitic capacitances. It should be noted that open channels on any grid including PtNRGrids can be problematic and these have been excluded (cut off of 100 kΩ at 1kHz) from our analysis. The detailed experimental investigation of cross-talk could be found in the supporting information (see Figs. S6-S11).

The PtNRGrids exhibited mechanical stability exceeding the American National Standards Institute / Association for the Advancement of Medical Instrumentation ANSI/AAMI CI86:2019 recommendations (Fig. S12).

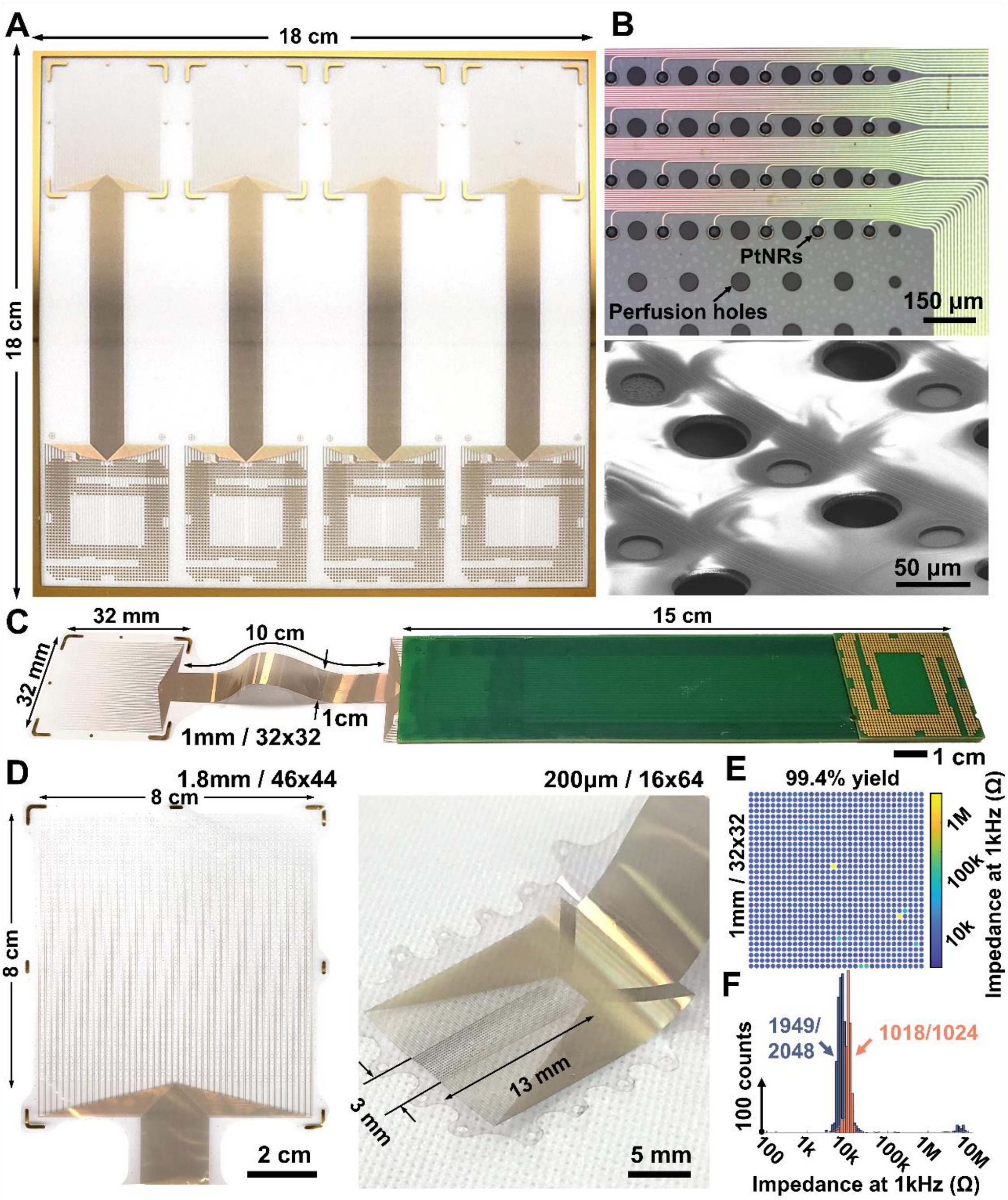

**Fig. 1. Multi-thousand channel PtNRGrids ECoG arrays.** (A) Scalable and large-area fabrication of electrode arrays on 18 × 18 cm$^2$ glass substrates. Microscale features of PtNRs contacts, metal leads, and perfusion holes shown by (B) optical (top) and electron microscope images (bottom). (C) A 1024 channel electrode bonded on the extender PCB that is compatible with CPU LGA sockets. (D) Reconfigurable electrode designs depending on the target placement on the human brain with a sensing area of 8 × 8 cm$^2$ (2048 channel) and 3 × 13 mm$^2$ (1024 channel). (E) Spatial mapping and (F) histogram of the impedance magnitude at 1 kHz for the 1024 and 2048 channel electrode. 1024 and 2048 channel electrodes had 1 kHz impedance magnitude of 11 ± 2 kΩ and 8 ± 4 kΩ with a yield up to 99.4 and 95.2%, respectively.

**PtNRGrids Isolate Functional Cortical Columns from the Surface of the Brain**

To test the broadband and high-resolution recording capabilities of the PtNRGrids, we mapped the primary somatosensory cortex of anesthetized rats. The rat cortex has well-defined organization of the somatosensory cortical structures, especially around the barrel cortex, where a series of sensory cortical columns map one-to-one with the whiskers.(*31, 32*) A square-shaped 1024 channel PtNRGrid with 150 μm pitch was implanted to record from the entire right primary somatosensory barrel cortex (Fig. 2). To evoke sensory activity, air-puffs were delivered through a microcapillary tube to individually stimulate the contralateral-side whiskers. (Fig. 2A; Fig. S13). We consistently observed large-amplitude raw evoked responses (N=50) for whisker (E4) stimulation (Fig. 2B). The raw waveforms exhibit localized, high amplitude responses as large as 500 μV with peak responses observed ~30 ms after the onset of the air-puff (Fig. 2C). These responses propagated as traveling waves across the cortical surface (Fig. S14). The spatial localization of individual stimuli is best represented in the gamma band (30-190 Hz), as expected, and observed in the root-mean-square (RMS) power of the measured responses when bandpass filtered (Fig. 2D). The high-gamma activity (HGA; 70-190 Hz) is known to be highly correlated to the location and timing of cortical activation with a strong link to spiking activities,(*33*) so we used HGA to map functional boundaries of the rat barrel cortex (Fig. 2E). We observed clearly distinguishable submillimeter sensory boundaries that classified the responses to different whisker rows and columns, revealing spatially organized barrels in a superior spatial resolution compared to the state of the art electrical imaging techniques(*31*). The locations of sensory-responsive areas were also identified by evoking cortical responses with air-puff stimulation of the neck, trunk and tail, and by electrical stimulation of the forepaw and hindpaw (Fig. 2E). The detailed signal processing procedures for HGA mapping (Figs. S18-S21) and the results across different rats (n = 4) could be found in the supplementary material (Fig. S23). Following completion of functional mapping, the implanted area was marked, and histochemical analyses were used to examine the anatomy under the implant.

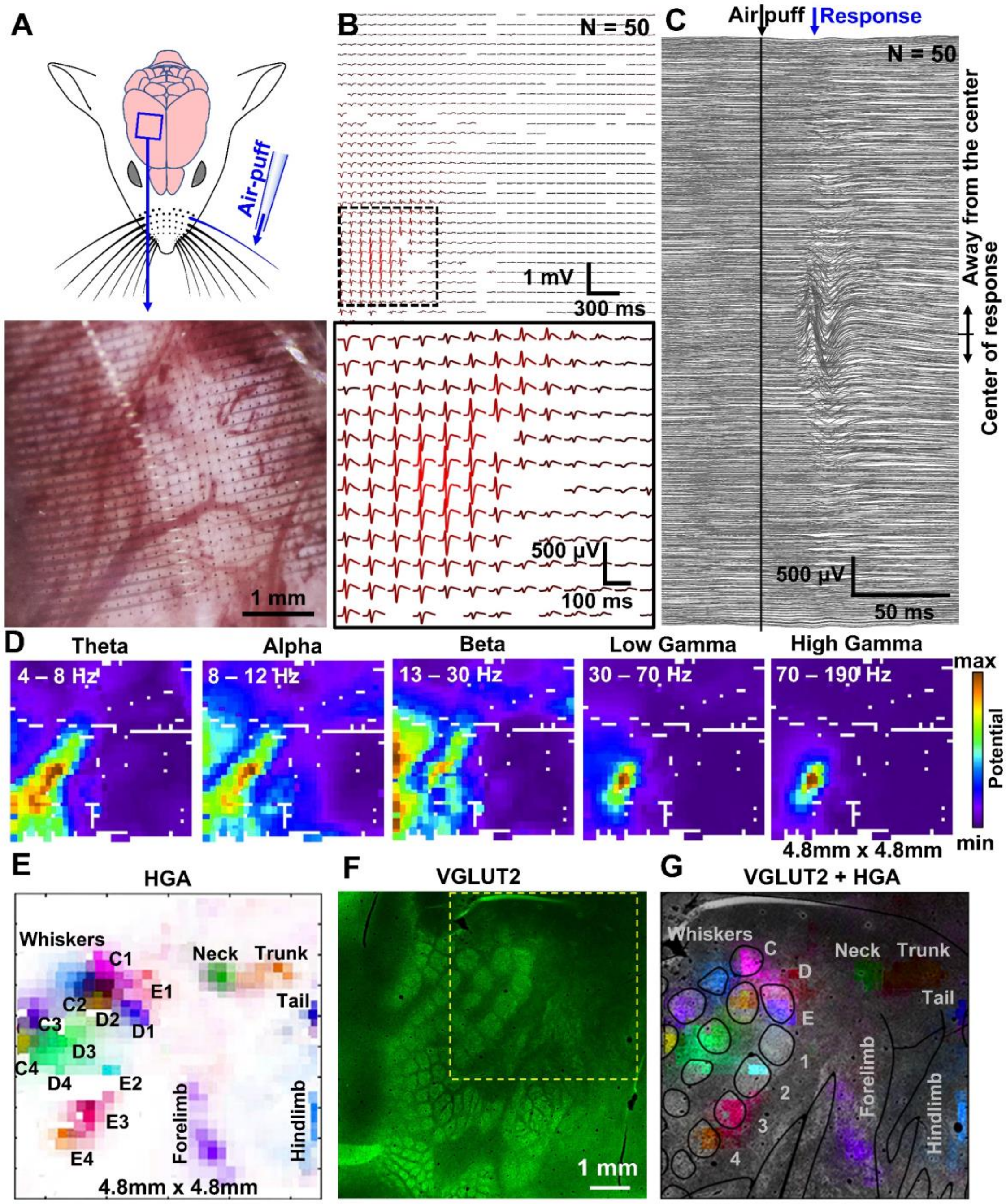

**Fig. 2. PtNR multi-thousand channel ECoG grids record somatotopic functional cortical columns with sub-millimeter resolution.** (A) Schematic of the rat brain implanted with 1024 channels, 4.8 mm x 4.8 mm array, and the air-puff stimulation of individual whiskers. The lower image shows the magnified microscope image of the electrode on the rat barrel cortex. (B) E4 whisker stimulation-evoked ECoG recordings (N=50, raw). (C) Stimulation locked response of all channels. (D) Spatial mapping of neural wave amplitude filtered at different frequency windows. (E) Spatial mapping of high gamma activity recorded by the high-density

PtNRGrid. Each label indicates the positions stimulated with air-puff. (F) VGLUT2 immunostaining of the rat barrel cortex. The electrode implantation location is marked with the yellow dotted box. (G) HGA superimposed on top of the histology image.

The anatomical and functional boundaries were in excellent agreement, as outlined using the vesicular-glutamate transporter 2 (VGLUT2), a well-established marker of thalamocortical afferents that compose the homunculus, including the barrels (Fig. 2F)(*34*)(*35*). The localized HGA responses to whisker stimuli agree remarkably well to the VGLUT2-labeled positions of the barrels as well as the homunculus-labeled positions of the forelimb, hindlimb, neck, trunk, and tail (Fig. 2G). Thus, this single-grid based mapping provides a reliable real-time high-resolution functional mapping of the rat brain, which contrasts with the traditional serial probing while recording evoked responses(*36, 37*).

**PtNRGrids Resolve the Curvilinear Nature of the Human M1-S1 Functional Boundary**

Precise intraoperative localization of the central sulcus, the boundary between primary somatomotor (M1) and somatosensory (S1) cortices is a necessary approach in several neurosurgical procedures, particularly in the resection of tumors. This anatomical boundary is identified by a functional phase reversal of somatosensory evoked potentials (SSEPs) at the boundary between M1 and S1(*38, 39*), an *a priori* assumed anatomico-functional relationship. Most commonly, these SSEPs – recorded in response to electrical stimulation pulses to the peripheral nerves – are evoked 20 ms post stimulus and demonstrate opposite polarity in their potentials across this boundary. The presence of pathological tissue can induce a shift in the functional organization and location far from its presumed anatomical localization(*40, 41*), and make traditional sulcal markers harder to discern with low spatial resolution of clinical ECoG grids.

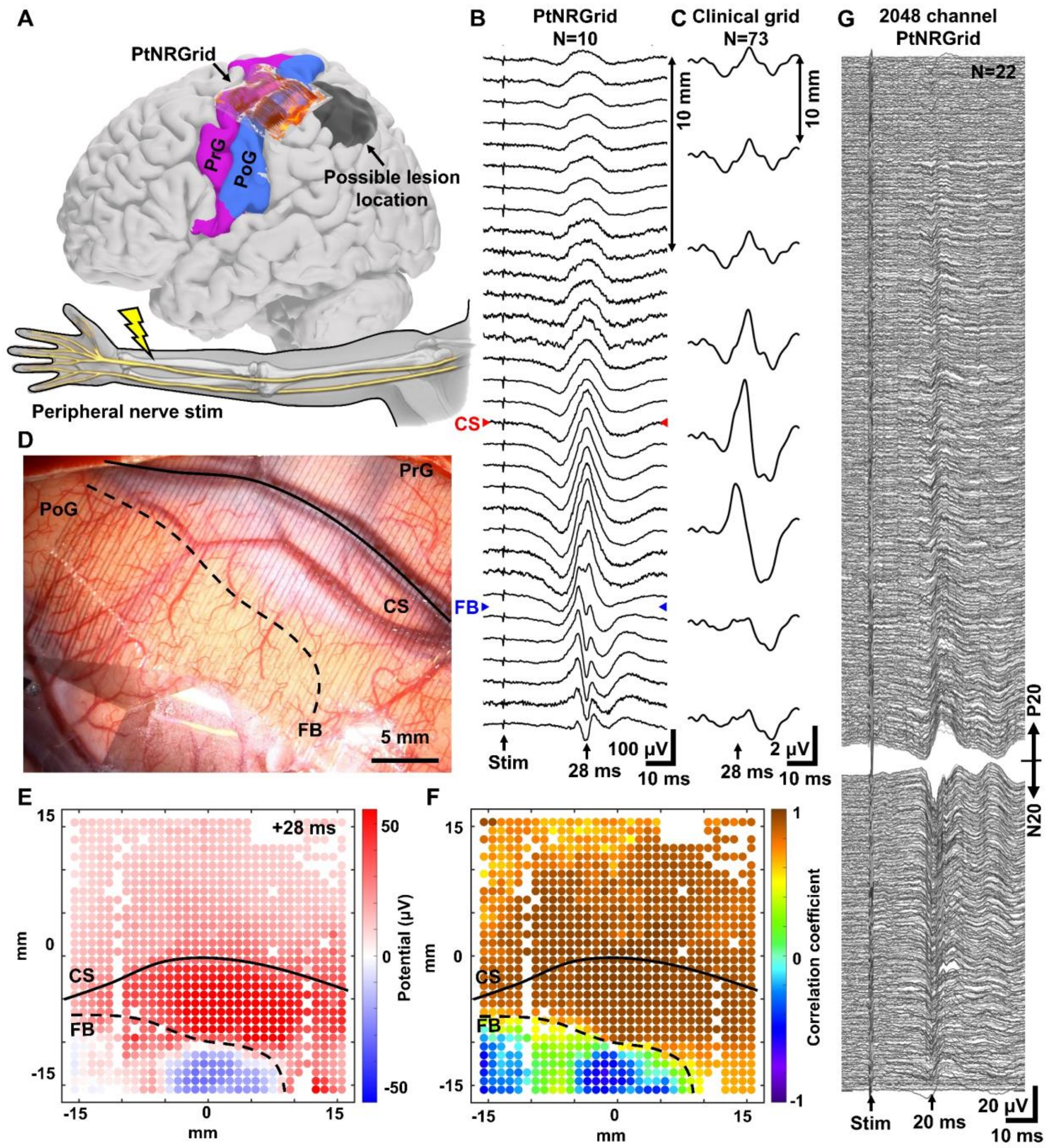

**Fig. 3. Mapping the curvilinear nature of the functional sensory/motor regions in the human brain with millimeter resolution.** (A) Reconstructed model of the patient's brain and the electrode implantation locations. Electrodes were implanted near the hand region, and the peripheral nerve was electrically stimulated. ~~fMRI images show the possible lesion location.~~ Somatosensory evoked potential (SSEP) waveforms along a line across the central sulcus (CS) and M1-S1 functional boundary (FB) recorded with (B) 32×32 PtNRGrid with 1 mm spacing and (C) 2×8 clinical grid with 10 mm spacing. ~~Potentials recorded with the clinical grid were multiplied by a factor of 50 to compare the waveforms on the same scale.~~ (D) Implantation photos of the electrodes near the hand region. (E) 1024-channels potential mapping of the stim-evoked waves 28 ms after the stimulation (as opposed to 20 ms due to distortion from brain lesion as shown in panels (B) and (C). (F) Correlation coefficient mapping with respect to the waveforms measured with respect to the channel in the center of the grid. (G) Human brain SSEPs from 2048 channel PtNRGrid. Channels are sorted according to the peak potential amplitude and polarity at 20 ms after the stimulation, and channels with high contact impedance

were excluded from the plot.

We recorded SSEPs from awake subjects (n = 4) undergoing tumor resection, each with a 1024-channel PtNRGrid placed across the central sulcus near the hand region of somatomotor sensory cortex while peripheral nerves were stimulated (Fig. 3A, 3D; Figs. S15-S16). The implantation site of the PtNRGrid was marked and identified on a reconstructed model of the patient's brain along with lesion location based on functional magnetic resonance imaging (fMRI) and structural MRI (Fig 3A). We observed a small stimulus artifact from peripheral nerve stimulation which was followed by high amplitude SSEPs 10~40ms (Fig. 3B). These waveforms revealed characteristic positive and negative peaks(*42*) that reversed phase at the functional boundary (FB) denoting the M1-S1 functional boundary (Fig 3B, Fig. S24). Similar SSEP waveforms were recorded with a conventional dual column 2 × 8, 16-channel clinical ECoG grid with 10 mm spacing and 2.3 mm diameter recording contacts (Fig. 3C, Fig. S25-26). ~~Significantly,~~ The maximum peak amplitude of SSEPs measured by the PtNRGrid were more than 20 times higher than that measured by the clinical ECoG grid (Fig. 3B). The lower amplitudes on the clinical grid can be attributed to both a spatial averaging on their ~6000X larger surface area than the PtNR contacts and a lower conformity to the brain surface resulting from the 1 mm thick substrate used in clinical grids (compared to the 6.6 μm thick parylene C substrate used in our PtNRGrids). It is also possible that the clinical ECoG grid that has sparsely distributed contacts (1 cm) missed the cortical region with the highest SSEP amplitude.

The PtNRGrids revealed the precise curvilinear nature of the M1-S1 FB with millimeter scale resolution at a sampling frequency of 20kHz, superior to the low-resolution boundaries identified with conventional ECoG grids(*42*) or the temporally limited fMRI(*43*). The subject's lesion (Fig. 3A) contributed to broadening and distorting the SSEP waveforms on both the PtNRGrids (Fig. 3B) and on the clinical ECoG (Fig. 3C), in agreement with the findings of prior clinical studies(*38*). Additionally, the SSEPs recorded with the PtNRGrid were minimally affected by the presence of underlying surface blood vessels ('CS' denoting the anatomical central sulcus in Fig. 3B) (*44*). Although the SSEPs were minimally affected by the blood vessel (Fig. S27), it is reported that higher frequency signals (30-70 Hz) could be attenuated by 30-40% by the presence of the blood vessel (*44*). To construct the two-dimensional maps for the curvilinear FB, we utilized the conventional potential-based phase reversal technique (Fig. 3E) and a correlation technique that we devised to identify the FB in diseased tissue (Fig. 3F). We calculated the Pearson correlation coefficients of the waveforms between 5 to 50 ms post-stimulus for all the working channels with respect to the channel in the middle of the grid. Channels with correlation coefficients above or below 0.5 were separated by a dotted line to depict the M1-S1 FB (Fig. 3F), which agrees well with that deduced directly from potential maps (Fig. 3E). Importantly, we observed a highly detailed spatial map depicting considerable offset, based on SSEPs (Fig. 3) and HGA (Fig. 4), between the M1-S1 FB and the anatomical central sulcus using the PtNRGrid (FB versus CS in Fig. 3D), consistent with functional reorganization with brain lesions.(*40, 41*). However, it is important to note that these SSEPs are projections of event related potentials (ERPs) from deeper layers that are often oblique to the plane of the cortical surface. Extension of this functional boundary below the surface must be validated with depth recordings(*45*). Nevertheless, the M1-S1 FB revealed by the PtNRGrid was concordant with the gold standard clinical mapping using conventional bipolar stimulation, and with higher resolution than conventional clinical ECoG grid passive gamma mapping using CortiQ system (G.tech medical engineering GmbH, Schiedlberg, Austria).

We also recorded SSEPs from the human brain using 2048 channel PtNRGrid (Fig. 1D). The waveforms (30-3000 Hz, N=22) exhibit clear P20-N20 peaks responses ~20 ms after the median nerve stimulation (Fig. 3G). The severe anatomical distortion and the limited time of recording precluded assessing a functional boundary in this return surgery (Fig. S28). Nevertheless, the results of Fig. 3G illustrate that the scalable PtNRGrids enable multithousand channel recordings from the human brain.

**PtNRGrids Reveal Novel Large Scale Spatiotemporal Dynamics of Motor and Sensory Activity in Humans**

Motivated by the rise of interest in using ECoG grids for brain-machine interfaces(*5, 8-10, 21-23*), we investigated whether the superior spatial and temporal resolution of the PtNRGrid could be used to map sensory- and motor-evoked activities. Following the phase reversal mapping of the functional M1-S1 boundary and with the same PtNRGrid placement on the same participant, we either stimulated individual fingers with vibrotactile stimulators, or asked the patient to perform specific hand movements (Fig. 4A; Fig. S17). Following individual fingertip stimulation, we observed a clear enhancement in HGA as large as 3 sigma ($\sigma$) from baseline (Fig. 4B), the largest of which were localized only in the primary somatosensory cortex (Fig. 4B). Vibrotactile stimulation of each fingertip evoked spatially distinctive HGA patterns, with some channels tuning to all fingertips with varying magnitudes. After superimposing HGA on an optical image of the implanted PtNRGrid, we could observe the fine spatial distribution of the neural correlates of vibrotactile stimulation and compare this with the M1-S1 boundary and the cortical anatomy using phase reversal (Fig. 4C).

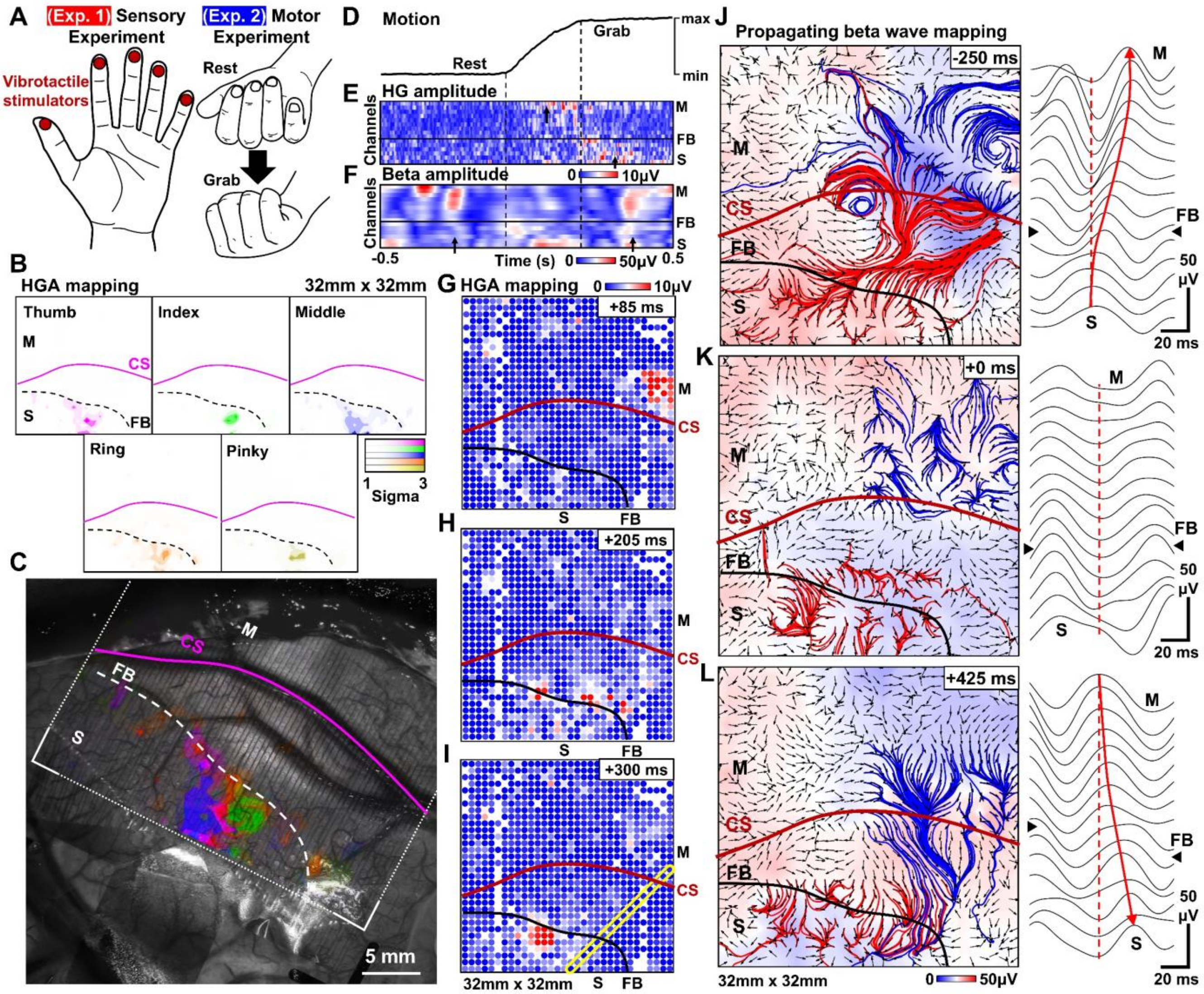

**Fig. 4. Functional mapping with millimeter resolution: PtNRGrid records detailed sensory and motor spatiotemporal dynamics in humans.** (A) Schematics of the sensory and motor experiments. For sensory experiment, individual fingers were stimulated by vibrotactile stimulation in sequence with 1 s stimulation at 2 s intervals. For the motor experiment, the patient was asked to perform a grasping task. (B) Spatial mapping of high gamma activity (HGA) of individual fingers in response to vibrotactile stimulations. (C) Overlay plot of HGA sensory responses for individual fingers superimposed on top of a photo of the surface of the patient's brain. (D) Motion of the hand recorded with the bending sensor. The amplitude of (E) HGA and (F) Beta activity of channels selected along the yellow diagonal rectangle in (I) plotted over a 1 s time window during the motion. Spatial mapping of HGA over three different time points during the hand grabbing motion. (G) Initially localized HGA appears on the motor region, (H) then both the motor and sensory region show HGA, and (I) eventually the HGA only appears on the sensory region. Propagating beta waves and waveforms across the CS in the (J) planning stage of the motion, (K) during the motion, and (L) after the completion of motion. The red and blue streamlines originate from sensory and motor cortex, respectively. The background color represents the amplitude of the beta wave potential, and the arrows indicates the propagating direction of the beta waves. Right panel plots are raw waveforms for the yellow box in (I) around the timestamps of (J)-(L).

We next demonstrated the high spatiotemporal capabilities of the PtNRGrids during a hand grasping task (Fig. 4D-L). As with the vibrotactile stimulation, we observed highly localized HGA on the PtNRGrid during the motion (Fig. 4G), near completion of motion (Fig. 4H), and for 100 ms after completion of motion (Fig. 4I) (See also Supplementary Movie S1). Furthermore, coordination between the S1 and M1 cortices during the grasping task could be seen at high resolution via the PtNRGrid. In a snapshot of the dynamics through time along a single line of electrodes cut across a corner of the grid (highlighted by the yellow rectangle in Fig. 4I), we illustrate the spatial dynamics for the selected 16 channels across the M1-S1 boundary for the high gamma (HG) (Fig. 4E) and beta (Fig. 4F) bands. These band-specific (spectrotemporal) dynamics showed remarkable correlation with the hand movement captured by time-locked flex sensors on the subject's hand (Fig. 4D). Distinctive HGA in the M1 cortex was observed during motor initiation, was seen in both M1 and S1 cortex during the hand closure onset, and finally, lingered only within the S1 cortex when motion was complete. Interestingly, a high amplitude beta wave in the M1 cortex was observed prior to the motion, during the planning stage, attenuated during execution of the motion, and increased once again after the motion was completed (Fig. 4D-F). Similar behavior was observed under trials of repeated hand grabbing motion (see Figs. S29, S32). These observations of alternating amplitude in HG and beta activity before, during, and after the motion agree with prior observations(*46-48*), though the PtNRGrid provides higher spatial resolution at similar coverage to reveal both the large and fine scale activity across the M1-S1 boundary.

Lastly, we illustrate that with the large channel count of the PtNRGrid, we can construct maps of brain wave propagation at an unprecedented spatial resolution within a physiologically relevant cortical coverage (Figs. 4J-L, Fig S22). Previous investigations of propagation characteristics of beta waves in the human brain were carried out with relatively small area coverage of $4 \times 4 mm^2$ using Utah arrays(*49, 50*), or with ECoG grids with sparse 1 cm spatial resolution.(*51*) During the hand grasping task, we calculated the spatial gradient of the phase of the beta waves(*49, 51*) recorded by our PtNRGrids to infer propagation direction (Figs. 4J-L, Figs. S30, S31, S33 and Supplementary Movies S2-3). We further overlaid streamlines originating from selected regions in S1 and M1 cortices on top of the vector fields for a visual aid of the long-range propagation directions. We found striking propagation dynamics across the M1-S1 FB which correlated with the hand grabbing motion. In the preparation stage of the motion, we observed noticeable long range beta waves propagating from the S1 cortex to M1 cortex (Fig. 4J). During the motion, the beta waves were suppressed and exhibited lack of coherence in the vector fields (Fig. 4K). After the motion was complete, the propagating direction reversed, as represented by the streamlines (Fig. 4L, atop a reconstructed brain model in Supplementary Movie S4). By detecting propagation dynamics of beta waves with high spatial resolution at physiologically relevant coverage using our PtNRGrids, we have enhanced functional mapping by revealing large-scale brain activity across frequency bands.

**PtNRGrids Record Pathological Wave Dynamics**

Finally, we determined the utility of PtNRGrids for high-resolution intraoperative neuromonitoring to detect ictal onset zones and patterns of seizure spread. PtNRGrids were placed over the cortex in a patient with intractable epilepsy related to a left anterior temporal lobe cavernoma who elected for surgical resection (Fig. 5A). Using PtNRGrids allowed passive recordings of local field potential (LFP) as well as active electrical stimulation through the perfusion holes (uniformly distributed at 0.5 mm diameter, 1 mm spaced, Fig. 5A) using a standard handheld clinical stimulator (Fig. 5A) to induce interictal epileptiform activity.(*52-55*) Control benchtop experiments on brain models made of gelatin confirmed that the PtNRGrid electrode and the electronics were not affected by the bipolar stimulation (Fig. S34).

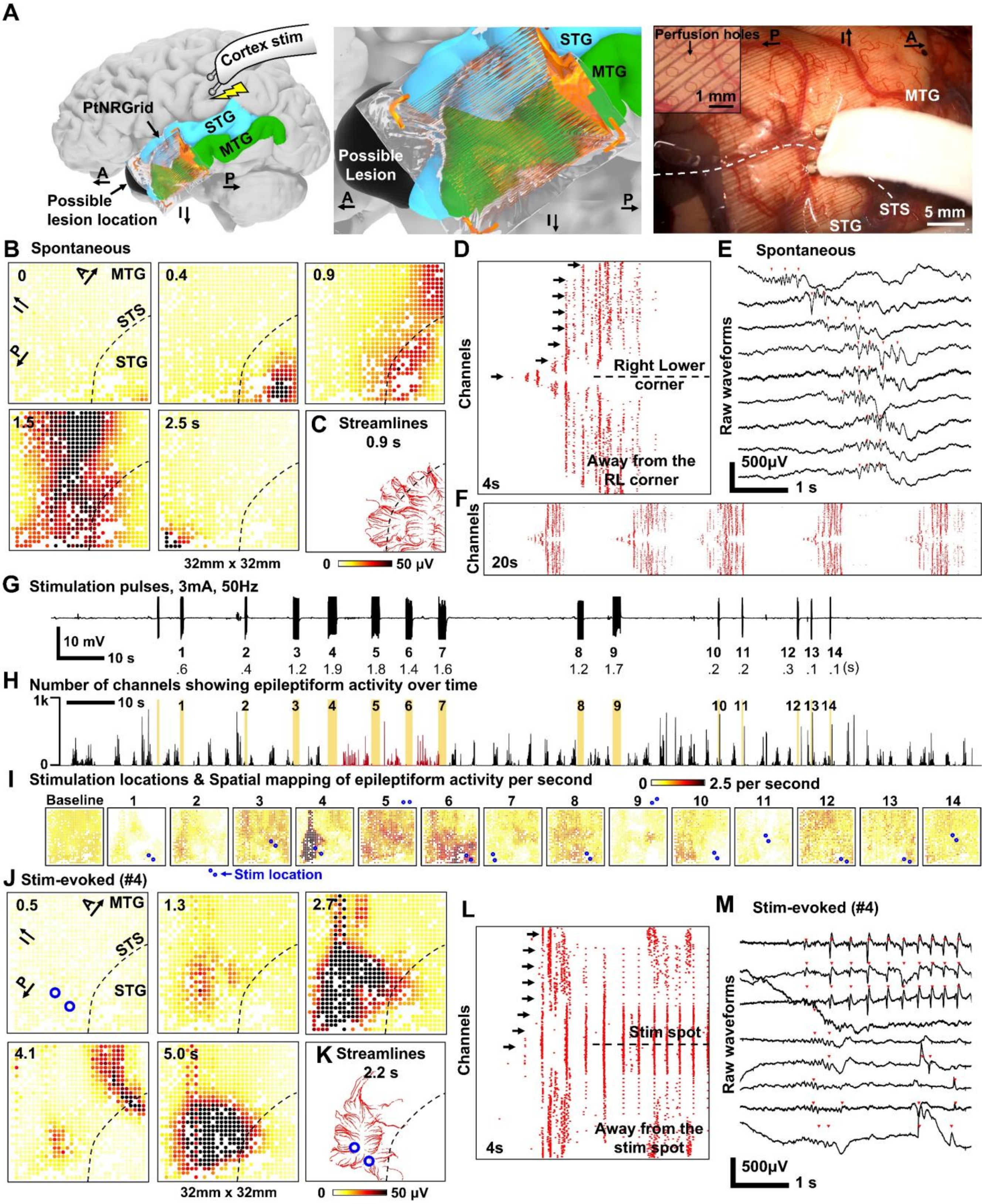

**Fig. 5. Pathological mapping with millimeter resolution: PtNRGrids reveal detailed spatiotemporal dynamics of spontaneous and stimulated epileptiform discharges from an epilepsy patient.** (A) (Left) Reconstructed model of electrode placement on the temporal lobe of the patient's brain, and the schematics of charge-balanced bi-phasic direct current stimulation with the bipolar (Ojemann) probe. (Middle) Magnified model near the electrode.

(Right) Photo showing the cortical tissues directly being stimulated through the electrode. Inset is a magnified image showing the 0.5 mm diameter perfusion holes that are distributed at a 1mm pitch on the electrode allowing direct current stimulation at any point on the grid. Positions of the superior temporal sulcus (STS), superior temporal gyrus (STG), and middle temporal gyrus (MTG) are marked on the photo. Anatomical orientation arrows indicate anterior (A), posterior (P), and inferior (I). **(B)-(F)** Spontaneous epileptiform discharges. (B) Spatial mapping of the 10-59 Hz spontaneous brain wave amplitude. Location of STS is marked with dotted lines. (C) Streamlines plot at 0.9 s depicting the spontaneous propagating wave. Automatically detected epileptiform discharges for all channels in (D) 4s and (F) 20s time windows. (E) Raw waveforms selected from arrow marked channels of (D). The channels are sorted according to the distance from the right lower corner of the electrode; the channel in the midline is closest to the lower right corner. (G) Time course and recordings of the stimulation pulse artifacts for time-locking with evoked response. The stimulation sequence number and duration of stimulation time is indicated below the waveforms. (H) Number of channels showing epileptiform discharges over time. The yellow color indicates the stimulation time points, and the red colored regions mark significant enhancement in epileptiform discharges detection for stimulations number 4, 5 and 6 ($p < .001$). (I) Spatial mapping of epileptiform discharge rate after each stimulation trial. Stimulation locations on or near the electrode are indicated by pair of blue dots for the 14 stimulation trials. **(J)-(M)** Stimulation evoked epileptiform discharges, similar to (B)–(F). (J) Spatial mapping of the 10-59 Hz stim-evoked brain wave amplitude. (K) Streamlines plot at 2.2s depicting the stim-evoked propagating wave. (L) Automatically detected epileptiform discharges for all channels in a 4 s time window. The channels are sorted according to the distance from the stimulation point. (M) Raw waveforms selected from arrow marked channels of (L). Propagating waves for (C) spontaneous epileptiform discharge and (K) stimulation-evoked epileptiform discharge. The red streamlines for (C) originate from the right lower corner while those for (K) originates near the stimulation location. The blue circles in (K) are the bipolar stimulator contact points on the cortex.

Passive mapping of the epileptogenic tissue with the PtNRGrid revealed ongoing epileptiform activity (Fig. 5B-F), also shown in the spatial mapping of RMS amplitude of the neural activity filtered in the frequency window of 10 to 59 Hz at different time epochs (Fig. 5E). The onset of epileptiform activity began near the lower right corner of the grid, then spread across the vertical midline, and terminated at the lower left corner (Fig. 5B and Supplementary Movie S5). To investigate whether the observed epileptiform waves were consistent over time, we applied automatic detection algorithms –generally used for detecting interictal discharges(*56*) – to all channels and generated a raster plot of the activity as a function of time (Fig. 5D, Fig. 5F). After sorting the channels according to the distance from the lower right corner, we found a 20 s window with repetitive epileptiform events occurring approximately every 4 s across the entire grid (Fig. 5F), which is magnified to show a single event of a recurring epileptiform waveform within a 4 s time window (Fig. 5D, Fig. 5E). These repeated epileptiform waves consistently originated from the lower right corner of the grid, spread across the entire grid within one second, and subsequently disappeared one second later. Automatically detected epileptiform events (Fig. 5E) exhibit clear temporal shifts between the spontaneous epileptiform waveforms from channel to channel (overlayed on a reconstructed brain model of the participant in Supplementary Movie S6). Collectively, these maps and videos provide further evidence of utility for the PtNRGrids for unprecedented large-scale high-resolution mapping of such pathologic activity, and its display in a meaningful and potentially medically informative way if displayed in real-time during the surgical procedure.

Active mapping was performed in the same patient (Fig. 5G and Fig. 5I). Utilizing the same epileptiform discharge detection algorithm as Fig. 5D, we counted the number of channels on the PtNRGrid that detected epileptiform discharges and plotted the number of events as a function of time (Fig. 5H). Spontaneous and repetitive epileptiform activity persisted across the entire recording. We observed that stimulation pulses with short duration (< 1 s) did not alter the underlying spontaneous epileptiform discharge activity. However, longer stimulation trials of 1.4 - 1.9 s increased the frequency of epileptiform discharge activity, particularly those after-discharges with waveform characteristics similar to interictal discharges. The duration of the pulse determines the total delivered charge density that is correlated with evoked responses(*57, 58*). The spatially resolved heatmaps of the stimulation evoked activity can be clearly observed where it can be noted that the longer trials (4 and 6) were characterized by significantly enhanced after-discharges detected around the stimulation positions ($p < .001$) (Fig. 5I).

In stimulation-evoked epileptiform discharges, we found a clear enhancement of the amplitude of detected epileptiform activity within the 10-59 Hz frequency window which persisted for more than 5 s after the bipolar stimulation ceased (Fig. 5J; see also Supplementary Movie S7). By sorting the automatically detected epileptiform events according to distance from the stimulation center (Fig. 5L), we observed that the first epileptiform events were initiated near the stimulated region. Following stimulation, bursts of epileptiform discharges occurred every 0.3 s and continued for a longer duration than the less frequent spontaneous epileptiform discharges. These phenomena are clearly exhibited in the raw waveform recordings from selected channels (Fig. 5L, Fig. 5M), and can be viewed atop a reconstructed brain model of the participant (Supplementary Movie S8).

Finally, we investigated the origin and spatiotemporal dynamics of both the spontaneous and stimulation-induced epileptiform activity using vector fields and streamlines. Immediately before epileptiform events, the vector fields are mostly incoherent (Figs. S35-S36), but become coherent near the larger amplitude epileptiform events. The characteristics of the spontaneous epileptiform activity can be inferred from the red streamlines that originate from the right lower corner near the location of the lesion in this patient (Fig. 5C). In contrast, the streamlines for stimulation-evoked epileptiform activity originate and spread away from the position of the stimulator (Fig. 5K), demonstrating high resolution spatial and temporal mapping of the sites of origin. The epileptogenic tissue tested by this experiment was removed as a planned left temporal lobectomy which included lesional (cavernoma) tissue and all the epileptic neocortical tissue discerned by prior sEEG seizure mapping and deemed resectable as well by the PtNRGrid. The patient remained seizure free to the date of this manuscript synthesis, which is approximately 6 months after surgery.

## DISCUSSION

Our studies demonstrate the range of utility of PtNRGrids for high spatial and temporal recording of neural activity for research and clinical intraoperative use. The PtNRGrids were built on thin, transparent, and conformal substrates and were reconfigured in pitch and total cortical coverage with 1024 and 2048 low impedance contacts over an area as large as 8 × 8 $cm^2$, scalable for rodent or human work. The fabrication was performed on a large-area 180 mm × 180 mm glass substrate with thin film processes significantly advancing the manufacture of neural probes beyond the conventional 100 mm and 150 mm Si substrates. The larger area manufacturing afforded the capability to connect to thousands of channels and the formation of long metal strips to isolate the sterile surgical medium from the acquisition electronics for patient safety. Additionally, the fabrication process afforded large area coverage of the PtNRGrids on the brain (up to 8 cm × 8 cm achieved in this work) and safety and sterility in the operating room for intraoperative monitoring purposes. Successful transition to the large-area glass substrates opens the possibility of integrating the display panel manufacturing technology with neurotechnology and indicates excellent scalability considering the size of the glass panel used in display industry (3,100 mm × 2,900 mm). The large area glass substrates also offer potential advances in manufacturing biomedical devices for use in humans which can leverage electronic (thin film transistors) and optoelectronic (light emitting diodes and imagers) advances achieved by the display industry for utility in human biomedical devices.

We show high-fidelity broadband recordings from rodents, where the high spatial resolution of the PtNRGrids enabled identification of individual cortical columns from the surface of the brain. We report the first human recordings with 1024 channel PtNRGrids from 19 subjects, and 2048 channel PtNRGrids from one subject (Figs. S37-S38). In the clinical setting, the PtNRGrids were easily integrated within the operating room and allowed sensory mapping and pathology epileptiform activity from the surface of the brain, and detected novel somatosensory dynamics.

PtNRGrids hold promise for superior mapping during neurosurgical intervention through high spatial resolution and coverage while maintaining excellent broadband temporal resolution compared to the clinical electrodes. PtNRGrid technology has the capacity to scale to significantly more than 2048 channels and to pave the way for a new era of enhanced neurosurgical mapping strategies, to enable new possibilities for therapies, brain-computer interfaces, and better patient outcomes as the technology is advanced for chronic applications.

**LIMITATIONS**

We demonstrated the first recordings from the surface of the human brain with novel 1024 and 2048 channel PtNRGrids, which enables high resolution intraoperative recordings in neurosurgical operations, estimated to be as many as 13.8M annually.(*59*) However, the system in its current form does not enable chronic recordings, which have different design considerations that are beyond the scope of this work. To achieve chronic recordings, either denser custom-made connectors that are slim enough to be externalized through the scalp (Fig. S39) – as done in epilepsy monitoring – or direct integration of integrated circuits and wireless transceivers to the implant become necessary.(*25, 60*) The engineering challenges for each part of the implant significantly increase and their safety as well as the durability of efficacy become critical.

We demonstrated recording density up to 4,444 $cm^{-2}$ (contact pitch: 150 μm) using a single metallization layer of 10-μm-pitch metal leads and 30-μm-diameter PtNR contacts. To increase the recording density, the width of the metal leads would need to decrease to enable tighter pitch and thickness would need to increase to maintain low metal lead impedance. There are practical lithographic limitations for patterning narrow and thick metal leads necessitating the use of multi-layer metallization which relaxes the constrains on the metal lead width. By making dual and triple layer electrodes, the electrode density can increase up to 17,776 $cm^{-2}$ (contact pitch: 75 μm) and 71,104 $cm^{-2}$ (contact pitch: 37.5 μm), respectively. Higher recording densities beyond 71,104 $cm^{-2}$ would require a smaller PtNR contact diameter. A trade-off between the recording density and the impedance magnitude of the individual contacts will need to be considered.

While the higher resolution mapping can carry significant implications for neurosurgical procedures, it is important to note that the surgical precision in current clinical practice does not meet mm-resolution. However, surgical resection boundaries obtained with current clinical electrodes are grossly determined with their 1 cm contact spacing. We anticipate that the PtNRGrids that determine surgical boundaries with mm-resolution can inform significantly better resection practices by delineating the curvilinear nature of the functional and pathological boundaries that is not possible otherwise. With the development of higher-precision resection methods such as laser ablation or robot-assisted surgery, mm-scale spatial resolution recording will be useful for performing mm-scale resection of the brain tissue.

Lastly, the PtNRGrids were used for passive mapping in this work. Although the PtNRGrid has perfusion holes that enabled bipolar stimulation with external devices, direct current stimulation through the grid is desired. Additionally, micro-stimulation of smaller tissue volume may be preferred, particularly for individual functional cortical columns that were isolated using the PtNRGrids. Extension of this work should enable stimulation through the PtNRGrids which were shown in our earlier work to hold one of the highest charge injection capacities for safe stimulation.(*27*)

## MATERIALS AND METHODS

### Study design

Objectives of the study were to: (i) demonstrate that 1024 and 2048-channel PtNRGrids have higher spatial resolution that better delineated functional boundaries in the human brain than the standard clinical electrodes; (ii) test the signal amplitudes of PtNRGrids compared to the standard clinical electrodes; (iii) isolate the neural correlates and dynamic activity during motor and sensory tasks with high spatial resolution with the PtNRGrids; and (iv) investigate the microscale dynamics of spontaneous and stimulation-evoked epileptic discharges with the PtNRGrids.

The subjects were undergoing awake or asleep craniotomy for resection of brain tumor or epileptogenic brain tissue. We selected the types of PtNRGrids and tasks depending on the location and size of the craniotomy determined by the clinical team. Twenty subjects were involved in the study. The data collected from subjects with severe brain tumor, low responsiveness due to age, anesthesia, or surgical procedures, or task-unrelated cortical exposure were excluded from the study. Experimental conditions and brief summaries of all subjects are presented in Table S1.

**Experimental methods are enclosed in detail in the Supporting Information.** These include the fabrication of reliable fine metal traces on parylene C on 7” × 7” × 0.06” glass plates, the formation of PtNRs with dealloying a PtAg alloy, the formation of via and perfusion holes and creation of electrode outline by dry etching. The electrodes were bonded to an extender printed circuit board and connected to custom acquisition electronics that interface with commercial Intan 1024 channel electrophysiology recording systems. PtNRGrids were characterized for noise and cross-talk and mechanical bending cycles. Detailed surgical procedures are also included in the supporting information for rat whisker barrel implantation and recording as well as details for histology. Analysis of propagating beta waves and high gamma activity are discussed. A Table that provides information about human participants and recording tasks are also discussed. Packaging and sterilization of the electrodes and the assembly and testing of stimulation capture system for use in the OR is presented. Signal processing for spatial mapping and channel selection and noise filtering is discussed as well as the localization of the neural responses in rat and in human experiments. Processing of the recorded data for epilepsy is also presented in detail. Comparison of PtNRGrid with other recent ECoG grid technologies published recently while this manuscript was under review is presented side-by-side in Table S2. The fabrication approach described in this work was compatible with other low-impedance organic electrochemical interface material such as poly-(3,4-ethylenedioxythiophene)-poly (styrenesulfonate) (PEDOT:PSS) (Figs. S40-41). The paired-sample t-test was used to analyze the spike rate differences between spontaneous and stim-evoked epileptiform activities.

**Supplementary Materials**

Materials and Methods

Table S1-S2, Fig S1-S41

References 1-8, Movie 1-8

**Acknowledgements**

The authors acknowledge insightful discussions with Prof. Eric Halgren of UC San Diego and Prof. Sydney Cash of Massachusetts General Hospital, and Prof. Anna Devor and Prof. Martin Thunemann of Boston University. The authors are grateful for the technical support from the nano3 cleanroom facilities at UC San Diego's Qualcomm Institute where the PtNRGrid fabrication was conducted. This work was performed in part at the San Diego Nanotechnology Infrastructure (SDNI) of UC San Diego, a member of the National Nanotechnology Coordinated Infrastructure, which is supported by the National Science Foundation (Grant ECCS1542148).

**Funding**

This work was supported by the National Institutes of Health Award No. NBIB DP2-EB029757 and NIDA R01-DA050159 and the National Science Foundation (NSF) Awards No. 1728497 and CAREER No. 1351980 and an NSF Graduate Research Fellowship Program No. DGE-1650112 to A.M.B. Any opinions, findings, and conclusions or recommendations expressed in this material are those of the author(s) and do not necessarily reflect the views of the funding agencies.

**Author contributions:** S.A.D. conceived and led the project. Y.T. fabricated the PtNRGrids, designed the tasks, and conducted all data analysis with S.A.D.'s guidance. A.M.B. designed the custom-made acquisition system and the motor task and stim-capture hardware with input from I.G. and participated in data analysis; D.R.C., B.S., D.S., A.C.P., J.Y., S.B-H. and A.M.R. performed tests and guided the electrode design for clinical translation. Y.T., A.M.B., K.L., K.J.T. performed the rat experiments and K.J.T. performed the histology. B.S., E.B, and A.M.R. designed clinical experiments and performed OR recordings; D.R.C., D.S., Y.T., A.M.B., and S.R. participated in some of the OR recordings. J.L. fabricated PtNRGrids used in the sensory/motor tasks and Y.G.R., and M.G. contributed to the fabrication process development. H.O. performed the bending cycle tests. A.C.P. and Y.T. composed the videos on the reconstructed brain models of subjects from fMRI and structural MRI. S.A.D. and Y.T. wrote the manuscript and all authors discussed the results and contributed to the manuscript writing.

**Competing Interests:** The authors declare the following competing interests. Y.T., A.M.R. and S.A.D. have equity in Precision Neurotek Inc. that is co-founded by the team to commercialize PtNRGrids for intraoperative mapping. S.A.D. and H.O. have competing interests not related to this work including equity in FeelTheTouch LLC. S.A.D. was a paid consultant to MaXentric Technologies. D.R.C., K.J.T, and D.A.S. have equity in Surgical Simulations, LLC. A.M.R. has an equity and is a cofounder of CerebroAI. AMR received consulting fees from Abbott Inc and Biotronik Inc.

**Data and materials availability:** Correspondence and requests for materials should be addressed to S.A.D. All data is in the paper and supplementary materials are available by contacting the corresponding author with reasonable requests while honoring patient, institutional, and funding agency guidelines.

# Supplementary Materials for Human Brain Mapping with Multi-Thousand Channel PtNRGrids Resolves Novel Spatiotemporal Dynamics

Youngbin Tchoe[1†], Andrew M. Bourhis[1†], Daniel R. Cleary[1,2†], Brittany Stedelin[3], Jihwan Lee[1], Karen J. Tonsfeldt[1,4], Erik C. Brown[3], Dominic Siler[3], Angelique C. Paulk[5], Jimmy C. Yang[5,6], Hongseok Oh[1], Yun Goo Ro[1], Keundong Lee[1], Samantha Russman[1], Mehran Ganji[1], Ian Galton[1], Sharona Ben-Haim[1,2], Ahmed M. Raslan[3], and Shadi A. Dayeh[1,2,7*]

[1]Integrated Electronics and Biointerfaces Laboratory, Department of Electrical and Computer Engineering, University of California San Diego, La Jolla, California 92093, United States
[2]Department of Neurological Surgery, University of California San Diego, La Jolla, California 92093, United States
[3]Department of Neurological Surgery, Oregon Health & Science University, Mail code CH8N, 3303 SW Bond Avenue, Portland, Oregon 97239- 3098, United States
[4]Department of Obstetrics, Gynecology, and Reproductive Sciences, Center for Reproductive Science and Medicine, University of California San Diego, La Jolla, California 92093, United States
[5]Department of Neurology, Massachusetts General Hospital, Boston, Massachusetts 02114, United States
[6]Department of Neurosurgery, Massachusetts General Hospital, Boston, Massachusetts 02114, United States
[7]Graduate Program of Materials Science and Engineering, University of California San Diego, La Jolla, California 92093, United States

* Corresponding author. Email: sdayeh@eng.ucsd.edu
† These authors contributed equally to this work

**This file includes Supplementary Methods, Supplementary Figures, and Supplementary References.**

## 1. Fabrication, Packaging, and Connectorization of PtNRGrids

### 1.1. Fabrication of Reliable Fine Metal Traces on Parylene C Films

Polished and cleaned photomask-grade soda lime glass plate (Nanofilm) with dimensions of 7” × 7” × 0.06” were used as substrates for the fabrication. Prior to coating Micro-90 and parylene C films on the glass, the surface was treated with 200 W oxygen plasma for 5 min (Plasma Etch, Inc. PE100) to make the surface hydrophilic. We then spin-coated a 0.1% diluted and filtered Micro90 layer on the glass substrate as a release layer for the parylene C film in the last step of the fabrication process. Then, a 3.7-μm-thick-parylene C layer was coated on the vertically loaded glass substates using a parylene deposition system (Specialty Coating Systems 2010 Labcoter) Metal leads with width, spacing, and length of 4 μm, 6 μm, and > 10 cm respectively were formed on the parylene C layers by a standard lithography, descum, metal deposition, and lift-off process using the AZ5214E-IR photoresist (MicroChemicals), maskless photolithography system (Heidelberg MLA150), UV flood exposure system (DYMAX), plasma etcher (Oxford Plasmalab 80), and e-beam evaporator (Temescal). Metal leads were composed of Cr/Au (10/250 nm) and the entire lithography, deposition, and lift-off process was repeated on top of the first metal lead layer to form Cr/Au/Cr/Au (10/250/10/250 nm) metal leads. This double-patterning process was employed to increase yield and reduce risk of photoresist particles from compromising the thin traces.

### 1.2. PtAg Alloy Deposition for PtNRs

After the metal leads were prepared a 30μm-diameter-PtAg alloy was formed on the individual recording sites by photolithography, descum, and PtAg alloy co-sputtering using the maskless photolithography system with NR9-6000 (Futurrex) photoresist, plasma etcher, and magnetron DC/RF sputter (Denton Discovery 18), respectively. The detailed fabrication methods and characteristics of PtNRs can be found elsewhere (*1*). Notably, this process involves a selective etching of silver in a dealloying process, leaving behind non-toxic platinum. A 50-nm-thick Ti capping layer was deposited on top of PtAg alloys to prevent oxidation in air or under oxygen plasma (see Fig. S1A).

### 1.3. Via Holes, Perfusion Holes, and Electrode Outline Etching

A second parylene C layer (3.1 μm) was then conformally deposited, preceded by an oxygen plasma treatment to enhance the adhesion between the layers. On top of the second parylene C layer, a Ti (50nm) hard mask was deposited and AZ5214E-IR photoresist was coated on top of the Ti layer and patterned to open via holes for recording sites, perfusion holes, and the outline of the electrodes (see Fig. S1B). An $SF_6$/Ar reactive ion etcher (Oxford

Plasmalab80) was used to pattern the Ti layer through the photoresist, and 200 W oxygen plasma was used to etch parylene C films all the way through the surface of the glass substrate or Ti-cap layers of PtAg. The outline of the electrode was defined by a dry etching process to produce much more clearly defined sidewalls compared to those defined by a laser cutter, which usually yielded rough sidewalls and left black carbon microparticles. After completing the etching process, the substates were hard baked at 150°C for 40min to release the stress built up in the parylene C layers. Ti hard masks on parylene C and PtAg alloys were removed by dipping the sample in 6:1 buffered oxide etchant (BOE) and rinsing in DI water. The electrodes were lifted-off from the substrate in DI water with the dissolution of the underlying Micro90 layer. The released electrodes were then dealloyed on the surface of 60°C nitric acid for 2min with the PtAg alloy facing downward, thereby forming the PtNRs, which were then immersed and rinsed in DI water (see Fig. S1C). The microscopic morphology of the electrode is shown by the SEM images in Fig. S2.

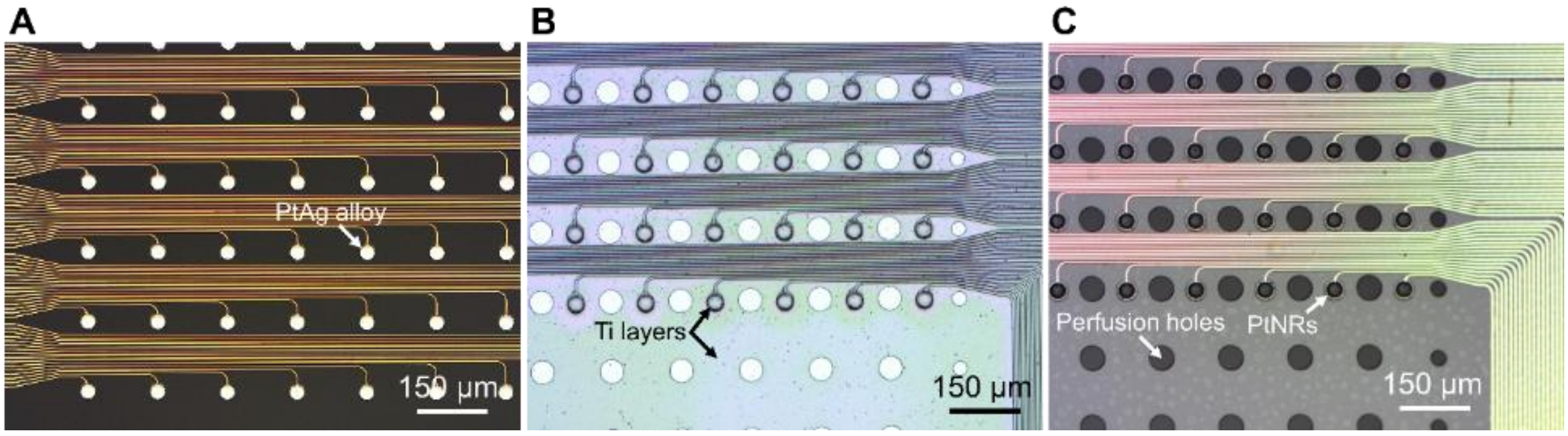

**Figure S1.** Microscope images during the fabrication processes. (A) Metal leads and PtAg alloy capped with Ti hard mask on 1st parylene C layer. (B) After 2nd parylene C deposition and Ti hard mask deposition. Photoresist is patterned for via hole etching. (C) After Ti hard masks removal and dealloying.

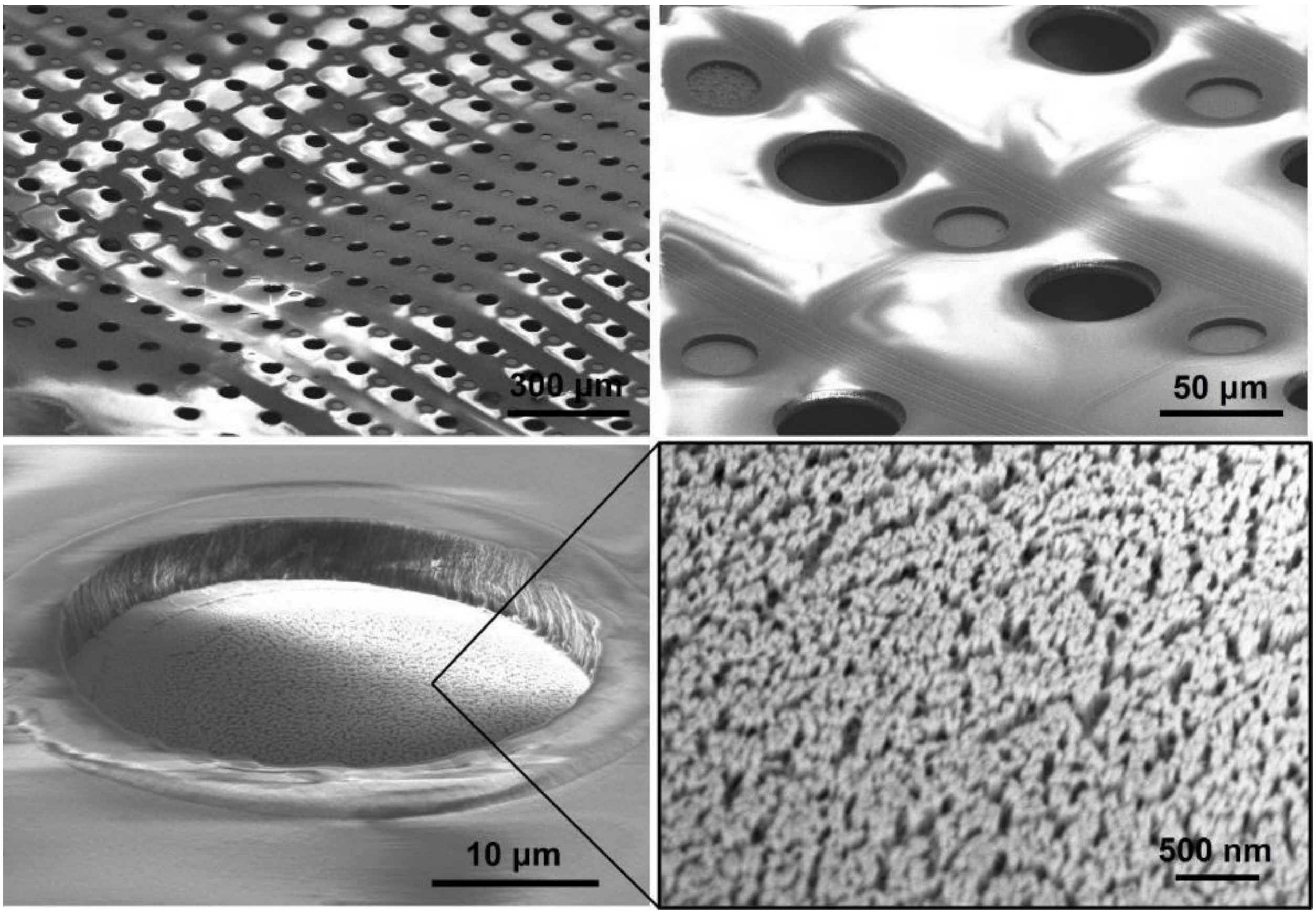

**Figure S2.** SEM images at different magnifications of the PtNRGrid and contacts.

1.4. Effect of Perfusion Holes in Achieving Highly Intimate Contact to the Brain Tissue

As the channel count and the area of the electrode scaled up to 3.2 cm × 3.2 cm, adherence of the electrode to the brain tissue was not fully preserved throughout the electrode area. Although the electrode was as thin as 6.6 µm, pockets of cerebrospinal fluid (CSF) as well as air bubbles were frequently trapped between the tissue and the electrode (Fig. S3F). These issues degraded the reproducibility of the recording due to the formation of dead zones (trapped CSF and possibly air bubbles). With the introduction of the perfusion holes, air bubbles or trapped CSF were not observed underneath the PtNRGrids in the OR (Fig. S3G), and the reproducibility of the human brain recording was greatly enhanced. We performed a benchtop experiment to demonstrate the effect of perfusion holes on adherence to the 3D objects (Figs. S3A-E). PtNRGrids with and without perfusion holes were placed on the red-colored gelatin droplet/PDMS mold and gently tapped with a gauze in the same manner. The PtNRGrid without the perfusion holes had trapped solution and air bubbles underneath it (Fig. S3B), and the cross-sectional image revealed that the separation between the electrode and PDMS can be as large as 200 µm (Fig. S3D). On the other hand, the PtNRGrids with the perfusion holes excellently conformed to the 3D PDMS mold (Fig. S3B), and the high proximity between the PtNRGrid and PDMS mold made it impossible to measure the thickness of the gap under the resolution of the cross-sectional microscope image shown in Figure S3E.

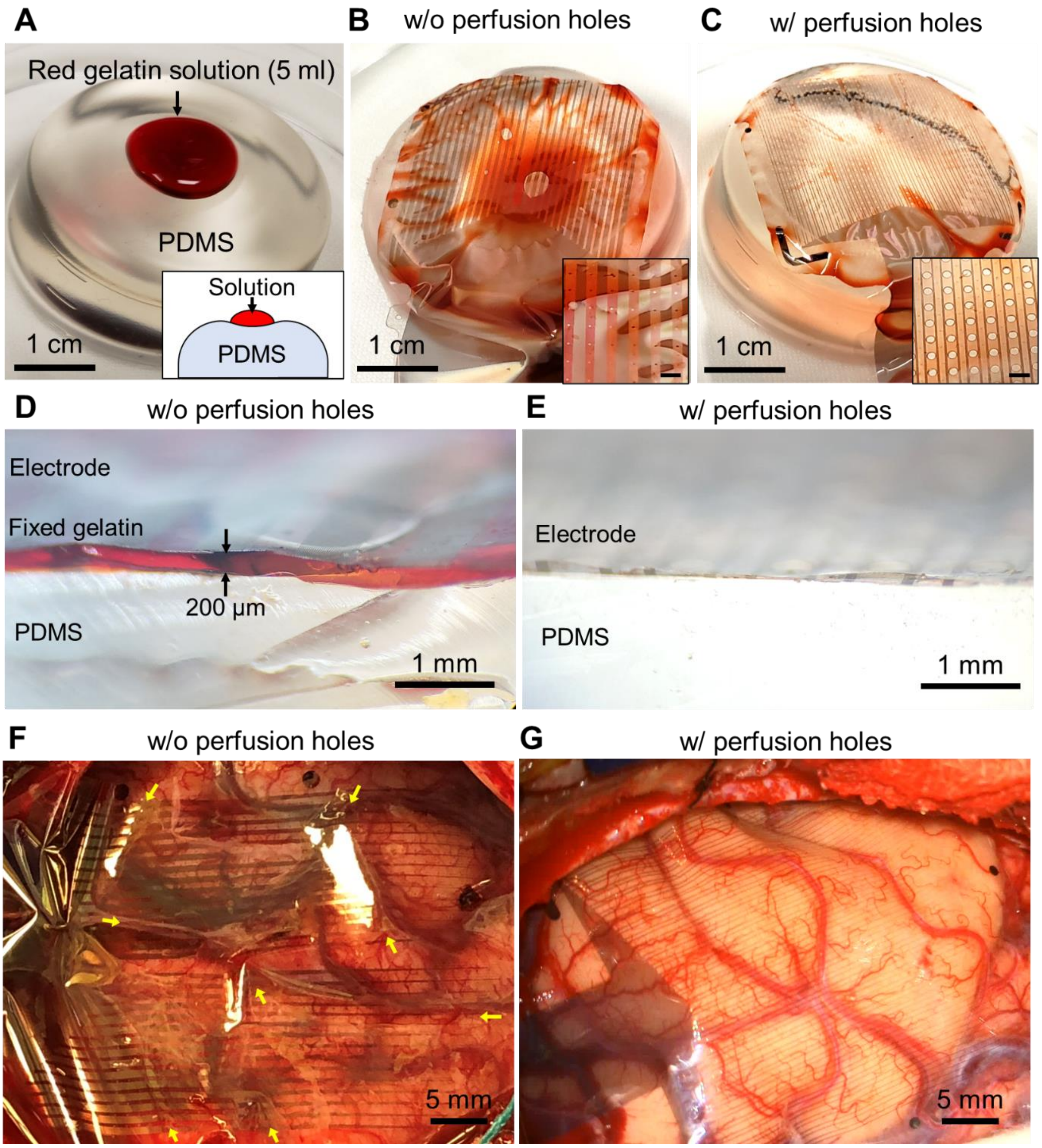

**Figure S3.** Effect of perfusion hole in establishing an intimate contact to the 3D objects. (A) Red gelatin solution droplet close to the human body temperature on top of the PDMS tissue model. The top surface of the PDMS model was a little recessed as depicted in the inset cross-sectional schematic. PtNRGrids (B) without and (C) with perfusion holes were placed on top of the red gelatin droplet/PDMS block and gently tapped with a gauze in the same manner. Inset figures in (B) and (C) shows the magnified photos with 1 mm scalebars. Cross-sectional images of the liquid (fixed red gelatin) trapped between the PDMS tissue model and the PtNRGrids (D) without and (E) with perfusion holes. PtNRGrids (F) without and (G) with perfusion holes on top of human brain surface. The positions where CSF is trapped underneath the electrode is marked with yellow arrows.

1.5. Bonding the Electrode to the Extender Printed Circuit Board (PCB)

The electrodes were then bonded on an extender PCB using silver epoxy (MG Chemicals 8331). Silver epoxy wasselectively deposited on the footprints of an extender PCB through a 50-μm-thick silicone adhesive PET tape (Advanced Polymer Tape) stencil mask where holes were defined using a laser cutter (Universal Laser Cutter, VLS 3.50). The amount of silver epoxy that was required for reliable bonding was optimized by adjusting the size of the holes in the stencil mask. Micro-alignment stages with four-axis degrees of freedom were used to precisely align the PtNRGrid (that was temporarily placed on 5” × 5” glass plate) and the extender PCB (see Fig. S4A). Once aligned and placed in contact, the PtNRGrid and extender PCB were cured on 85°C hotplate for 15min under 5-10 N force to fully cure the silver epoxy and ensure electrical connection across all bonding contacts. The electrically bonded PtNRGrid was then released from the glass plate, and benchtop characterization of the electrodes was performed. Prior to the bonding, the PCB edge was grinded and smoothed to minimize the potential damage on the thin metal leads in parylene C films as shown in Fig. 4C. The bonding interface and the photos of the electrodes on both sides are shown in Figs. 4B, D, and E.

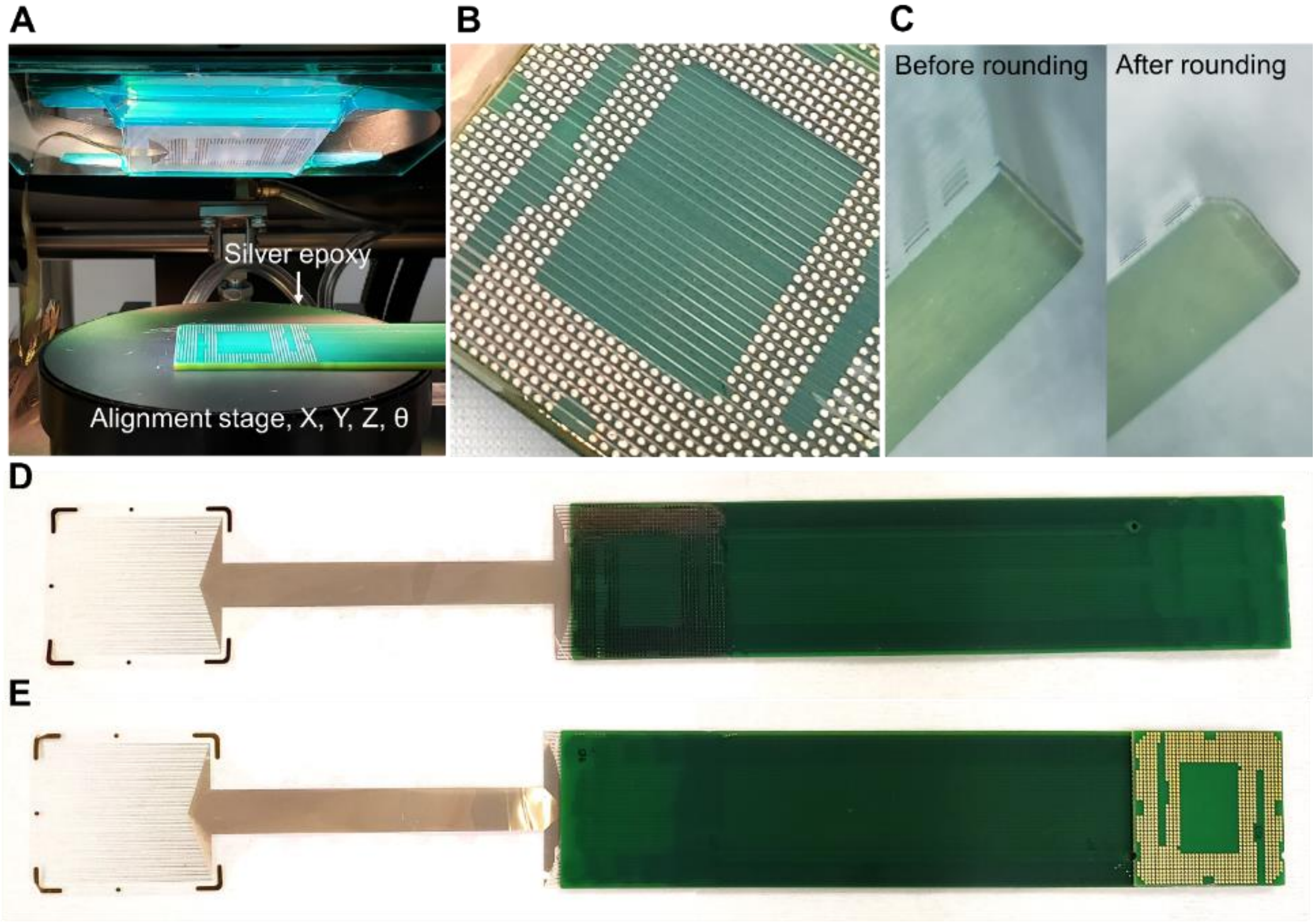

**Figure S4.** Bonding process of the flexible electrode on the rigid PCB. (A) Electrode and extender board placed on the alignment state. Silver epoxy bumps are deposited on the pads of the PCB side. (B) The bonding interface between the parylene C electrode and the PCB after curing the silver epoxy. (C) The PCB edge rounded to minimize potential damage of the metal leads during handling. Photos of the flexible electrode bonded on the PCB with PtNRs facing (D) downward and (E) upward.

1.6. Connectorization and Acquisition Electronics

We developed a custom acquisition board using Intan Technologies' RHD2164 chipset to simultaneously record from up to 1024 channels with up to 30 kilo-samples per second (ksps) per channel. These chips contain an AC coupled differential analog front end (AFE) which applies a programmable bandpass filter (typically from 0.09 Hz to 7.6 kHz) and samples these signals with a 16-bit ADC. The total data throughput of the system was roughly 2.4 GB/min when sampled at 20 ksps. Given the proposed applications of this platform for use in micro-electrocorticography, it is important to capture the full bandwidth of neurophysiological signals - from local field potentials to spiking and high gamma activity. To this end, our platinum nanorod microelectrodes exhibited an average 1 kHz impedance magnitude in the tens of kilo-ohms which resulted in a high signal-to-noise ratio (SNR) for this particular AFE.

While Intan Technologies offers an excellent platform for acquisition of ultra-high-resolution neurophysiology, they currently do not offer any scalable connectorization methods. It is known that the connectors are a major bottleneck in the effort to scale to thousands or hundreds of thousands of recording channels, thus we adapted connectorization techniques developed by the microelectronics industry. The first version of our acquisition board (denoted ORB1024 V1, see Fig. S5) used an off-the-shelf LGA1155 CPU socket to make an electrical connection to our sterilizable electrode arrays. Adopting this land grid array socket allowed us to temporarily and robustly connect to our flexible and sterilizable electrodes without having to integrate expensive acquisition chips into each electrode. The socket was placed in the center of the ORB1024 V1 to optimize routing and separate analog traces from digital. Additionally, a central cutout was made in the board to allow for back-side access to the center of the connector for in-vitro studies.

The ORB1024 V1 also includes a latching relay which selects between an on-electrode reference and an external needle reference. This latching relay is driven by a digital output from one of the RHD2164 chips which can be set through a software interface, allowing researchers to choose between these two options without physically touching the board (i.e. in an intraoperative setting, where the board is covered by a sterile drape). The board also includes a physical switch to select between shorting reference to ground or keeping them separate, which was often useful in animal and benchtop testing to optimize baseline noise performance prior to conducting experiments. Finally, touch-proof connectors are included to allow for clinical twisted-pair needle reference and ground electrodes to be connected, and an LED indicates power to the board.

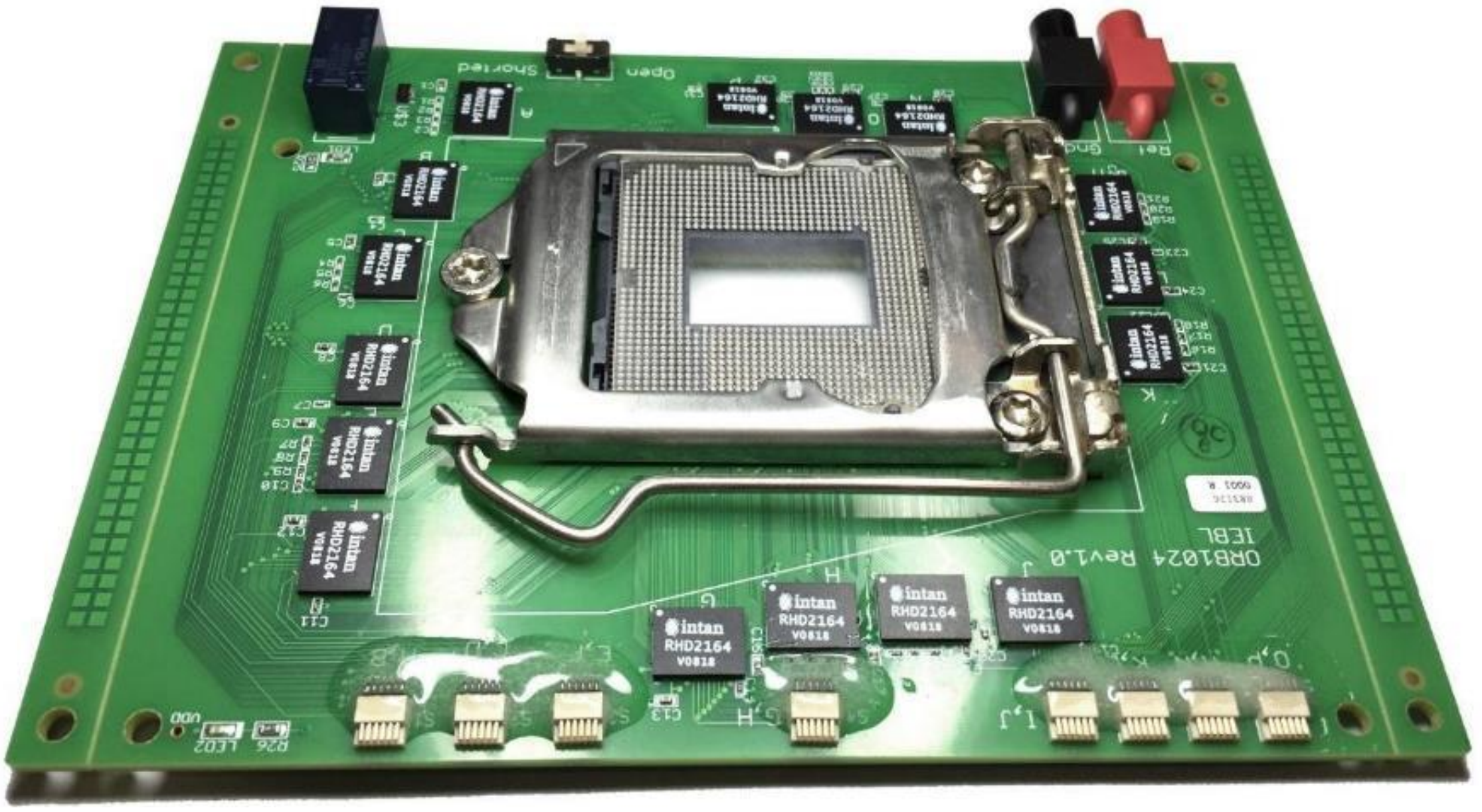

**Figure S5.** 1024 channels amplifier board populated with LGA1155 connector and Intan RHD2164 chips. 8 SPI cables connect the amplifier board to the recording controller.

## 2. Electrical and electrochemical characterization of PtNRGrids

### 2.1. Crosstalk and Noise of Ultra-High-Density Gold Interconnects

A commonly voiced concern over increasing the density of microelectrode arrays is the potential for electrical crosstalk to introduce artifacts into the neurophysiological recordings. This electrical crosstalk is primarily a result of parasitic capacitance between neighboring leads and thus will scale directly with increasing trace length and inversely with their trace pitch. Thus, traces should be kept short to reduce these parasitic paths. The termination impedance of neighboring channels to tissue (i.e. the electrolytic interfacial impedance) also needs to be accounted for, especially for conventional high-impedance electrochemical interfaces which can affect crosstalk through parasitic capacitance paths. However, this is not a concern for the low impedance PtNR contacts which maintain 1 kHz impedances that are at least 10 million times lower than the impedance of the parasitic capacitances. It should be noted that open channels on any grid including PtNRGrids can be problematic and these have been excluded (cut off of 100 kΩ) from our analysis.

It is exceedingly difficult to measure crosstalk directly in a benchtop experiment in part because there is inherent shunting between neighboring channels at the electrode-tissue interface. Furthermore, the electrochemical impedance acts as a terminating impedance for neighboring channels (i.e. victim channels), meaning that simply grounding these channels would not be an adequate method of replicating a real-world setting. Thus, to investigate the crosstalk in our electrodes, we performed several benchtop measurements to isolate these parasitic capacitances as well as the electrochemical impedances of the PtNR microelectrodes in saline to generate an electrical model for the signal pathways.

Below is a simplified circuit model detailing the relevant impedances of our PtNR microelectrode arrays. This model includes lumped models of each electrochemical impedance (denoted Z_ex), coplanar parasitic capacitance between neighboring channels (denoted Cpx), parallel-plate parasitic capacitance between the electrolytic solution and each metal trace (denoted Cs), and finally, the amplifier input impedance which is reported by Intan Technologies to be 12pF.

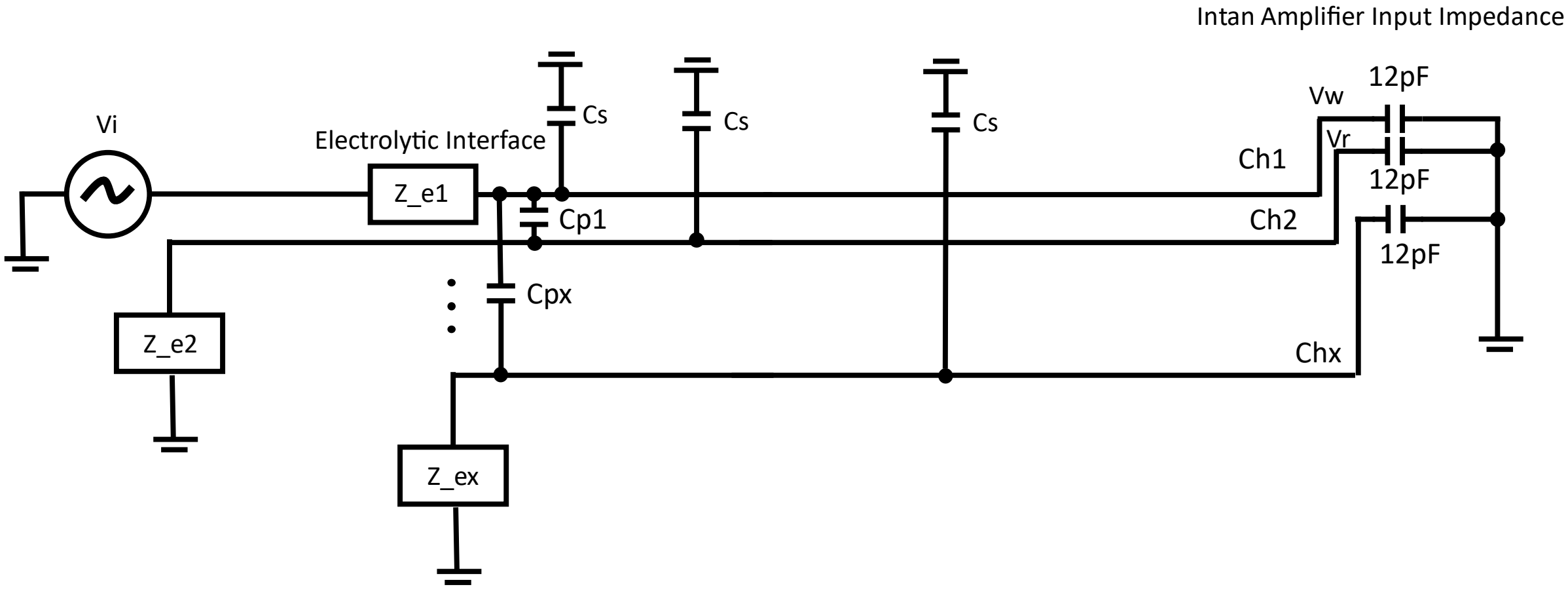

**Figure S6.** A simplified circuit model detailing the parasitic capacitances present in the high-density thin-film interconnects on our PtNRGrids. Here, we demonstrate the concept of an ideal isolated signal, Vi, and the paths this signal can take to couple on to neighb oring channels (i.e. crosstalk). It is insightful to view this system as a current divider, w here most of the leakage current in adjacent channels will be directed back through the l ow impedance electrolytic interfacial impedance (denoted Z_ex) rather than through the a mplifier.

We first measured the 1kHz electrochemical impedance of all 1024 channels on a freshly prepared microelectrode array and plotted the data spatially to find a region in the array with working neighboring channels whose leads ran adjacent to one another. Using a calibrated benchtop LCR meter, we then measured the impedance magnitude and phase (and subsequently extracted the equivalent parallel capacitance and resistance) over a frequency range of 1Hz to 1Mhz.

First, we measured the impedance between a platinum reference needle placed in phosphate buffered saline (PBS; Gibco® DPBS (1X) 14190-144), and the metal contact of the PCB bonded to our microelectrode array (which was also placed in PBS). Note that this measurement represented not only the electrolytic impedance, but also the trace impedance, contact resistances, and saline spreading resistance. We repeated this measurement for several neighboring electrodes and found similar measurements that matched well with our group's previously reported electrochemical impedance spectra of 30um diameter PtNR microelectrodes,(*1*) indicating that these resistive components were small compared to the primarily capacitive electrolytic impedance of the PtNR electrode.

Next, focusing on the same set of neighboring channels (whose leads ran parallel to one another), we measured the impedance between adjacent channels, channels separated by one, and channels separated by two. We repeated these measurements under three conditions: The electrode floating in air (infinite termination impedance), the electrode floating over grounded PBS, but face up so as to only form a parallel-plate capacitance with the metal traces without forming an electrolytic interfacial impedance, and finally, with an equivalent but un-bonded PCB in air. The measurements with the electrode floating in air represented the coplanar capacitances, Cp1, Cp2, and Cp3. The measurements with the electrode flipped over grounded PBS represented the parallel combination of the coplanar capacitance with the parallel-plate capacitance. Finally, the measurement of the un-bonded PCB was done to separate the parasitics of the PCB from the parasitics of the thin film metallization.

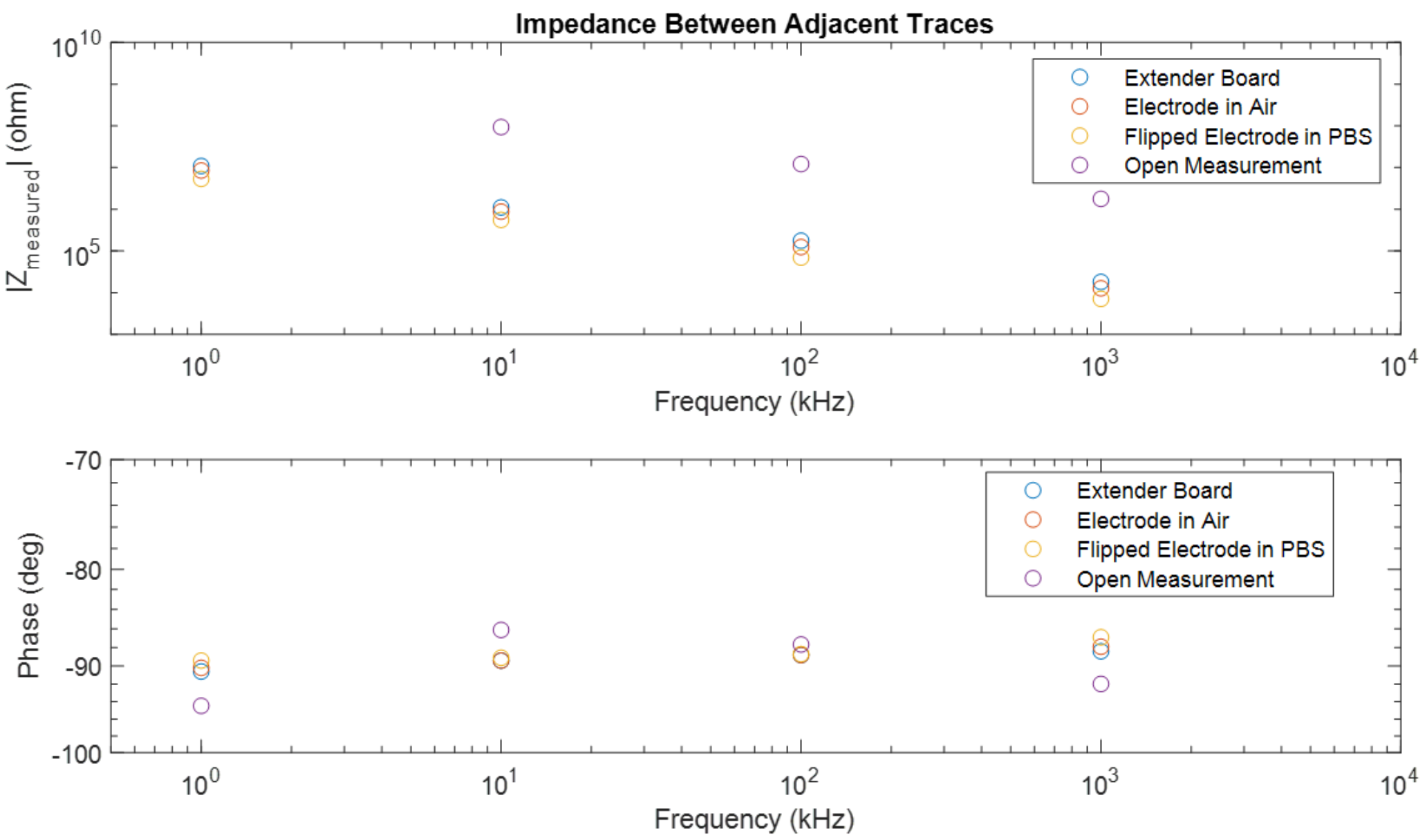

**Figure S7:** Magnitude and phase plots of the complex impedance between two adjacent channels in various measurement configurations showing capacitive characteristics. A system open measurement is shown as a reference. Values for co-planar capacitance, Cp, and parallel-plate capacitance, Cs, were extracted from the above measurements and used to generate bode plots for crosstalk (see Fig. S10).

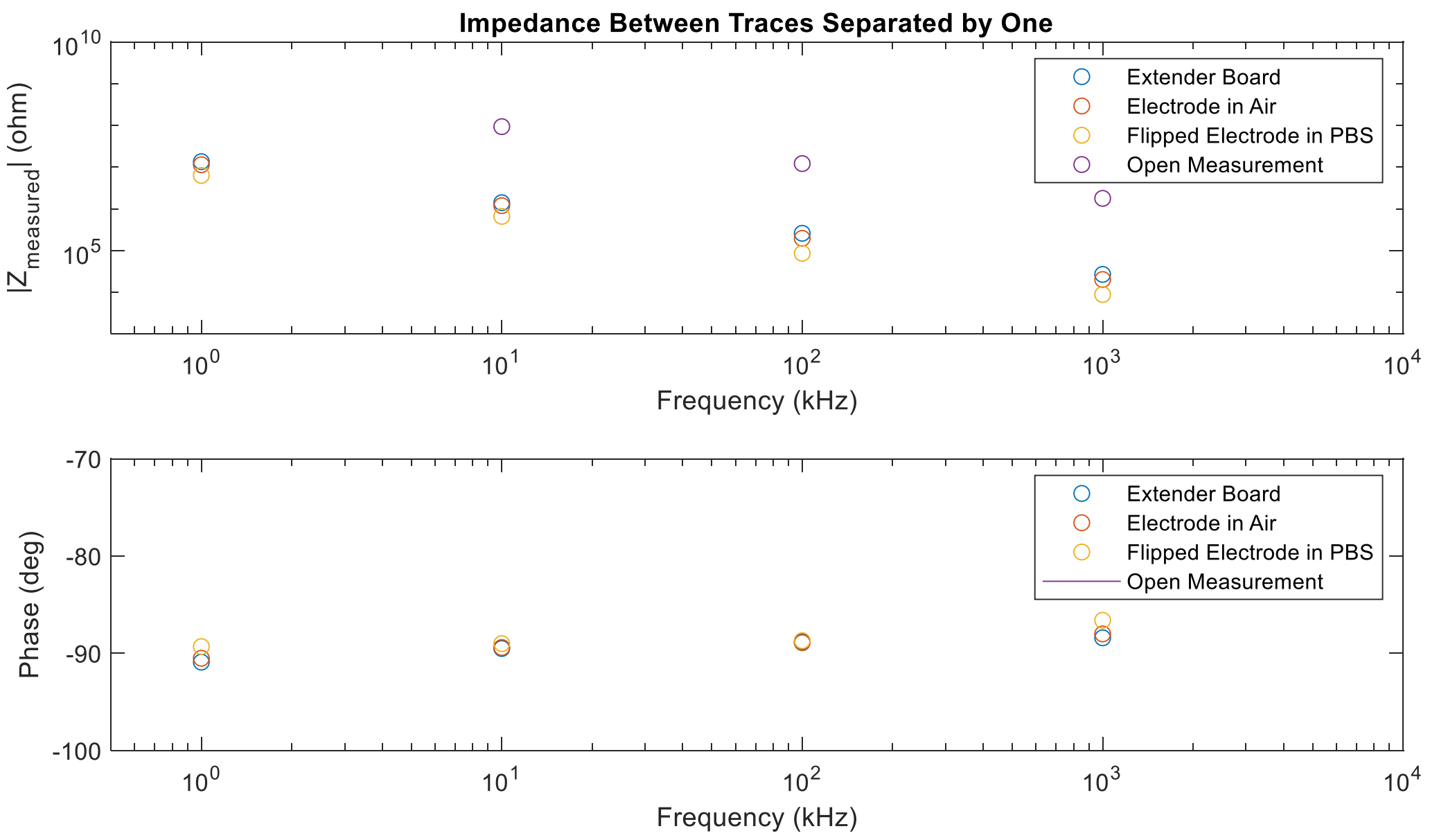

**Figure S8:** Magnitude and phase plots of the complex impedance between two channels separated by one (i.e. there is one channel between each channel) in various measurement

configurations showing capacitive characteristics. A system open measurement is shown as a reference.

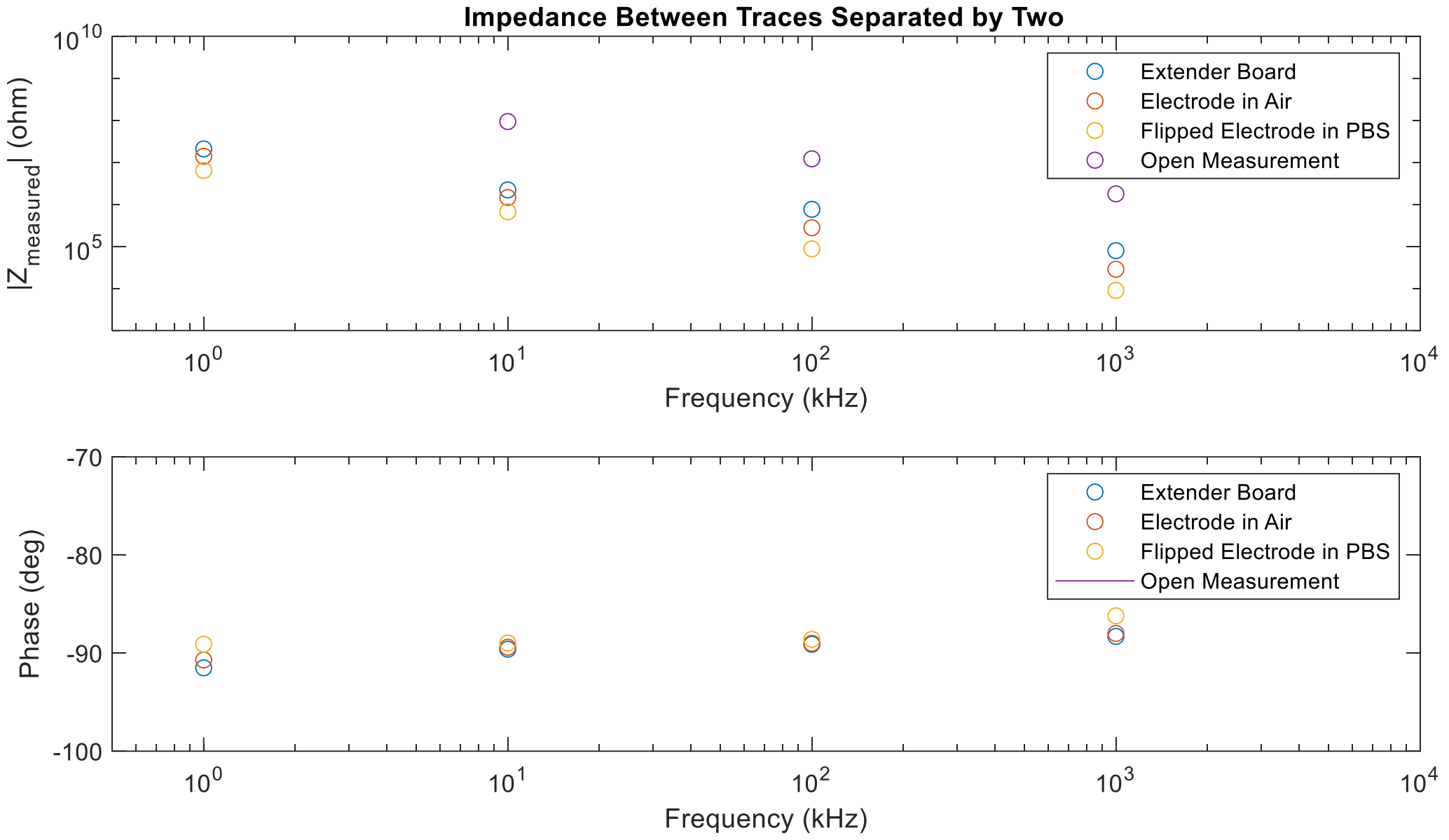

**Figure S9:** Magnitude and phase plots of the complex impedance between two channels separated by two in various measurement configurations showing capacitive characteristics. A system open measurement is shown as a reference.

From the above measurements and model, we then derived a transfer function under two conditions: first, with a working neighboring channel whose electrochemical impedance, Ze2, is on par with the mean across the array (Fig. S10), and second, with an open neighboring channel whose electrochemical impedance, Ze2, is several orders of magnitude higher than the mean across the array (Fig. S11).

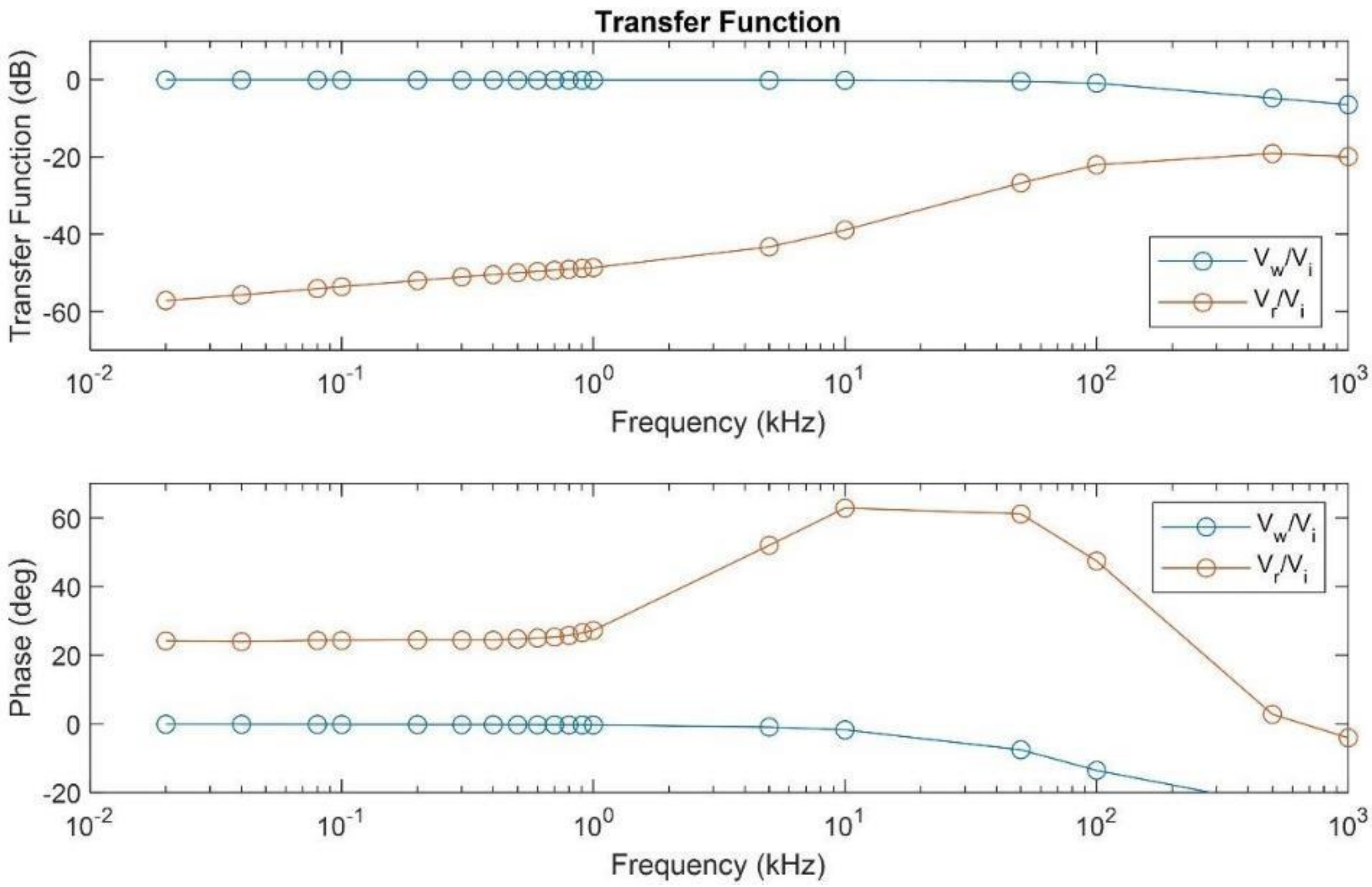

**Figure S10.** Transfer functions between input and output for a working channel and nearest neighboring channel. Note that the signal bandwidth of interest for neural signals is typically below 1 kHz, thus we can say that the PtNR microelectrodes introduce negligible signal attenuation in the signal path. Additionally, we see that neighboring channels will have near -60 dB crosstalk isolation at signals of interest. For a 500 µV amplitude signal, this corresponds to a 0.5 µV amplitude signal which is below the noise floor of the Intan acquisition system.

We found that the amplitude of an input signal in adjacent channels (denoted Vr) is several orders of magnitude lower than the working channel (Vw) amplitude so long as the adjacent channels have proper termination impedance at the electrolytic interface. Importantly, this analysis is consistent with observed baseline recordings which clearly demonstrate low crosstalk.

We also observed that in the limit as Ze2 → ∞, the adjacent channel sees a capacitive voltage divider between Cp, Ca, and Cs, resulting in the bode plot (Fig. S11).

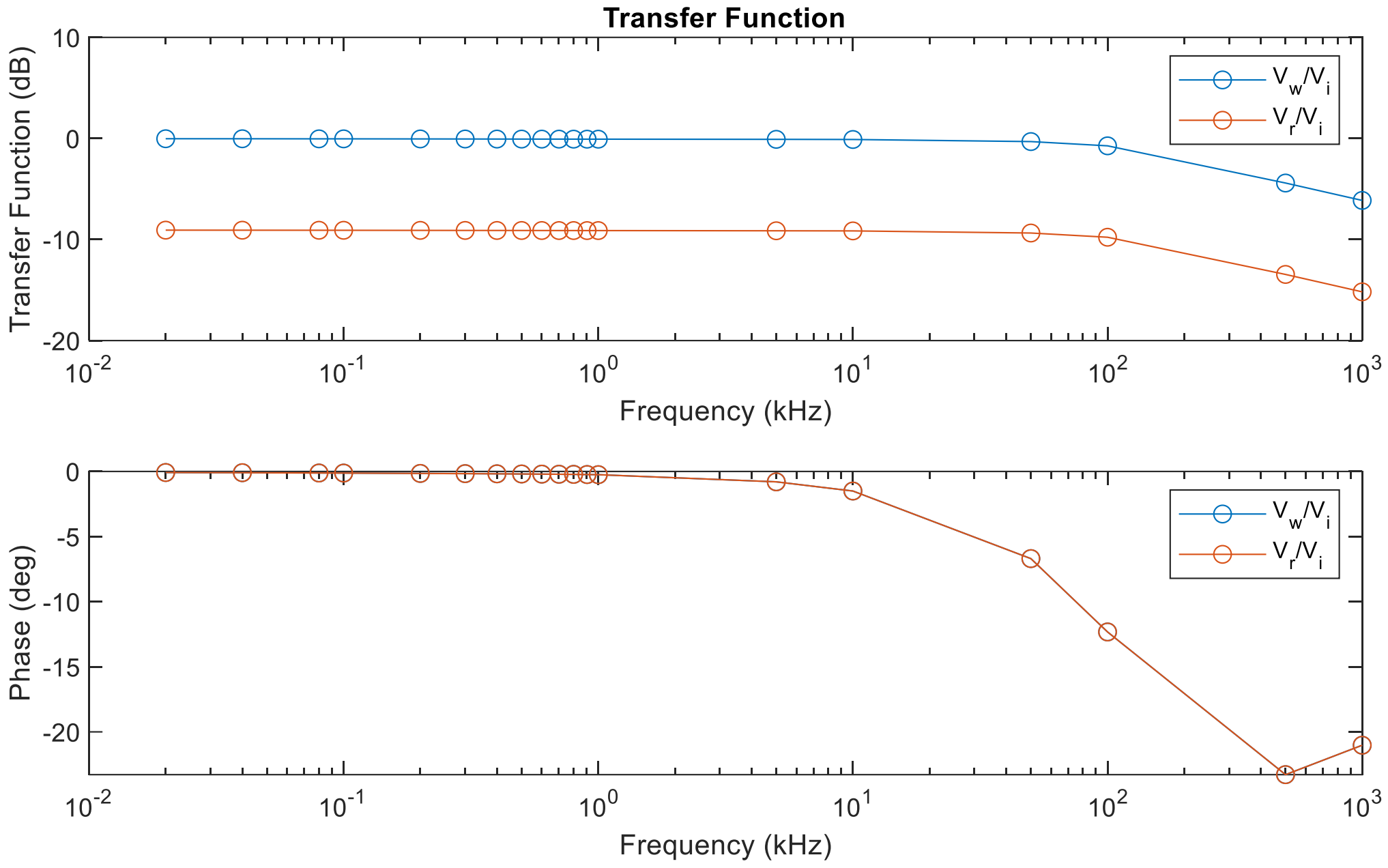

**Figure S11:** Bode plots showing the transfer functions between 1. an ideal voltage source, Vi, and the working channel intended to measure said voltage, and 2. An ideal voltage source, Vi, and the adjacent channel which has been electrically disconnected from the electrolytic solution or tissue (denoted Vr). Note that when an adjacent channel is disconnected, the only path for leakage current to flow is through its respective amplifier input impedance. Thus, we see significantly larger crosstalk in the open-channel configuration.

The above investigation shows that crosstalk is much worse when a channel has infinite electrochemical impedance. Of course this is intuitively obvious, because if a channel is not working (i.e. is disconnected from the electrophysiologic medium), it essentially becomes an antenna that will only pick up noise. A very similar situation arises when thin-film metal traces crack or are disconnected by particles during fabrication near the electrode site. Thus, when we say “proper termination impedance”, we essentially mean that the electrochemical impedance is non-infinite.

## 2.2. Bending Cycles Test

We performed 84,000 cycles of lead bend testing exceeding 90° bends (EN 45502) using a robotic gripper model 2F-140 (Robotiq) and custom made sample holders as shown in Fig. S12. We measured 755 functional contacts before bending (average impedance was 23±3

KΩ for 30µm diameter contacts before and after all bending tests). After bending, 759 contacts were functional. The same device was then subjected to 205,000 lead bend cycles of –40° to +40° bends (ANSI/AAMI CI86:2017, recommended –15° to +15°, 100,000 cycles). 752 contacts were functional after bending tests. The less than 1% change in number of functional channels is attributed to different contact latching between the device extender and the LGA socket on the acquisition board and is minimal indicating resilience of our electrodes to bending cycles.

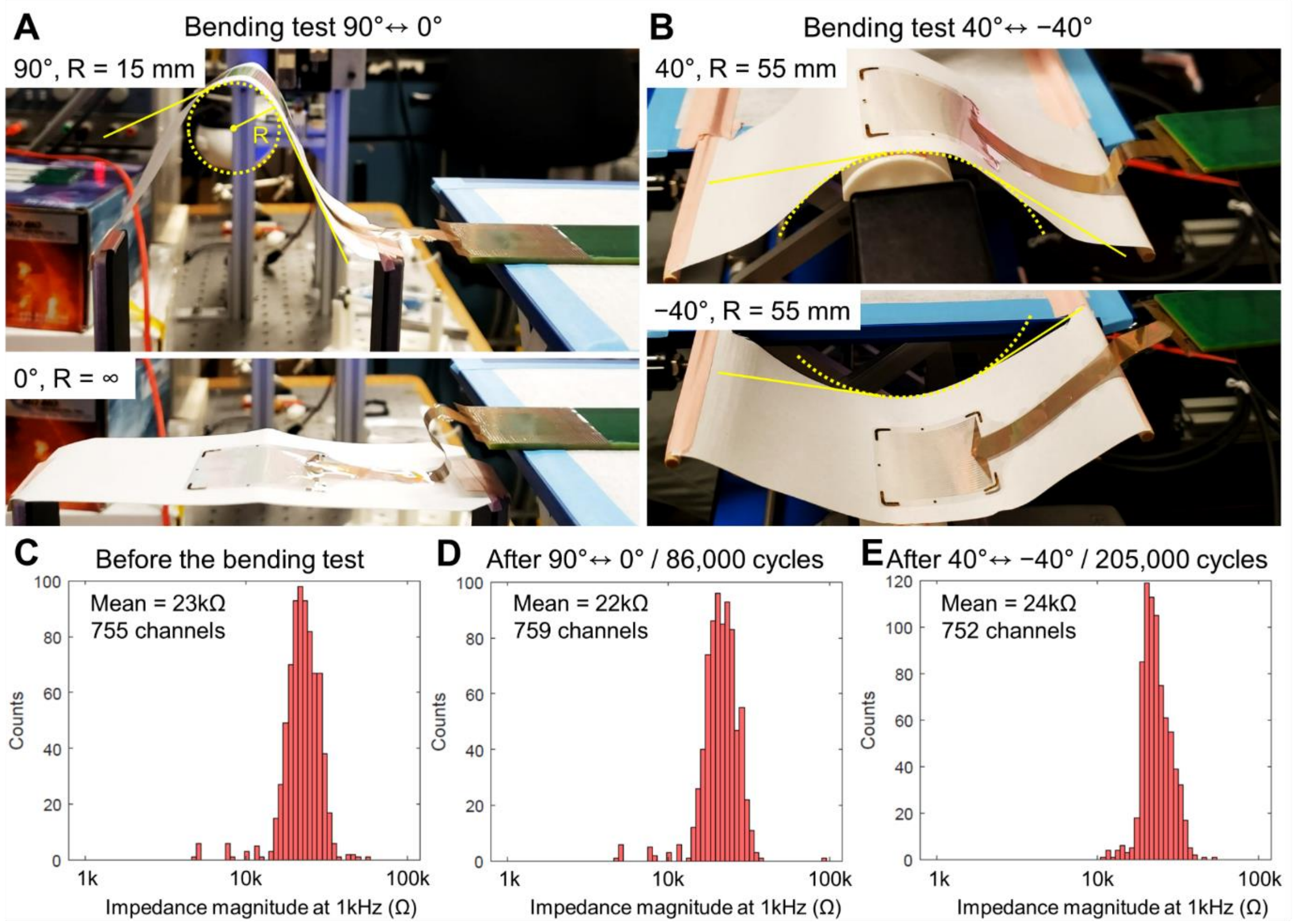

**Figure S12.** Bending cycles test. Photo of (A) 90°↔0° and (B) 40°↔–40° bending tests. Impedance histogram of the electrode (C) before, (D) after 86,000 cycles of 90°↔0° bending, and (E) after 205,000 cycles of 40°↔–40° bending.

## 3. Rat Whisker Barrel Recording

3.1. Surgical Procedures of Anesthetized Rat Craniotomy

All procedures were approved by the UCSD Institutional Animal Care and Use Committee. Rats (300–600 g) were sedated with 3-4% isoflurane and fixed in a stereotaxic frame (Kopf Instruments). Once stable, rats were reduced to 3% isoflurane for maintenance, while monitoring heart rate and breathing (Mouse Stat Jr, Kent Scientific). Prior to the craniotomy, contralateral side individual whiskers that were to be stimulated were colored with Sharpies to easily distinguish them in the air-puff stimulation experiment, and the remaining whiskers were trimmed off. A craniotomy was made on the right skull 1 cm lateral and 2 cm posterior from the bregma, exposing the somatosensory (including barrel) cortex, over the right motor cortex. The dura was carefully opened and retracted from the brain, and a small piece of saline-soaked gauze was placed over the brain until the implant was ready. Once the craniotomy was complete, the rat was transitioned from isoflurane to ketamine/xylazine (100 mg/kg ketamine / 10 mg/kg xylazine) and re-dosed every 20-30 min for the duration of the experiment. Temperature, heart rate, and oxygen concentrations were monitored for the entirety of the experiment to ensure adequate anesthesia.

### 3.2. Implantation of PtNRGrid in Rat Model Undergoing Sensory Stimulation

The typical size of the craniotomy was 6 × 6mm$^2$, and the 5 × 5mm$^2$ 1024 channel PtNRGrid was implanted covering nearly the entire exposed area of the brain. The reference needle electrode was typically implanted near the neck of the rat, and ground was typically connected to a surrounding faraday cage or stereotaxis frame. Individual whiskers were stimulated with an air-puff stimulator using the Pneumatic PicoPump (WPI, PV830). Air-puff was delivered through a 1 mm diameter glass microcapillary tube with a pressure of 20 psi for single whisker stimulation and 40 psi for whole whiskers, neck, trunk, tail, and limbs stimulations. After a 10 s baseline recording, each whisker or organ was stimulated 50 times, once every 1 s. To minimize the chance of stimulating multiple whiskers other than the whisker of interest, whiskers were subsequently trimmed off after each recording. Electrical stimulations of the hindlimb and forelimb were done by inserting a pair of subdermal needle electrodes into the limb muscles and injecting bi-phasic current pulse (2.55 mA, 1 ms positive/1 ms negative) using the Intan RHS system. Both the air-puff and electrical stimulation were time locked to the recording system by sending TTL signals to both the stimulator and the Intan recording controller. From the animal model, we performed experiments with the PtNRGrid bonded to either a small PCB (Fig. S13D) or to an extender board (Fig. S13E). Both the small PCB (Rat 1, 2, and 3) and extender PCB (Rat 4) recorded excellently localized HGA that is presented in Fig. S13.

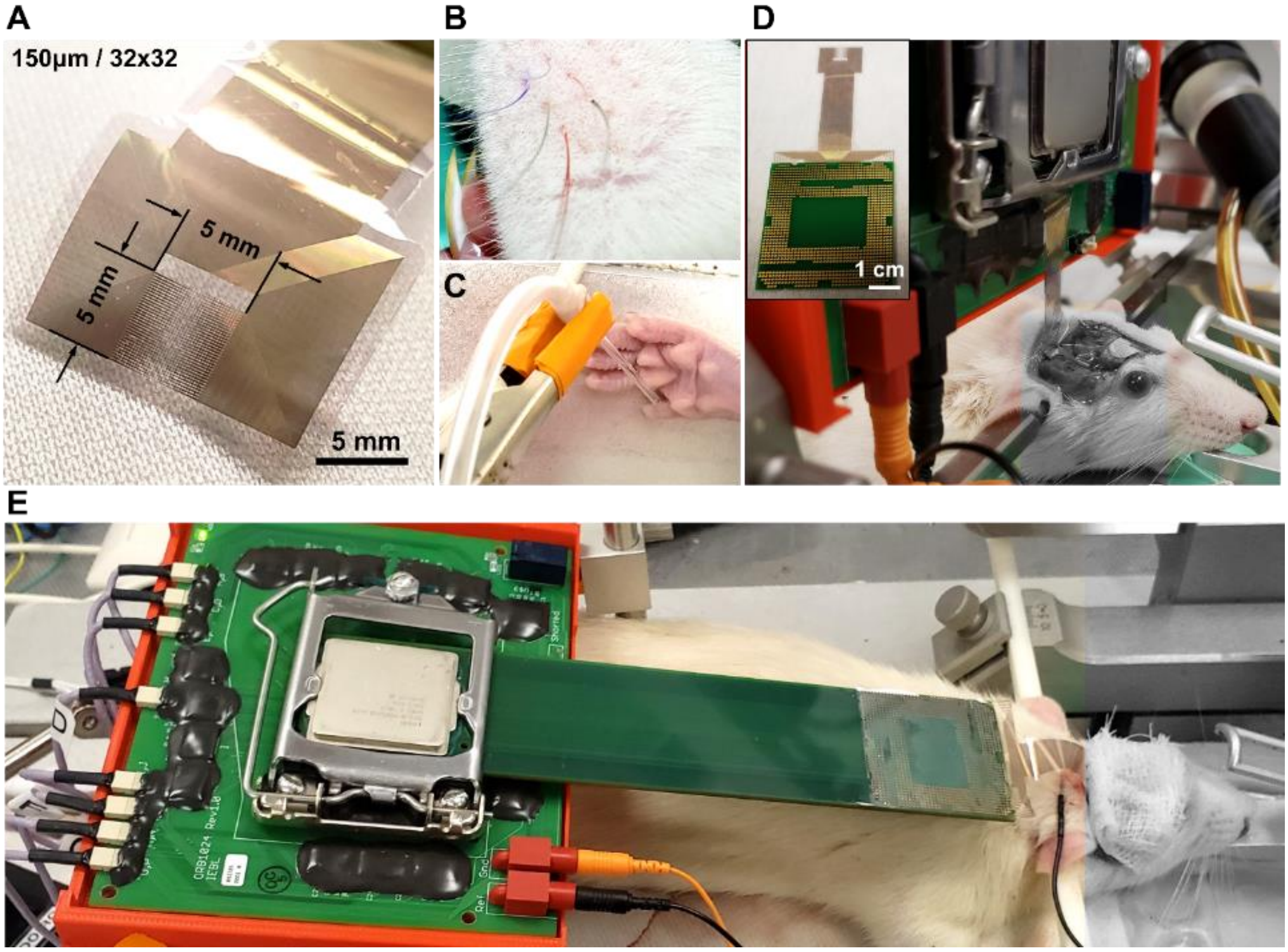

**Figure S13.** Rat whisker barrel recording experimental setup. (A) Magnified photo of the rat electrode near the recording sites. (B) Individual whiskers colored and (C) hindlimb of the rat. Microcapillary tube shown in (C) is used for an air-puff stimulation. Rat barrel cortex recording with PtNRGrids bonded on (D) a small PCB and (E) extender board.

3.3. Histology of the Rat Brain

After recordings, the sedated rat was euthanized and the corners of the PtNRGrids were marked by vertically puncturing the brain tissues with Dil (Invitrogen$^{TM}$)-coated ultrafine tungsten probe tips (The Micromanipulator Company). The brain was then rapidly dissected and placed into 4% paraformaldehyde (PFA) solution in phosphate-buffered saline (PBS) at room temperature for 2 hours. Each hemisphere of the cortex was then peeled away and flattened per Lauer *et al*. (*2*). Briefly, cortices were flattened between two glass slides in 0.1 M PB for 24 hours at 4ºC, to the thickness of 2 mm (using two other standard slides as supports). Flattened cortices were fixed overnight in 4% PFA at 4ºC and then transferred to 30% sucrose. Once the tissue had sunk, they were sectioned at 50 µm on a cryostat and serial sections were collected in a 24-well dish in PBS. Alternate sections were washed in PBS and underwent antigen retrieval for 20 min in heated 1X Citra buffer (Biogenex Laboratories). Slices were washed in PBS and underwent a protein block in PBS-T (0.03% Triton-X) with 5% goat serum (Vector Labs, S-1000) for one hour. Slices were then incubated in rabbit anti-goat VGLUT2 antibody (Abcam, ab216463, 1:1000) in PBS-T overnight at 4ºC. The next day, slices were washed in PBS and incubated for 30 min in secondary (1:100 goat anti-rabbit 488 (Thermo-Fisher, A11008)) in PBS-T. Slices were washed, mounted, and cover-slipped using ProLong Gold (Thermo-Fisher). Images were acquired at the UCSD Nikon Imaging Core using a Nikon Eclipse Ti2-E equipped with a DS-Qi2 CMOS camera, controlled by NIS-E-Elements (Nikon). Images were processed using Image J.

3.4. Propagating Beta Wave in Rat Barrel Cortex Together with HGA

We applied sensory stimulus evoked propagating beta wave analysis to the rat whisker barrel experiment data. Vector fields express the propagating directions of beta waves, with blue and red colors showing beta and high gamma activity amplitudes, respectively (Fig. S14).

We observed well-differentiated regions for the sources of the beta waves and their destinations, as well as sub-millimeter scale spiraling waves.

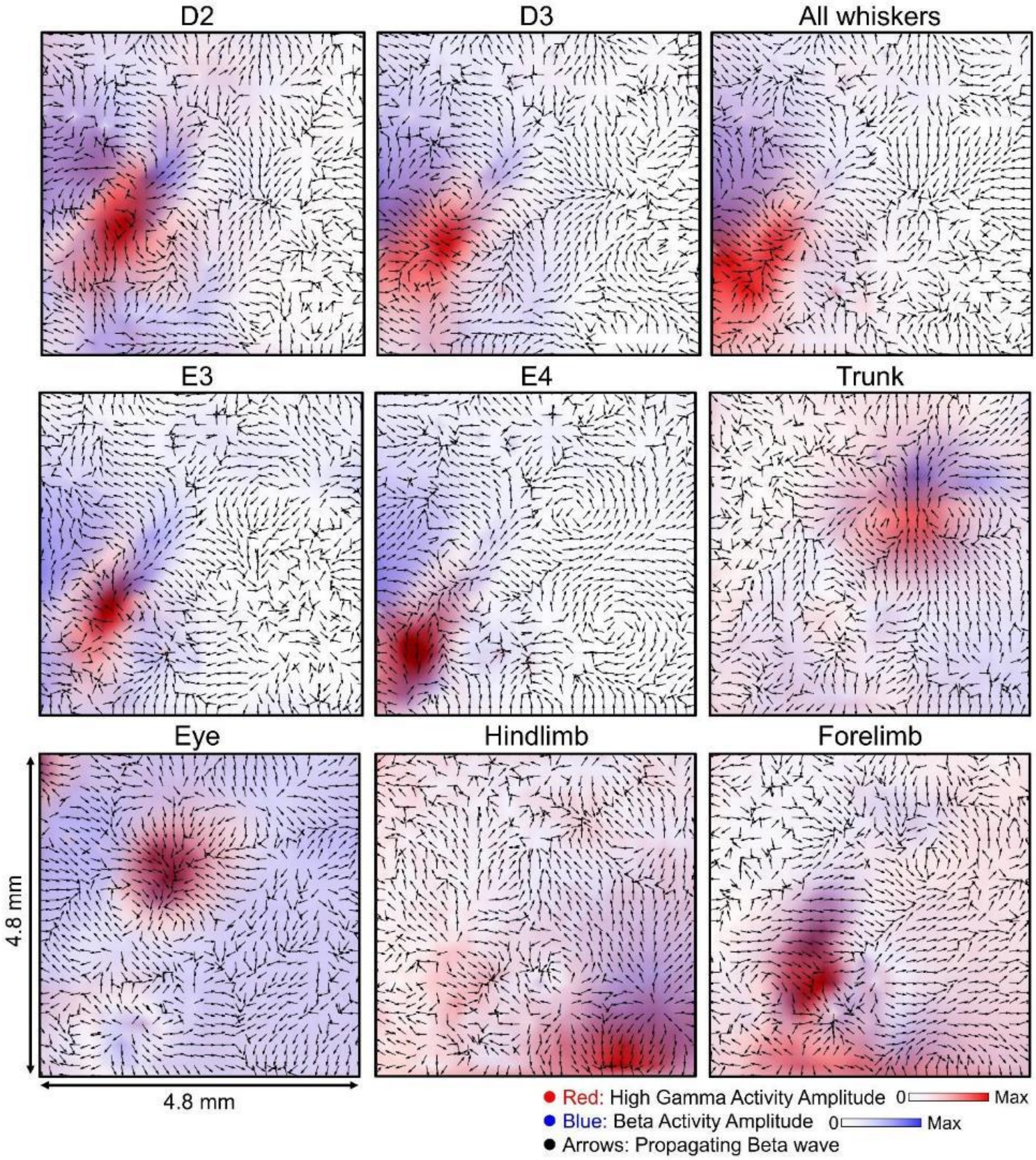

**Figure S14.** Propagating beta wave vector field overlaid with beta amplitude (blue) and high gamma amplitude (red) for (A) E4 whisker, (B) eye, and (C) trunk.

## 4. Human Brain Recording

4.1. Participant and Task Information

**Table S1.** Table summarizing patient participants with corresponding electrode types (PEDOT, PtNRs), electrode area, sterilization method, case and task details, and the summary of the recording.

| Participant # | Location | Case Info | Electrode | Sterilization method | Task Info | Summary of the recording |
|---|---|---|---|---|---|---|
| **1** | UCSD - Hillcrest | Implantation setup | PEDOT 32mm×32mm | Heat (pre-vac) | Baseline recording | Validated the recording noise level |
| **2** | OHSU | Awake Craniotomy - Precentral gyrus | PEDOT 3mm×18mm | Heat (pre-vac) | Motor tasks | Hand-motion related LFP were detected |
| **3** | OHSU | Awake Craniotomy - STG | PEDOT 32mm×32mm | Heat (gravity) | Auditory task | Difference between sense/non-sense data were detected |
| | | Awake Craniotomy - STG | PEDOT 3mm×18mm | Heat (gravity) | Auditory task | |
| **4** | OHSU | Awake Craniotomy, recurring tumor (third resection near motor cortex) | PEDOT 32mm×32mm | Heat (gravity) | Hand motion capture task | Hand-motion related LFP were detected |
| **5** | UCSD - Jacobs Hospital | Anesthetized Craniotomy | PEDOT 32mm×32mm | Heat (gravity) | Phase reversal | Phase reversal in SSEPs (Figure S41) |
| **6** | OHSU | Awake Craniotomy - Right hemisphere motor strip | PEDOT 32mm×32mm | Heat (gravity) | Hand motion capture task | Captured hand-motion related LFP |
| Perfusion holes were made / PEDOT electrodes -> PtNRGrids | | | | | | |
| **7** | OHSU | Right craniotomy (epileptic tissue removal) | PtNRs 32mm×32mm | Heat (gravity) | Hand motion and vibrotactile tasks | Electrode impedance increased after the heat sterilization (tap water issue) – Noisy recording |
| **8** | OHSU | Left Craniotomy | PtNRs 32mm×32mm | Heat (gravity) | Hand motion and vibrotactile tasks | |
| **9** | OHSU | Awake Craniotomy, M1/S1 | PtNRs 32mm×32mm | Heat (gravity) | Hand motion and vibrotactile tasks | |
| Sterilization method was changed to V-PRO | | | | | | |
| **10** | OHSU | Awake Craniotomy, Posterior edge of craniotomy, pSTG | PtNRs 32mm×32mm | V-PRO | Phase reversal | CS mapping of facial region |
| | | Awake Craniotomy, Posterior edge of craniotomy | PtNRs 32mm×50µm (1024-ch dual column linear grid) | V-PRO | Phase reversal | Linear CS boundary (agreed with above result) |
| **11** | OHSU | Awake Craniotomy, recurring tumor, M1/S1 | PtNRs 46mm×44mm (2048-ch) | V-PRO | Hand motion and vibrotactile tasks | Brain tissues seriously affected by the large tumor. SSEPs detected (Figure 3) |
| **12** | OHSU | Anesthetized Craniotomy, Temporal Lobe | PtNRs 32mm×32mm | V-PRO | Epileptogenic activity monitoring, Cortex stimulation | Spontaneous and Stim-evoked epileptic discharge observed (Figure 5) |
| **13** | OHSU | Awake Craniotomy, 2 positions on temporal lobe | PtNRs 32mm×32mm | V-PRO | Epileptogenic activity monitoring, Cortex stimulation | Spontaneous epileptic discharge observed |
| **14** | OHSU | Awake Craniotomy | PtNRs 32mm×32mm | V-PRO | - | Interrupted the research due to bleeding |
| **15** | OHSU | Awake Craniotomy, hand knob, M1/S1 | PtNRs 32mm×32mm | V-PRO | Phase reversal, Hand motion and vibrotactile tasks | CS mapping (Figure 3), Motor/sensory mapping (Figure 4) |
| **16** | OHSU | Awake Craniotomy, facial region, M1/S1 | PtNRs 32mm×32mm | V-PRO | Phase reversal, Facial motor task | SSEPs not detected from |
| | | Awake Craniotomy, pSTG | PtNRs 32mm×32mm | V-PRO | Auditory task | neuromonitoring (elder patient) / Very low amplitude SSEP recorded from PtNRGrid |
| **17** | OHSU | Awake Craniotomy, Left prefrontal lobe | PtNRs 32mm×32mm | V-PRO | Phase reversal, Hand motion and vibrotactile tasks | Clear SSEP & CS mapping. Implantation location was a little off from the fingers |
| **18** | OHSU | Awake Craniotomy, Left temporal lobe | PtNRs 3mm×18mm | V-PRO | Auditory task | Clear auditory HGA response |
| **19** | OHSU | Awake Craniotomy, Left prefrontal lobe | PtNRs 32mm×32mm | V-PRO | Phase reversal, Hand motion and vibrotactile tasks | Clear SSEP & CS mapping. Implantation location was a little off from the fingers |
| **20** | OHSU | Awake Craniotomy, M1/S1 | PtNRs 32mm×32mm | V-PRO | Phase reversal, Hand motion and vibrotactile tasks | |

### 4.2. Sterilization of the Electrodes

The PtNRGrid was packaged in DuraHolder Instrument Protection System (Key Surgical) pouches that kept the electrodes flat during the sterilization process for Steam Sterilizers (Steris) and V-PRO (Steris) as shown in Fig. S15A. The electrodes sterilized by STERRAD® (ASP Global Manufacturing, GmbH) were packaged with Plasma-Cel™ Instrument Foam Protection (Healthmark Industries). The pouch and foam effectively protected the PtNRGrid during the sterilization process by providing non-stick cushions and keeping them flat under multiple cycles of chamber pressure changes while allowing steam or aerosolized hydrogen peroxide to freely access the electrodes. Even after standard interstate shipping (FedEx Air Freight), the characteristics of PtNRGrids packaged in DuraHolder pouches were unaltered, confirming the robustness of the PtNRGrids as well as the packaging method. The packaged electrodes were placed inside a sterilization tray on top of an autoclave-compatible silicone mat and underwent a sterilization process. For steam sterilizers, we used Gravity mode 121°C, 30 min process, and, for V-PRO and STERRAD®, we used a default sterilization mode. The PtNRGrids were compatible with all the sterilization systems mentioned above. However, the electrode sterilized by the steam sterilizer often experienced an increase of electrochemical impedance depending on the maintenance state of the heater unit or the purity of the tap water used for steam generation. We believe that the observed increase in impedance magnitude has to do with the adsorbates and impurities from the tap water that incorporates into the electrochemically active sites in PtNRs. Since both V-PRO and STERRAD® use reagent grade hydrogen peroxide, we did not observe an increase in impedance for PtNRGrids after these sterilizations. Most of the results presented in this work were recorded with the PtNRGrids sterilized by V-PRO.

### 4.3. Sterile Procedures in the Operating Room and the Implantation

The sterilized PtNRGrids were transferred onto a sterile Mayo Stand, and 2/3 of the extender board was inserted through an opening cut in a Situate$^{TM}$ Sterile Drape (Medtronic 01-0020). Another hole was opened on the Sterile Drape to insert the touch-proof connectors of a sterile twisted pair of subdermal needle electrodes. Tegaderm$^{TM}$ films (3M) were applied on both sides to hermetically seal the gaps between the extender board, wires of needle electrodes, and the sterile drape (Fig. S15B). While the sterile person was holding the outer surface of the Sterile Drape together with the PtNRGrid, the non-sterile person accessed the inner surface of the Sterile Drape and established a connection between the extender board and the amplifier board that was connected with the recording controller through SPI cables (Figs. S15C and D). After connecting the electrode on the amplifier board, the non-sterile person pulled the 2-meter-long Sterile Drape from the inner side to cover all the non-sterile items. This setup allowed us to bring the sterilized PtNRGrids connected to the amplifier board close to the surgical field. Prior to implant, the connectorization of PtNRGrid was first validated by measuring the impedance magnitudes in sterile saline (Fig. S15E), and then implanted onto the patient's brain (Fig. S15F). The Greenberg® Retractors (Symmetry Surgical GmbH) clamped on the surgical table stabilized the position of the amplifier board. After the PtNRGrid was implanted, subdermal needle electrodes that served as reference and ground were implanted on the temporalis muscle near the recording field.

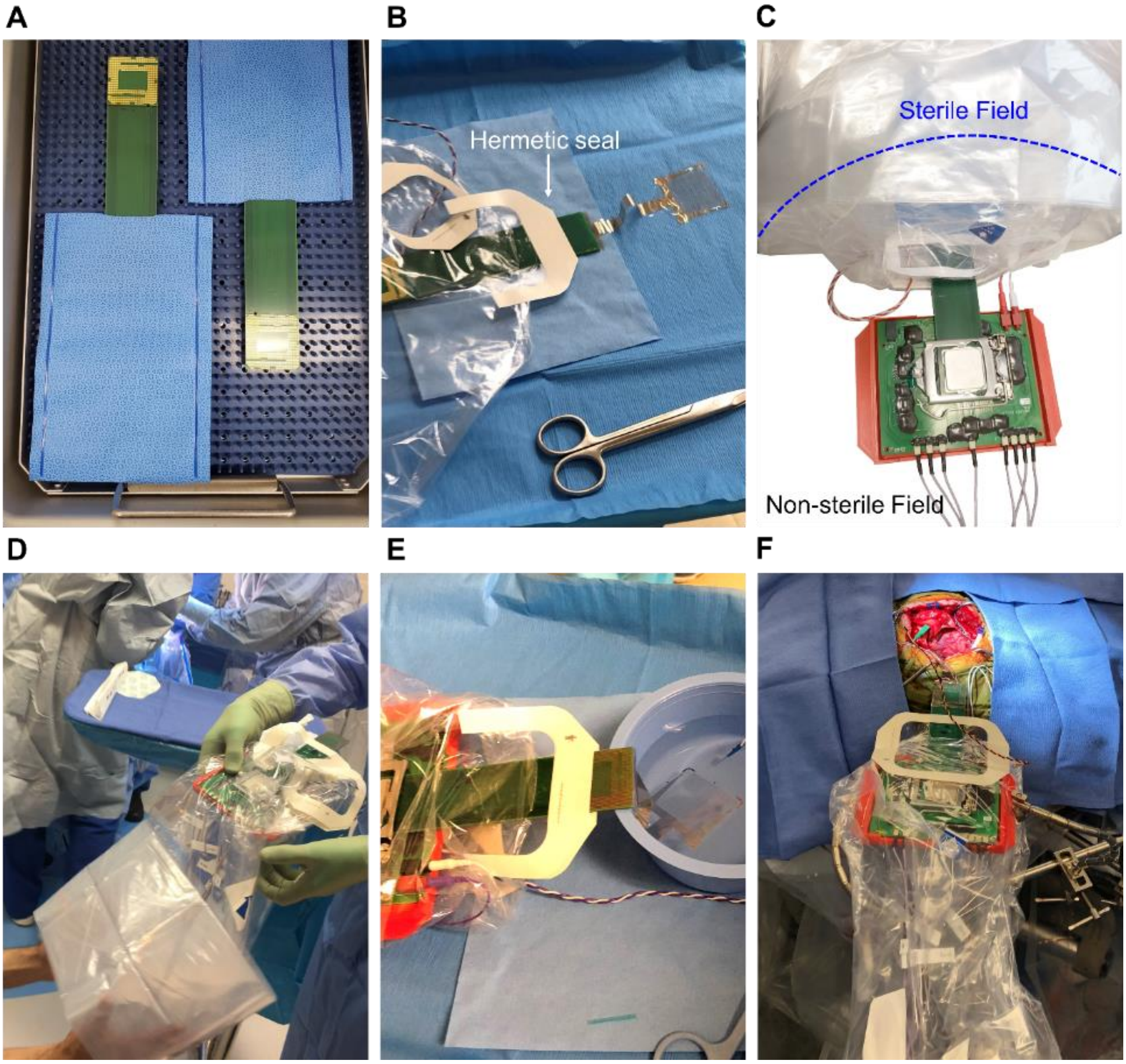

**Figure S15.** Sterilization and sterile procedures in the operating room. (A) Electrode packaged in DuraHolder pouch placed inside the sterilization tray with silicone mat. (B) Extender board's connector region placed inside the Situate$^{TM}$ Sterile Drape. Tegaderm$^{TM}$ films were applied for hermetic sealing. (C) Establishing a connection in a non-sterile zone. (D) Sterile person holding the PtNRGrid and the amplifier board, and non-sterile person pulling the sterile drape from the inner surface. (E) Electrode impedance measurement in saline prior to implantation. (F) Implantation on the patient's brain. Greenberg® Retractors are holding the amplifier board.

### 4.4. Stimulation Capture System

A common experimental paradigm we used took advantage of neuromonitoring procedures wherein a patient's median or ulnar nerves are electrically stimulated while somatosensory evoked potentials (SSEPs) are monitored in order to map the central sulcus functional boundary. The time delay between stimulation and SSEP detection is critical for this paradigm, thus we needed to capture the clinical stimulation signal in parallel with the neurophysiologic signals. This posed a challenge of capturing the timing of stimulation without modifying the clinical medical equipment in any way. The clinical equipment used for bipolar stimulation did not provide digital means of synchronization, thus we developed a custom differential capacitive sensing system which wrapped around the stimulation leads and captured the timing and amplitude information and provided a TTL synchronization signal which was split to the Intan 1024-channel recording controller (see Fig. S16).

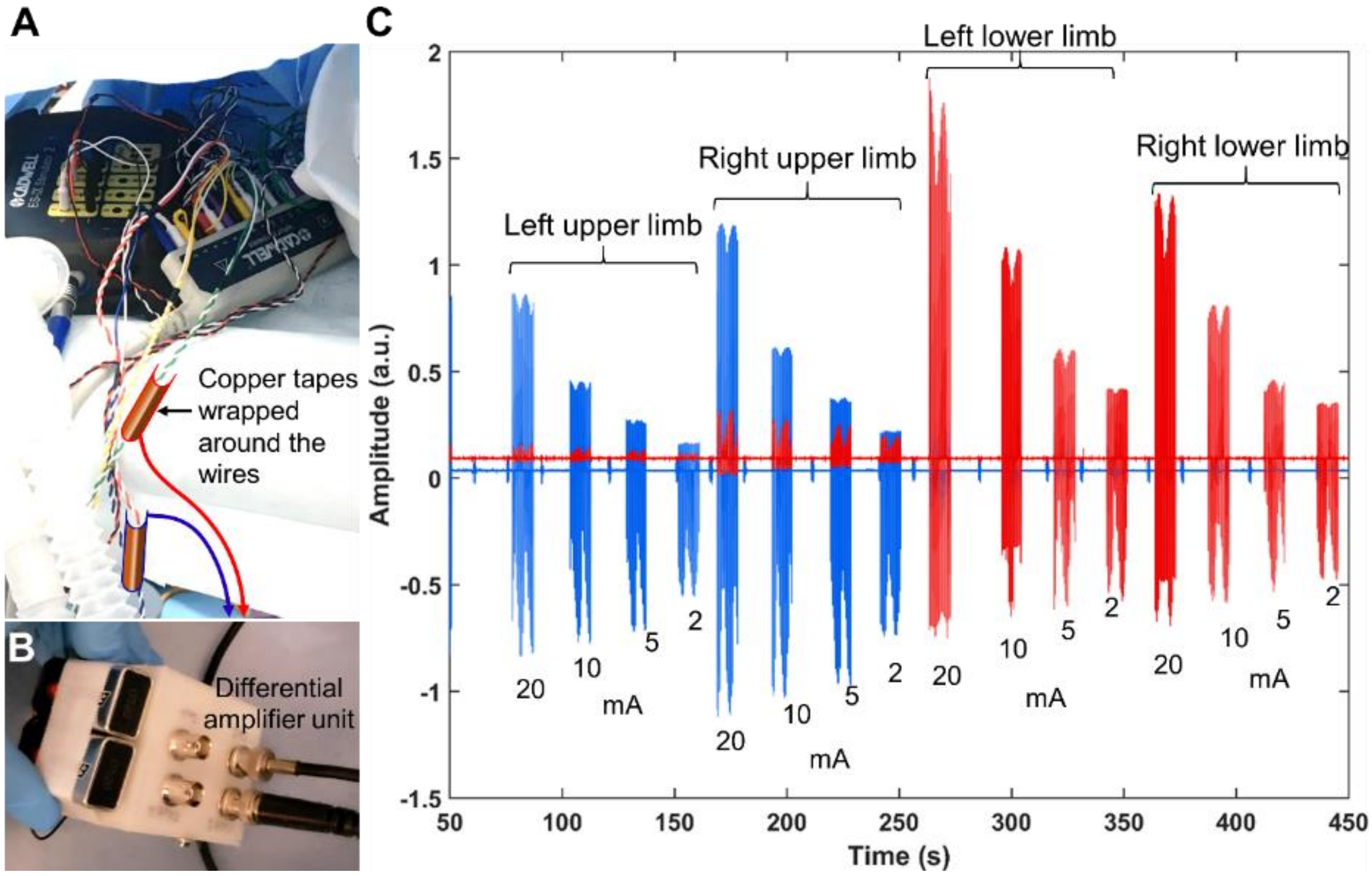

**Figure S16.** Electrical stimulation signal capture system. (A) Photo of the clinical stimulation controller and the schematics showing how the copper foils were wrapped around the wires. (B) Differential amplifier. (C) Signals captured with this system.

4.5. Motor Glove Experimental Platform

Many intraoperative cases exposing motor cortex provided us with a unique opportunity to investigate neural correlates between surface recordings and motor tasks in awake patients. In order to take full advantage of these opportunities, we developed a custom glove system that was capable of capturing 9 degrees of freedom (flexion/extension of each finger and wrist, and 3D acceleration of the hand) and could provide vibrotactile stimulation to each fingertip (see Figs. S17A-B). The system connected to a dedicated task tablet, providing and capturing the timing of visual and audio instructions to the patient. The system also included a hand-held button which allowed researchers to capture gross timing information of improvised tasks. Each buffer of data sent from the glove system to the task tablet was marked by a TTL synchronization pulse which was split to the Intan Technologies recording controller unit to time-lock this information to the neural data.

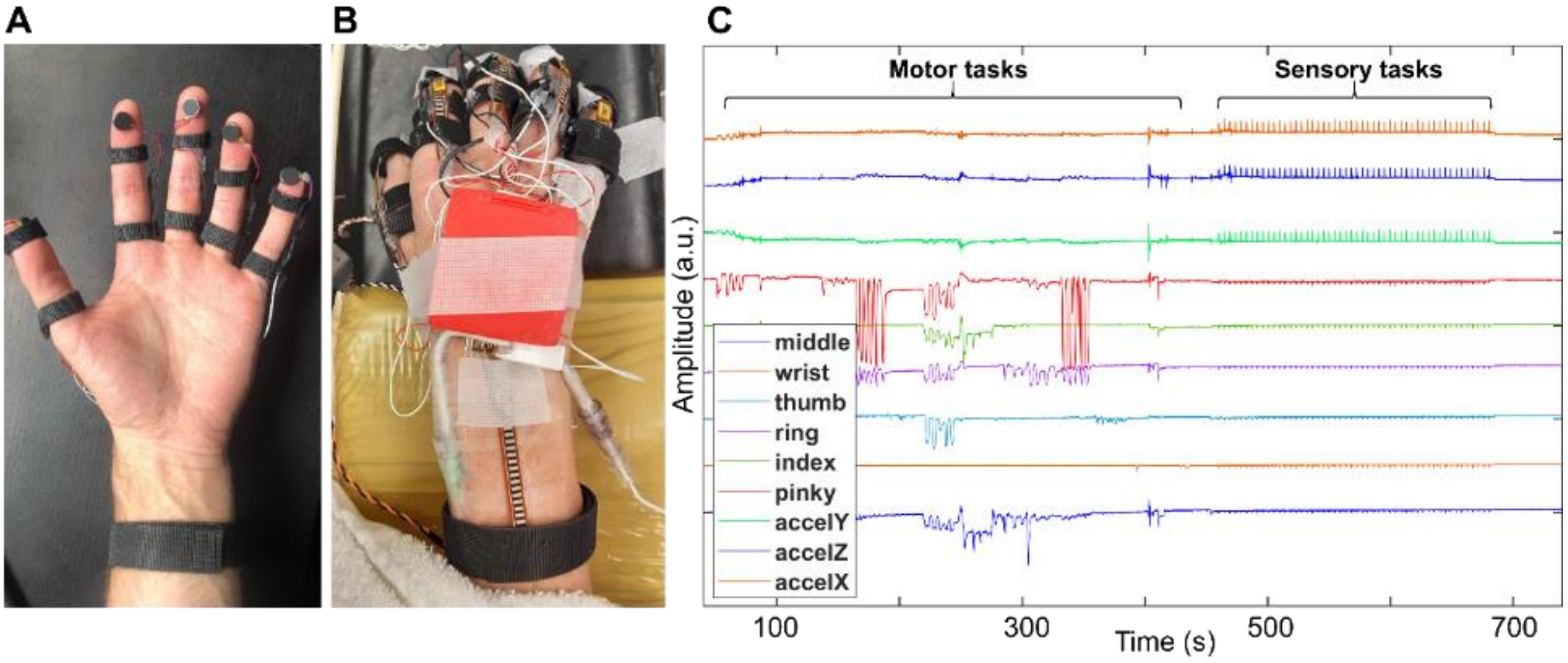

**Figure S17.** Motor glove that could (A) deliver vibrotactile stimulation on individual fingertips and (B) capture the motion of individual fingers and wrist flexion as well as acceleration of the hand. (C) Motor and sensory task captured with the motor glove system. Vibrotactile stimulus was clearly captured in the accelerometers.

## 5. Signal Processing

### 5.1. Spatial Mapping, Channel Selection, and 60 Hz Noise Notch Filter

The 1024 and 2048 channels recorded with the customized amplifier boards and the Intan recording controllers were mapped to the spatial coordinates of the individual recording sites on the PtNRGrids. We tabulated this mapping in a spreadsheet for each type of electrode and used this information to spatially display the impedance magnitudes, waveforms, and potentials. The channels with above 100 kΩ *in vivo* impedance magnitudes at 1 kHz were deemed open channels and excluded from the analysis. Neighboring channels with exceedingly low impedance magnitude from the norm were also evaluated as potential shorts. All recorded signals, unless specified as 'raw' signal, were first processed by eliminating 60 Hz and their noise harmonics with digital notch filters.

5.2. Localization of the Neural Responses in Rat Experiments

The waveforms presented in Figs. 2B and C are the N=50 trial average of the raw waveforms based on the TTL pulses time-locked to the air-puff stimulation. We then re-referenced the recorded signals by subtracting the common-averaged signal across channels. The common-average was calculated either by taking the average of all the working channels or by taking the average of a few selected channels. The problem of taking common-average from all the working channels was that, due to the high SNR of PtNRGrids, clear stim-evoked neural response was present in the common-average even after averaging signals from all the channels. Because of this, the re-referencing process caused negative-potential deflection on the channels that recorded no neural response which we believed was misleading. For this reason, we performed common-average referencing by averaging signals from ten channels that did not show ECoG responses. This effectively removed the motional artifacts, electrocardiogram, and low frequency noise. The signals were digitally filtered using a Butterworth 4th order filter under selected frequency windows, and 50 trials were aligned and averaged based on the TTL pulses that triggered air-puff or electrical stimulation. All digital filters were implemented in Matlab using the zero-phase distortion filtering function, "filtfilt", which effectively doubled the filter order to 8. The amplitude of the signals in each frequency window were calculated by taking root-mean-square (RMS) of the absolute value of the Hilbert transformed signal in a 10~100ms time window after the stimulation.

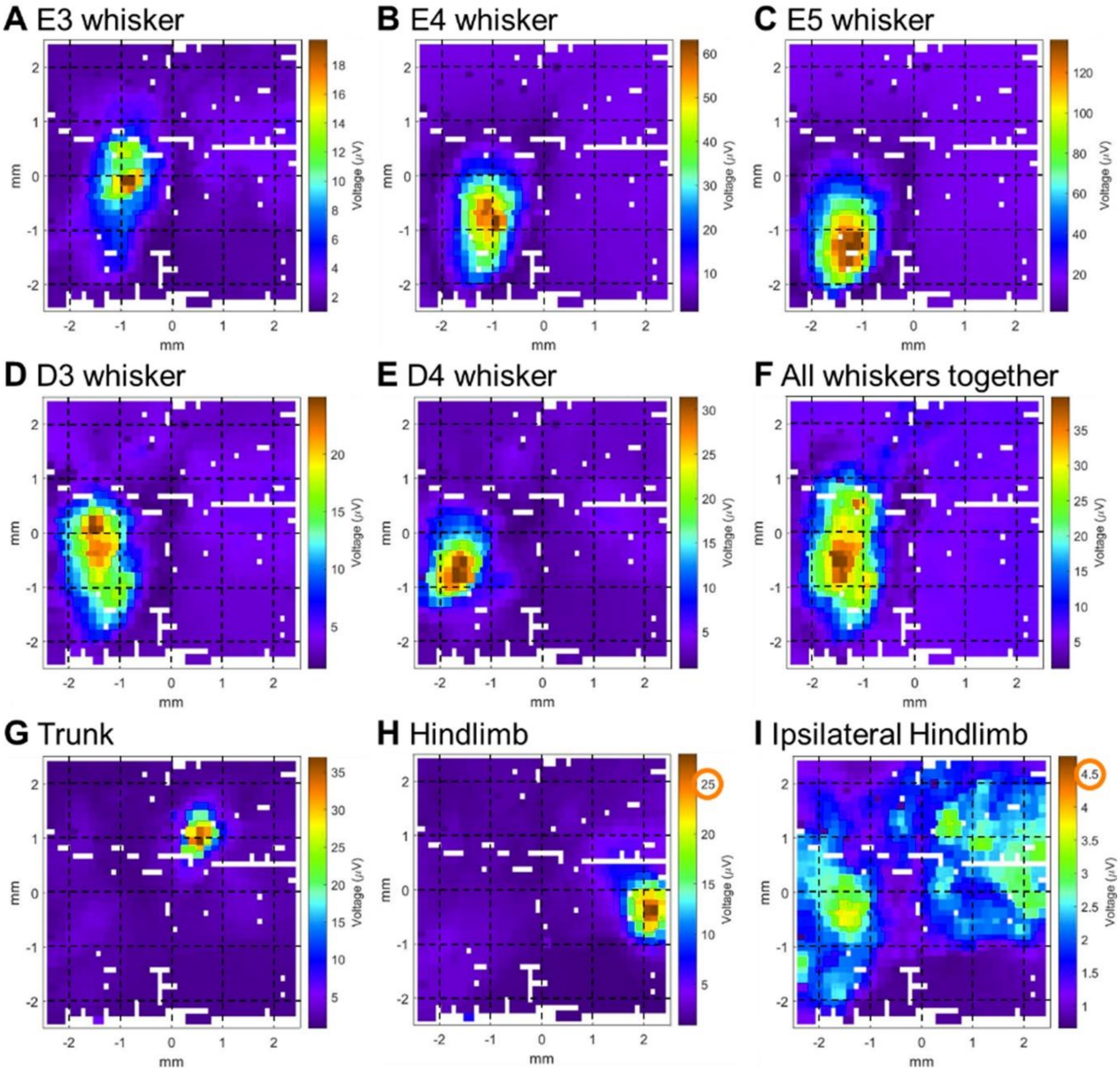

**Figure S18.** Localized high gamma response to air-puff stimulation of contralateral side (A)~(E) individual whiskers, (F) all whiskers, (G) trunk, (H) hindlimb, and (I) ipsilateral side hindlimb.

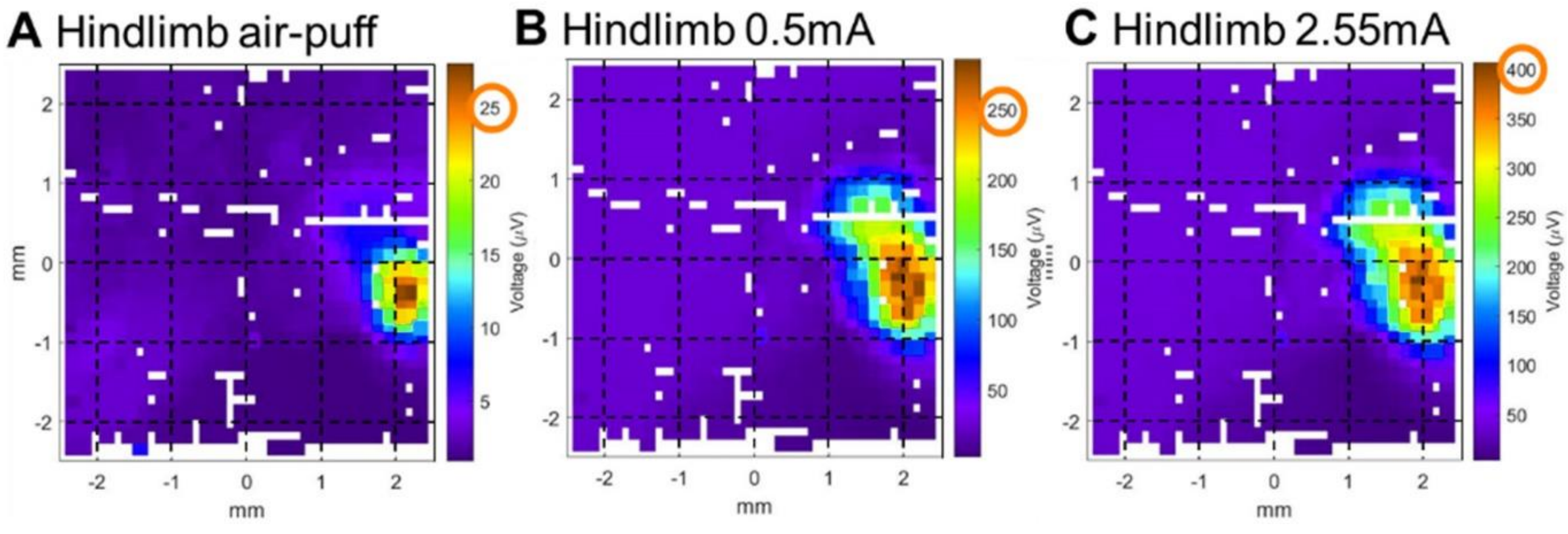

**Figure S19.** Comparison of (A) air-puff and (B)-(C) electrical stimulation of contralateral side hindlimb of the rat. Electrical stimulation of rat's hindlimb thigh muscles evoked broader and more than 10 times larger amplitude response on the sensory cortex compared to that of hindlimb air-puff stimulation.

Since the stim-evoked peak responses usually appeared a few milliseconds offset from each other, we also calculated trial averaging by finding the maximum correlations between the stim-evoked responses between trial epochs (see Fig. S20). Compared to the trial average taken by the TTL pulses, the trial average taken by the maximum correlation approach showed higher peak response and better represented individual trials in terms of pulse width due to the reduced jitter between trial epochs. The spatial mapping of high gamma activity (HGA) was calculated by taking RMS of the HGA in 2 ms time windows near the peak response for all the channels. We used customized colormaps to further define the spatial extent of HGA colormap (see Fig. S21). For superimposing the spatial mapping of HGA on top of desaturated VGLUT2 histology image, we used the 'Color' blending mode in Adobe Photoshop that preserved the grey levels in the background histology image while adding hue and saturation of the colored localized HGA (Fig. 2E).

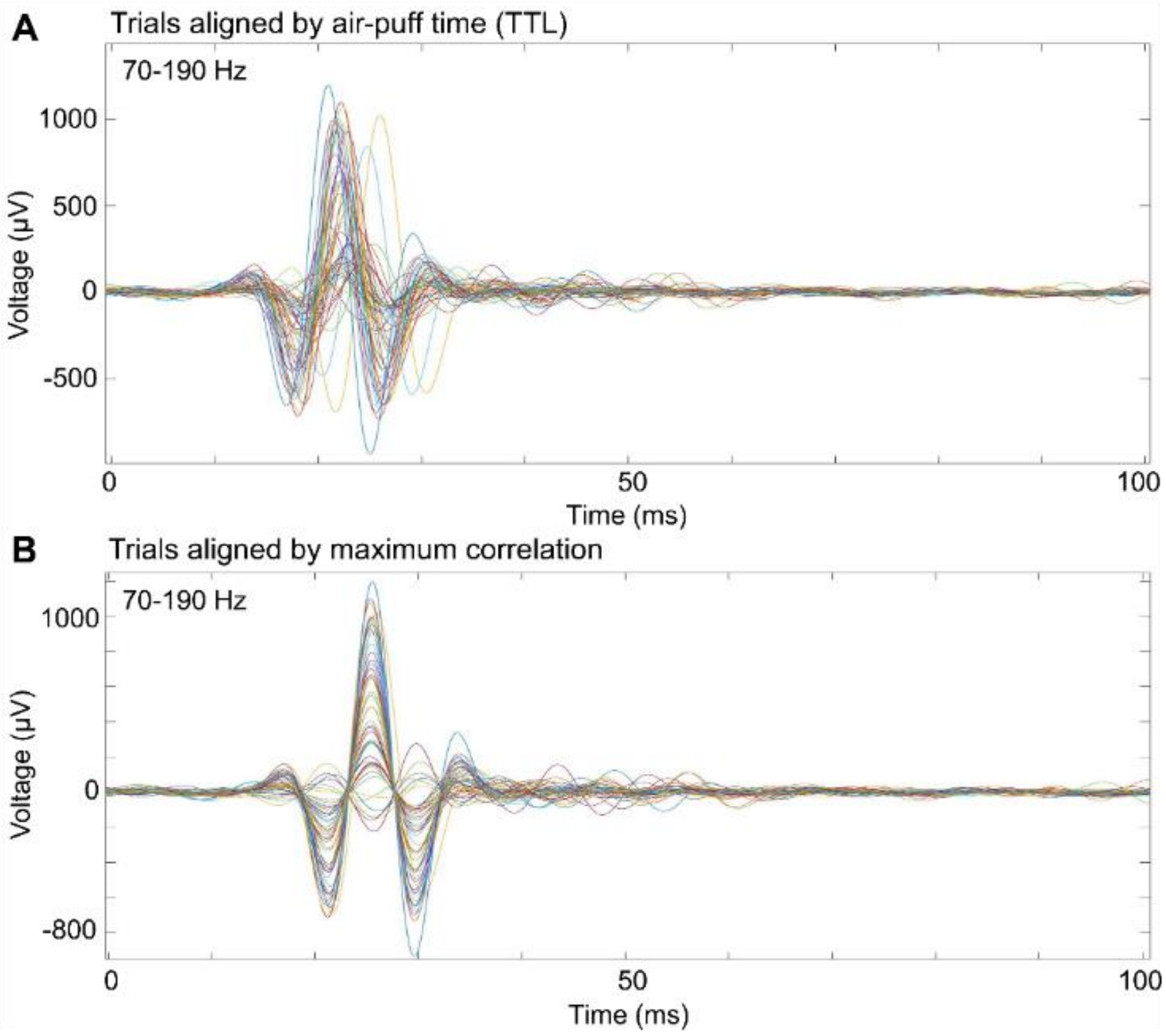

**Figure S20.** High gamma activity aligned by (A) air-puff stimulation time and (B) maximum correlation of the responses.

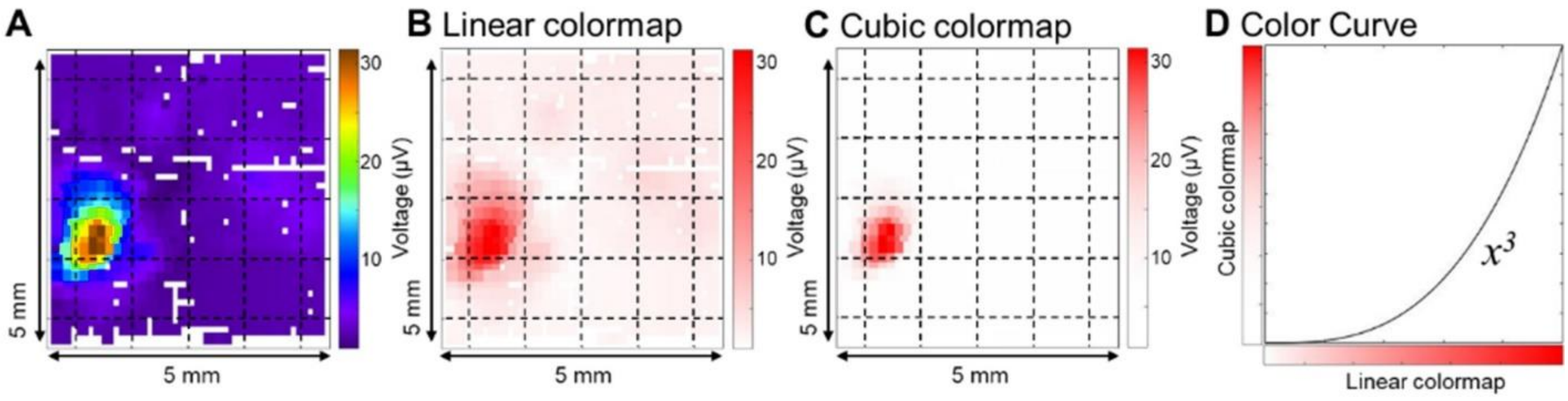

**Figure S21.** Localized HGA under different colormaps. (A) Purple-green-orange color map, (B) linear color map, (C) cubic color map. Cubic colormap was used to overlay high gamma responses to different whiskers. (D) Color curve used to generate cubic colormap.

5.3. Central Sulcus Localization

Signals were digitally filtered in the frequency window of 10-3000 Hz using a Butterworth 4$^{th}$ order filter with Matlab's "filtfilt" function. No re-referencing was used since this could potentially cause an undesirable offset in SSEPs. The train of peripheral nerve stimulation pulses was captured by our differential capacitive sensing system and was used to determine time epochs for trial averaging (N=10~20) the SSEPs. The peripheral nerve electrical stimulation pulses also appeared as sharp stimulation artifact peaks in the ECoG signals, which equivalently worked well to indicate timing of each trial. Since the electrical stim-artifacts agreed so well with the signals captured by the stim capture system, in the later experiments we did not set up the stim capture system and fully relied on the stim-artifacts to align trials.

5.4. Human Brain Somotosensory HGA Localization

Recorded data were re-referenced to ten selected channels as discussed in the rat signal processing section, and were digitally filtered in the high gamma frequency band of 70-190 Hz using a Butterworth 4$^{th}$ order filter with Matlab's "filtfilt" function. The action of the vibratotactile stimulators was clearly captured on the accelerometers in the glove system, which was used to time lock the sensory input event to the ECoG recording. The amplitude of HGA over time was calculated by taking the absolute value of Hilbert transformed data. We then took the standard deviation, σ, of the baseline noise level for each channel with an assumed average of zero, which is equivalent to calculating the RMS value. The RMS values were then taken from the high gamma responses evoked by individual finger stimulations. By dividing the RMS HGA with the noise floor RMS value, we obtained a normalized metric for how many σ the signal deviated from the noise floor for each channel. This method provided more reliable spatial mapping results compared to the potential mapping method since the noise amplitude significantly varied over channels for this particular recording due to the variation in *in vivo* impedance. ECoG response of sensory stimulation of each finger was spatially mapped with a color range of 1~4σ, overlayed on top of each other by 'Multiply' blending mode and superimposed on top of a desaturated craniotomy photo using 'Color' blending mode in Adobe Photoshop.

5.5. Human Brain Motor High Gamma and Beta Wave

The data was re-referenced and digitally filtered in a high gamma frequency band of 70-190Hz and beta band of 9-18Hz using a Butterworth $4^{th}$ order filter with the "filtfilt" function in Matlab. Amplitudes were calculated by taking the absolute value of the Hilbert transformed data. The hand motion captured by the flexion sensor was used to time lock the event with the ECoG signals. The spatial mapping of HGA in the sigma unit was obtained by dividing the RMS signal from RMS baseline noise. The high gamma amplitudes in Figs. 4E and F are shown without any baseline noise subtraction.

5.6. Human Brain Propagating Beta Wave Signal Processing

The propagating dynamics of the beta waves were calculated by taking the spatial phase gradients of the beta waves following the methods described in Rubino *et al.* (*3*) and Muller *et al.* (*4*). The phase angle of the beta wave for each channel was calculated by taking the inverse tangent of the imaginary part over the real part of Hilbert transformed data, and the phase was unwrapped over time. The propagation directions of the beta waves calculated from the spatial phase gradient were represented as a vector field and red and blue streamlines originating from SI and MI cortex, respectively, were used to visualize the long-range propagation directions of the waves. The streamlines were plotted using the streamline function in Matlab with a default setting. To validate if the propagating wave model applied to the 1024 channel grid could properly interpret the traveling direction of the waves, we plotted the vector fields for artificial gaussian wave models rotating, linearly moving, expanding, shrinking, and changing in amplitude as shown in Fig. S22. The vector fields effectively expressed the propagating direction of the artificial gaussian waves. The vector field became incoherent for a stationary gaussian wave that only changed in amplitude (Fig. S22G), which is in agreement with the previous works (*4*).

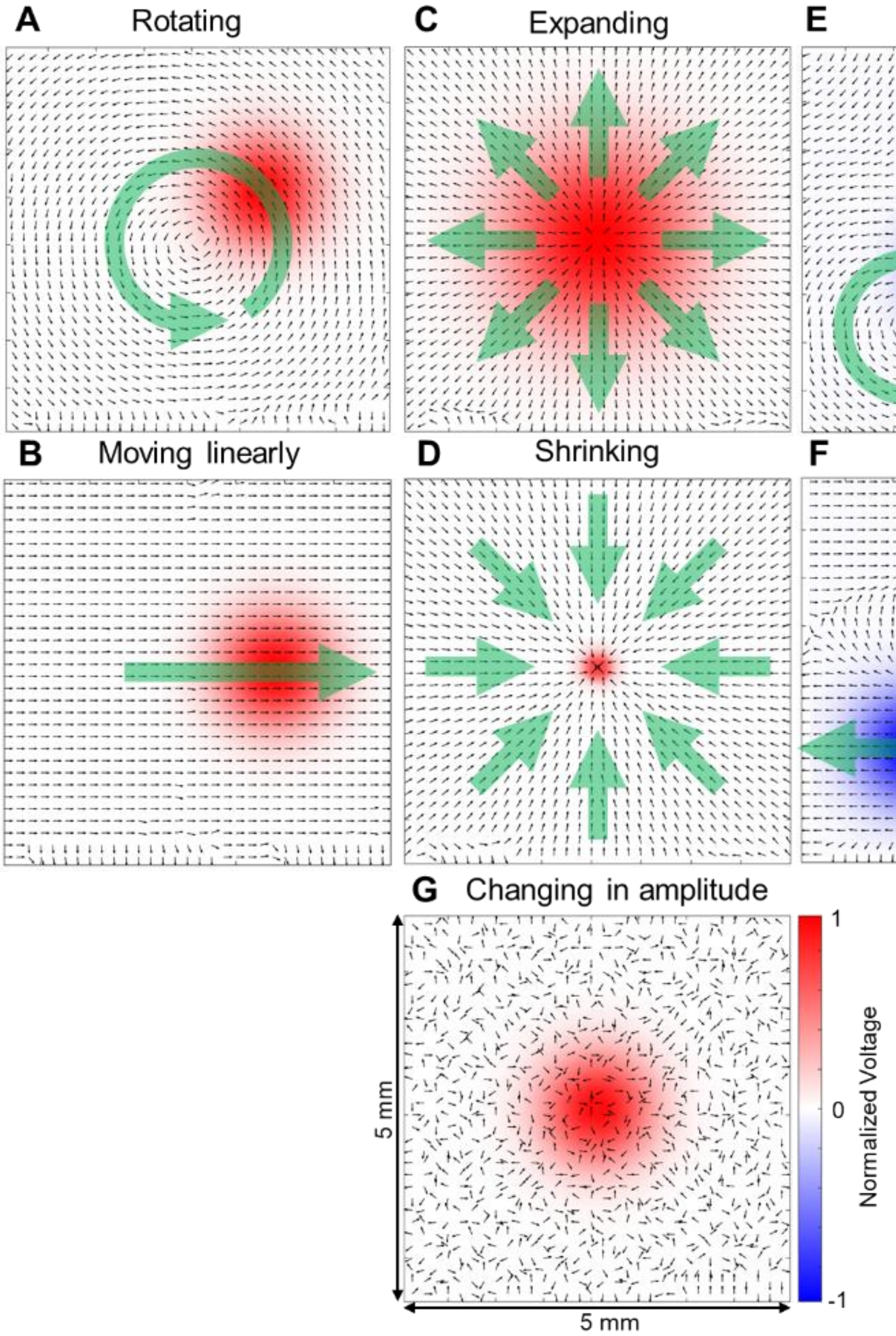

**Figure S22.** Vector field plots of the propagating beta wave of artificial traveling gaussian wave. Single gaussian wave (A) rotating, (B) linearly moving, (C) expanding, and (D) shrinking. Two gaussian waves (E) rotating and (F) linearly moving. (G) Gaussian wave only changing in amplitude.

## 5.7. Human Brain Epilepsy Monitoring Processing

The data was digitally filtered with a passband of 10-59 Hz using the Butterworth 4$^{th}$ order filter and the "filtfilt" function in Matlab. No re-referencing was done for this set of data to preserve the epileptiform activities happening throughout the entire surface of the grid.

Spatial mapping of the amplitudes of the epileptiform activities were calculated by taking the RMS of the potentials in 50 ms time windows. Epileptiform activity was automatically detected by employing an automatic detection algorithm developed to capture interictal discharges in Janca *et al.*(*5*). The channels were sorted according to the 2D Euclidean distance from the suspected onset zone of the epileptiform activities. The bi-polar stimulation events were clearly captured as large amplitude artifacts by the PtNRGrids, and the stimulation positions were recorded by the surgical microscope video. Adopting the same phase gradient and streamline approach described above, we estimated the propagating direction of the epileptiform activities near the onset zone.

## 6. Supplementary Data

### 6.1. Rat Whisker Barrel Experiment

We recorded sub-millimeter scale localized HGAs from four different rats as shown in the overlaid HGA mapping evoked by different whiskers or limb stimulations as shown in Figure S23. All stimulation were air-puff, except the forelimb and hindlimb electrical stimulation for Rat 4. Rats1-3 were recorded with PtNRGrid bonded on small PCBs, while Rat4 was recorded using a PtNRGrid bonded on an extender board.

We performed different number and type of tasks for each rat since the conditions of the rats were variable. For example, Rat 3 was not doing well after three whisker stimulations, and we had to terminate the experiment after C2, C3, C4 whisker stimulation. We note that for every stimuli and task we performed, we did observe a corresponding submillimeter region of neural correlates.

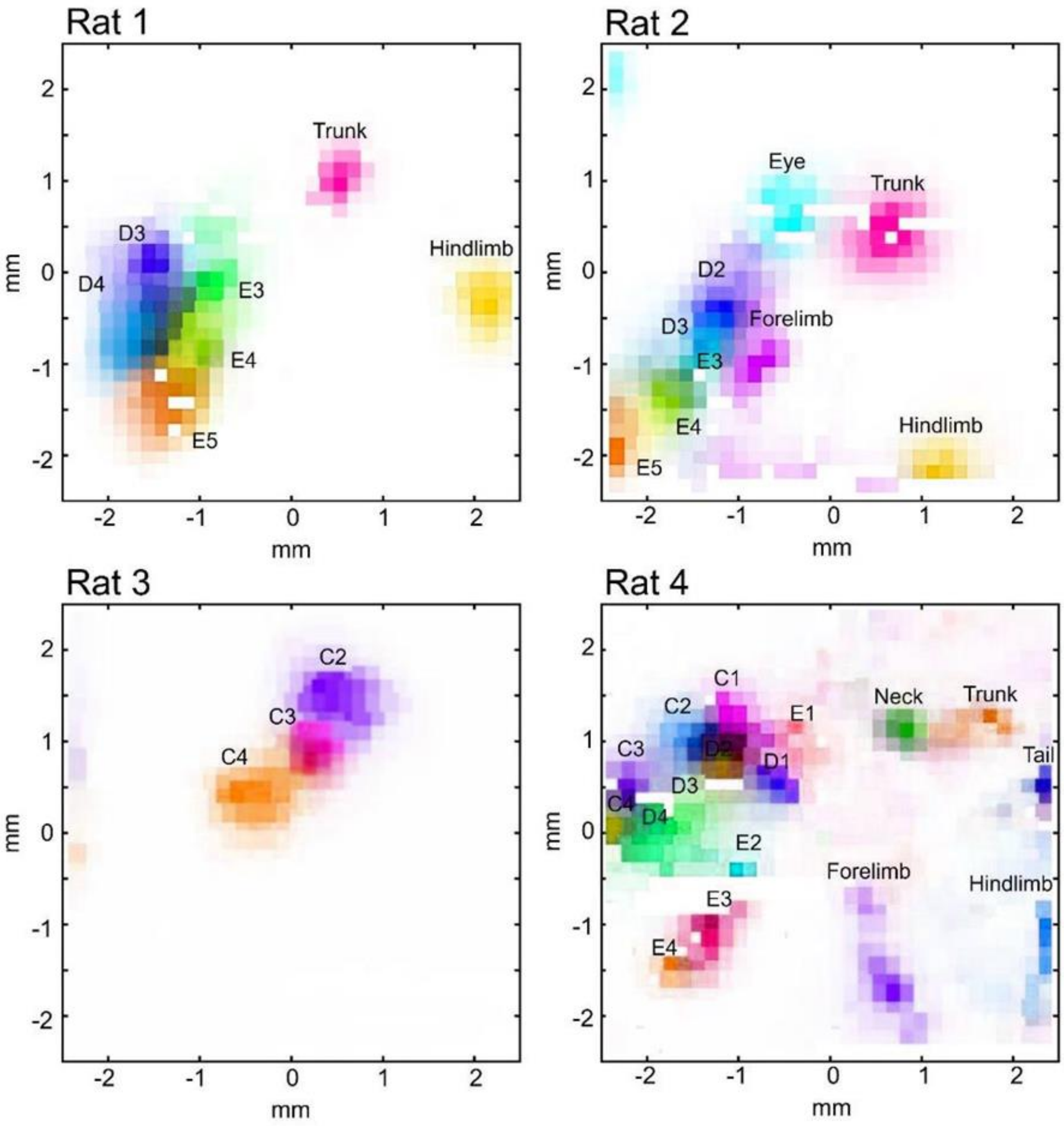

**Figure S23.** Localized HGA evoked by different whiskers or limbs stimulation for four different rats.

6.2. Central Sulcus Mapping

The SSEP waveforms corresponding to Fig. 3 captured by the entire PtNRGrid are shown below.

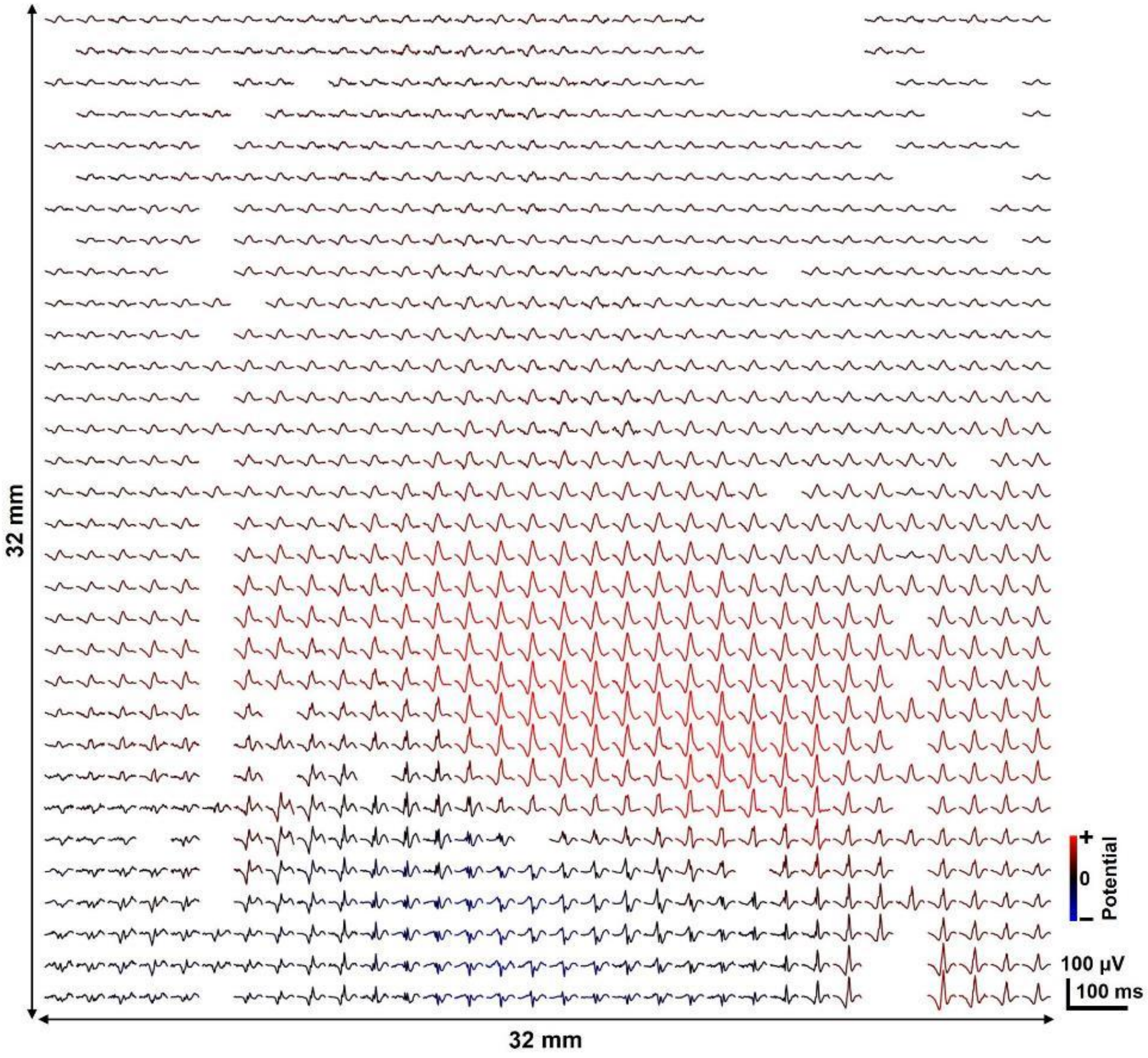

**Figure S24.** SSEP waveforms recorded with PtNRGrids implanted near the hand region. The waveforms are colored according to the polarity of the potential at 28ms post stimulus.

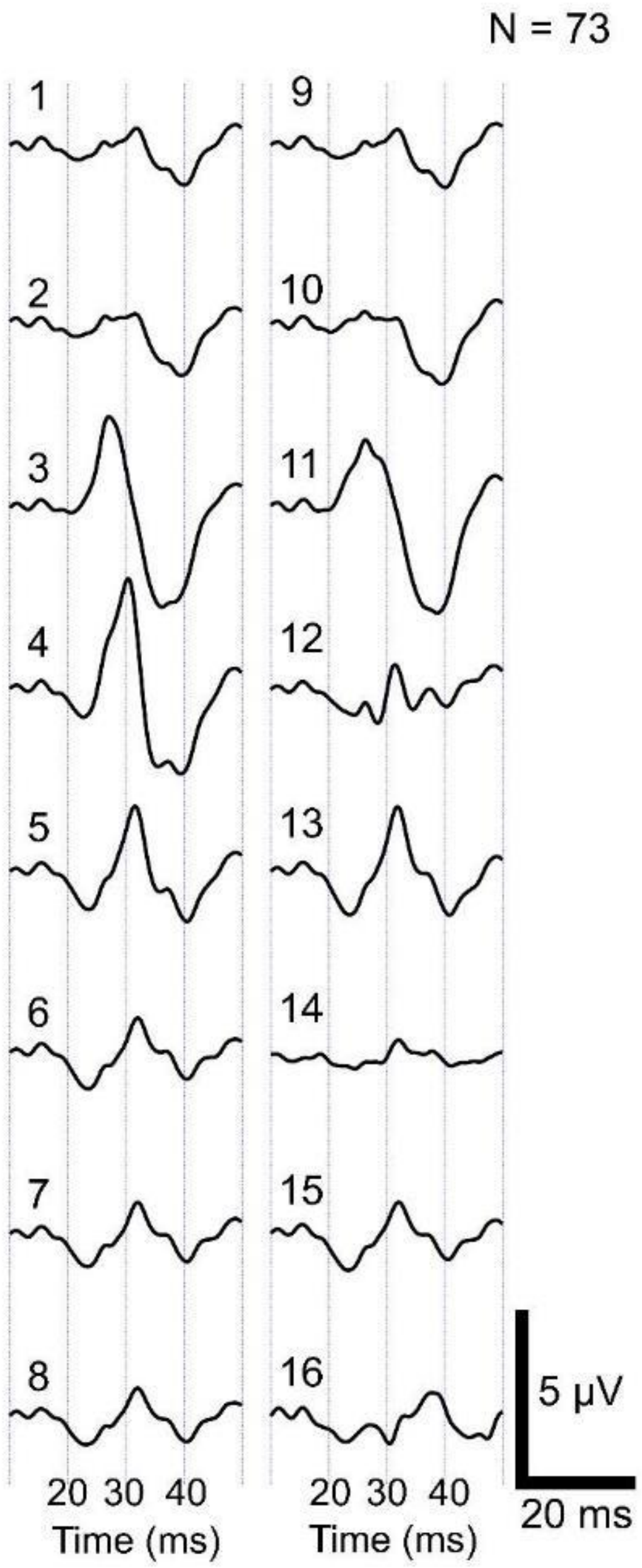

**Figure S25.** Waveforms of the SSEPs recorded with conventional dual column 2×8, 16 channels clinical grid implanted near the hand region.

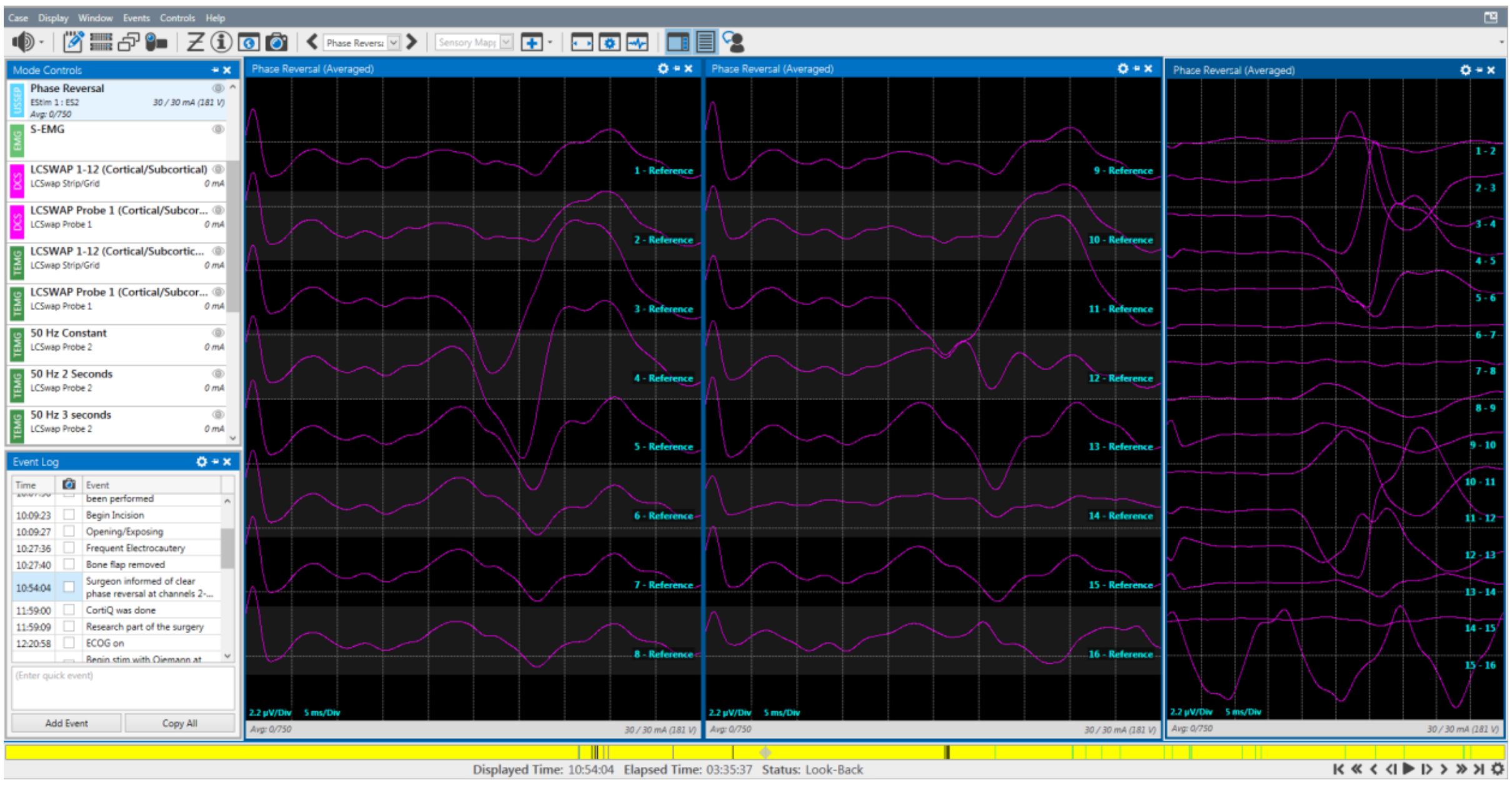

**Figure S26.** Snapshot of the neuromonitoring software showing trial-averaged phase reversal waveforms measured with the 2×8 clinical ECoG grid.

Although prior studies reported that the presence of a blood vessel underneath the clinical grid could attenuate the ECoG signal by 30-40% especially in 30-70Hz frequency window (*6*), the SSEPs recorded with PtNRGrid were minimally affected by the presence of a blood vessel underneath it: The waveform shapes and amplitudes ('CS' in Fig. 3B, and Fig. S27) of the channels on top of the blood vessel did not show any noticeable difference compared to those of the adjacent channels.

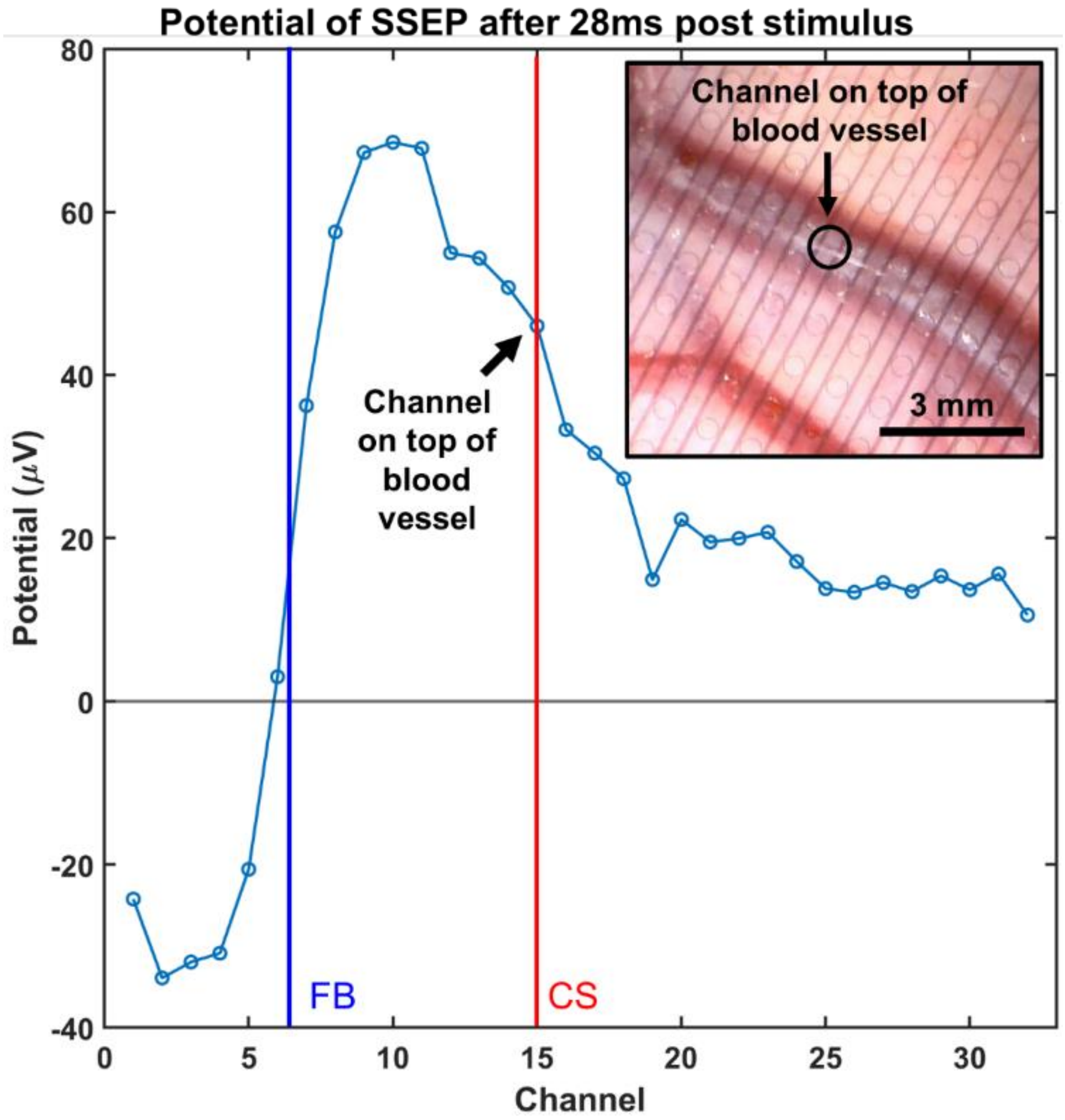

**Figure S27.** Potential of SSEP after 28 ms post stimulus. Inset figure indicates the channel placed on top of the blood vessel.

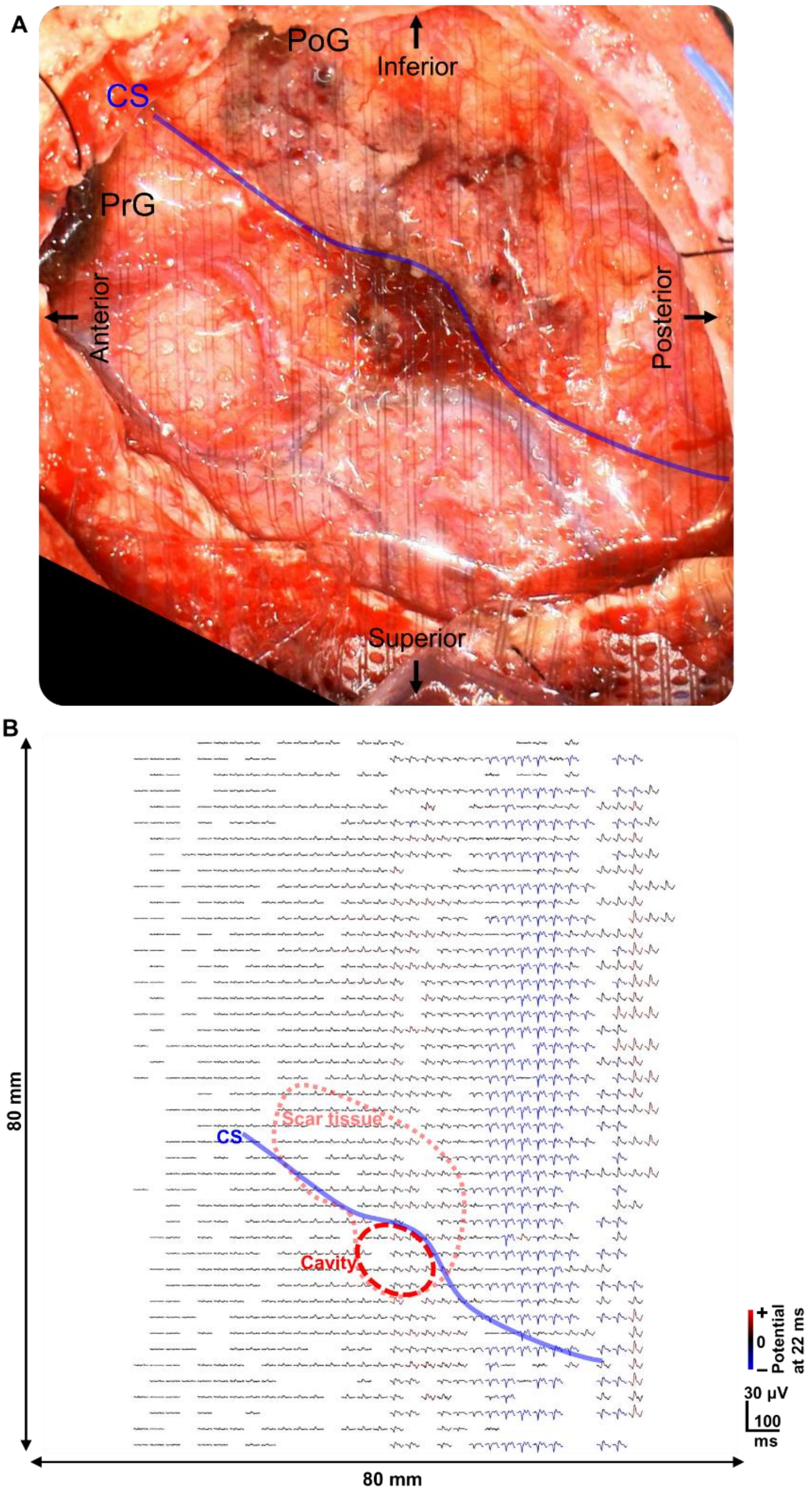

**Figure S28.** Human brain SSEP recording with 2048 channel PtNRGrid. (A) Electrode placement on return surgery patient with scar tissues and cavity from previous resective surgery. (B) Spatial mapping of SSEP waveforms recorded from 2048 channel PtNRGrid.

6.3. More trials for hand grabbing motion

We present high gamma and beta wave activities from additional hand grabbing trials.

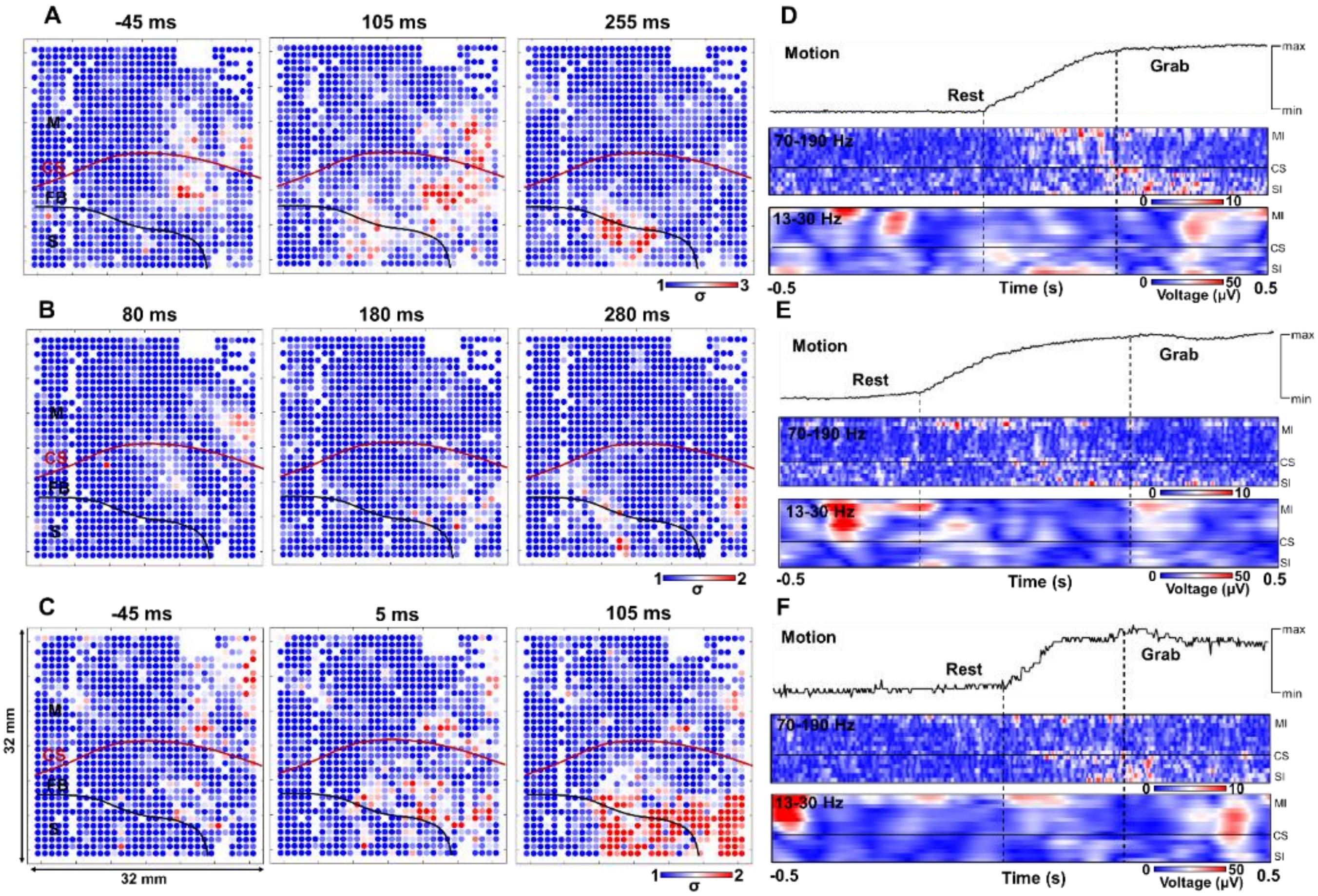

**Figure S29.** Multiple trials of hand grabbing motion. (A)-(C) Spatial mapping of HGA during the motion. (D)-(F) Captured motion, high gamma and beta amplitudes.

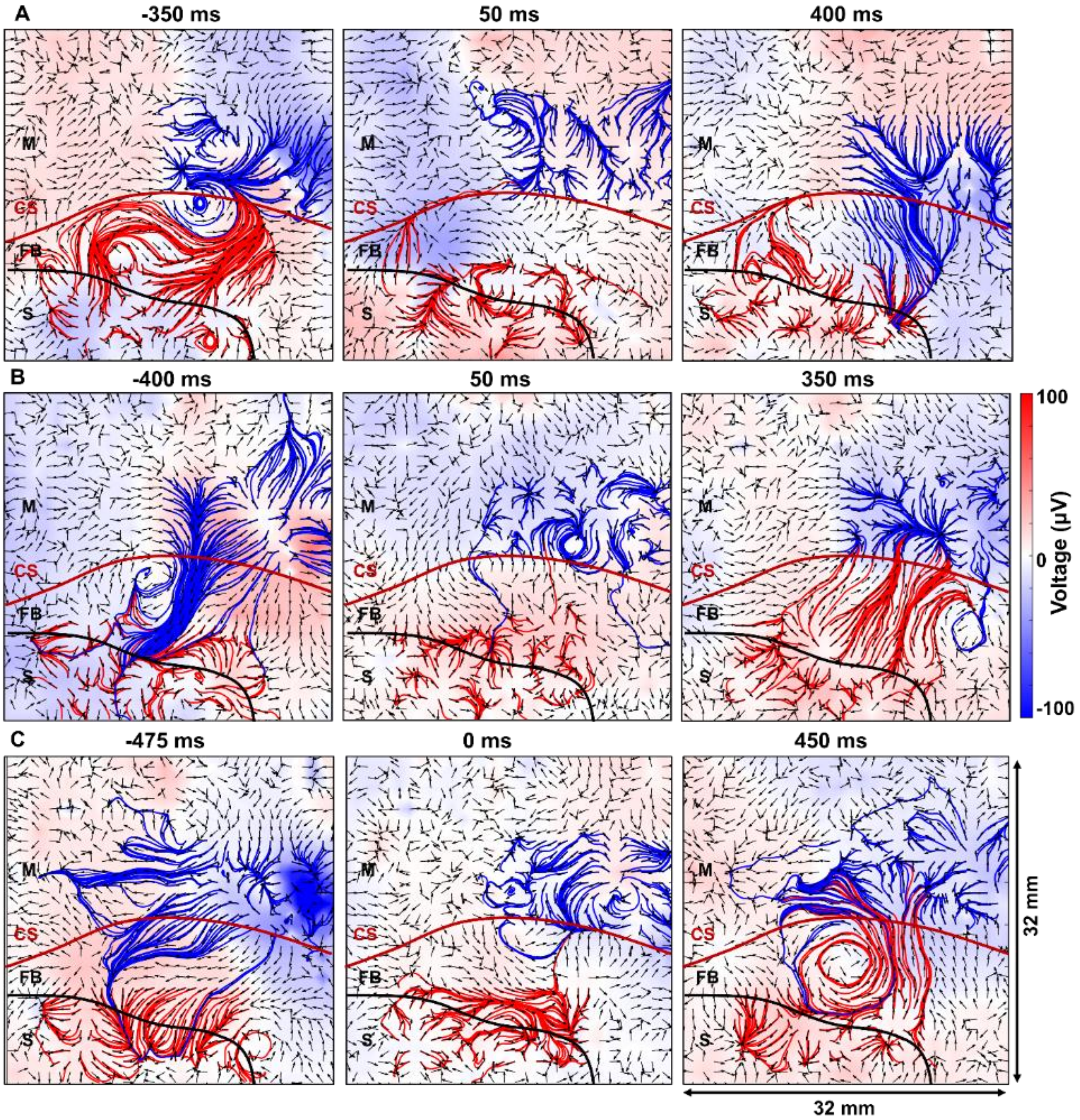

**Figure S30.** Propagating beta waves for multiple trials of hand grabbing motion before, during, and after grabbing. Different trials shown in (A)-(C) corresponds to the trials in Figs. S29A-C.

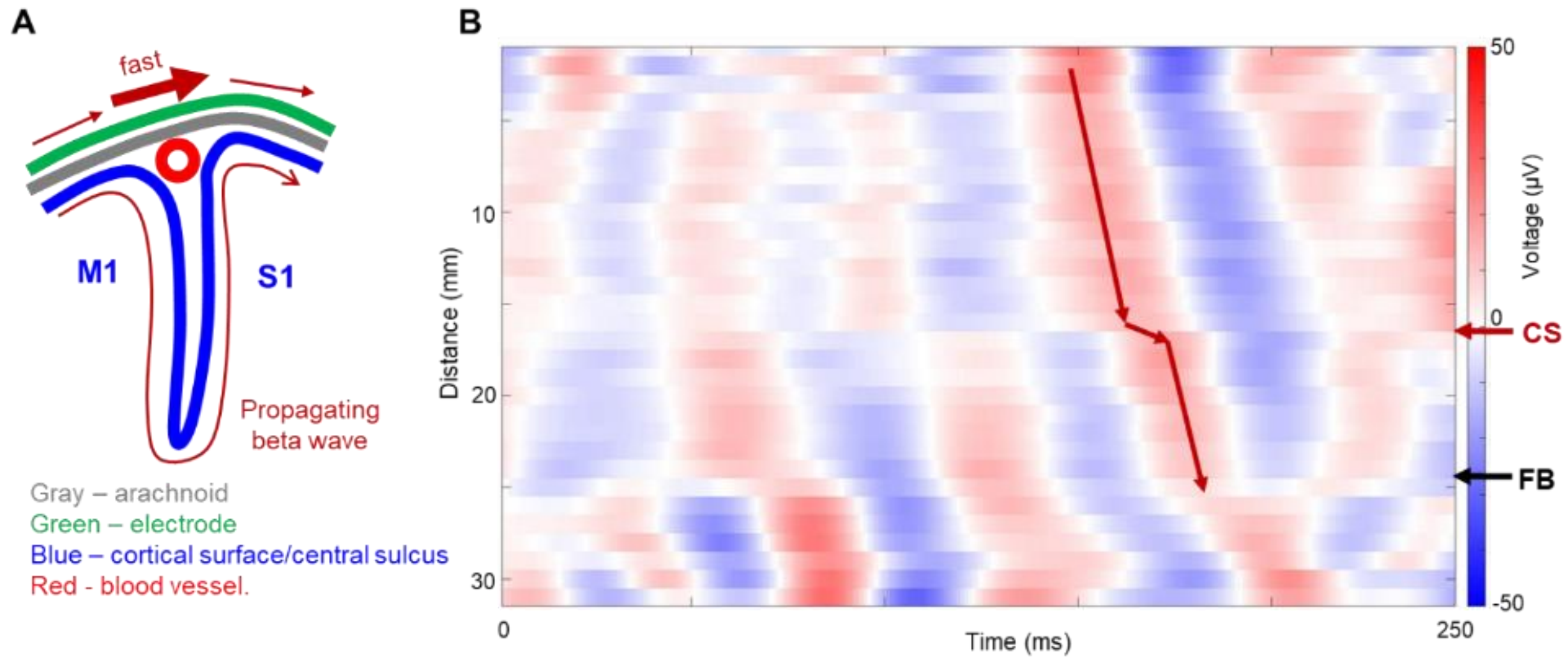

**Figure S31.** Propagating beta waves measured across the sulcus. (A) Cross-sectional schematics of the measurement configuration. (B) Beta wave potentials along the channels perpendicular to the central sulcus. The beta waves measured from the electrodes across the CS typically showed higher propagating speed, which is an artifact, since the adjacent electrodes placed across the sulcus are actually measuring activity happening on that scale of a few mm to tens of mm distance on an unfolded cortical surface.

6.4. Beta Waves Filtered at Different Frequency Windows

We analyzed beta waves filtered at a frequency window of 13-30Hz. Since different frequency windows of 9-18Hz(*4*) and 10-45Hz(*3*) were selected in prior works, we present the beta waves presented in Fig. 4 filtered at different frequency windows.

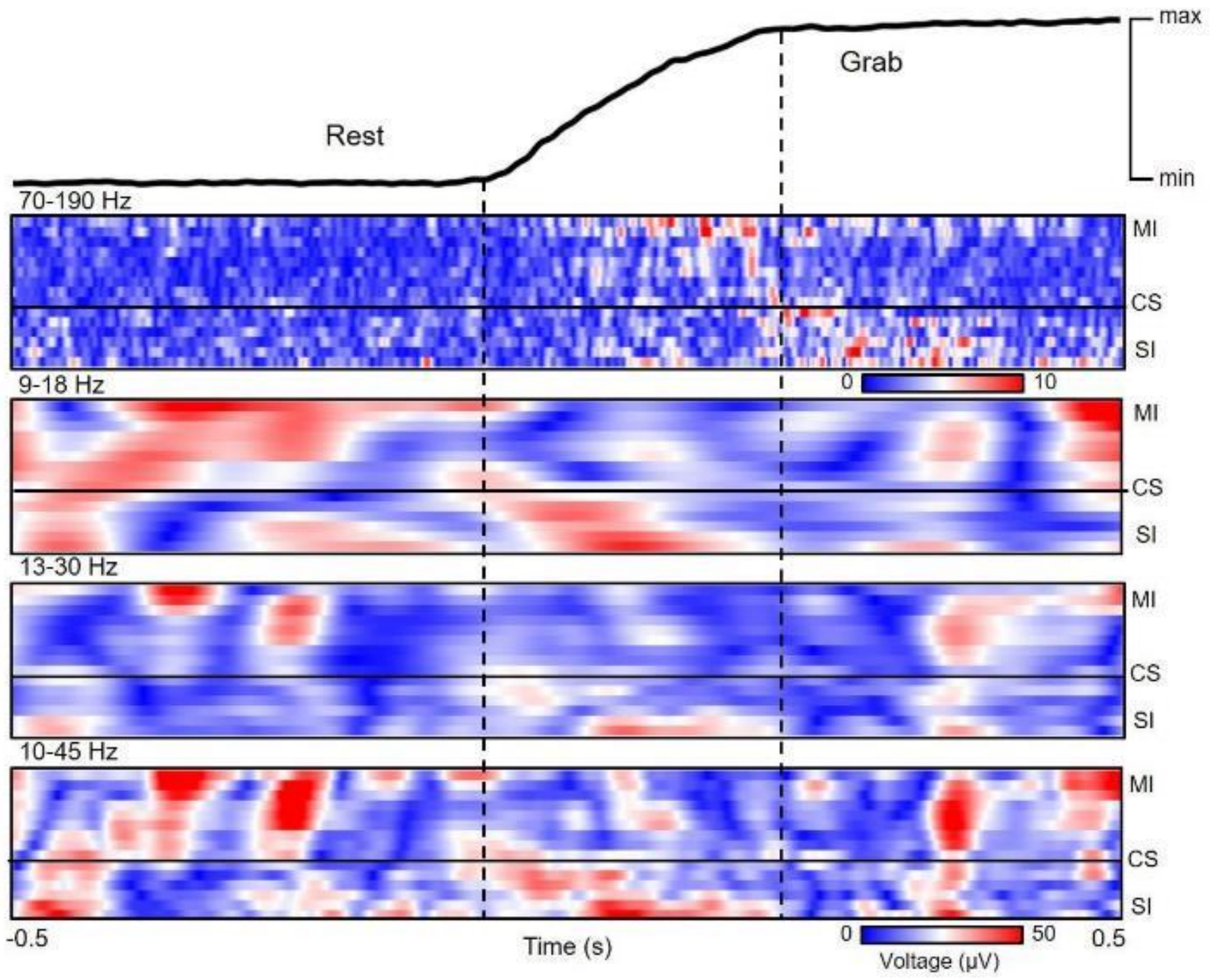

**Figure S32.** Beta wave amplitudes under different frequency windows of 9-18Hz, 13-30Hz, and 10-45Hz.

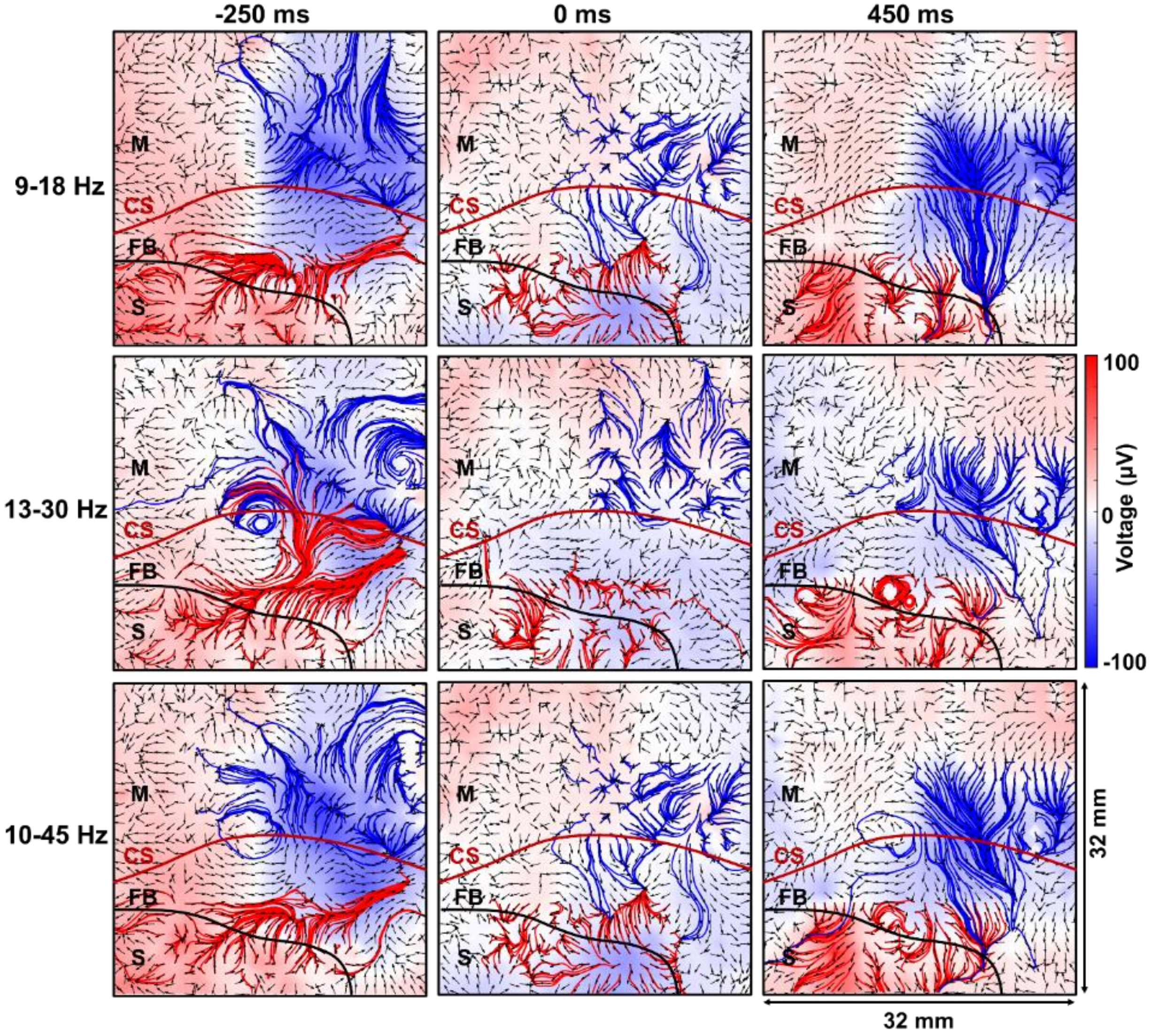

**Figure S33.** Propagating beta waves for the same trial under different frequency windows of 9-18Hz, 13-30Hz, and 10-45Hz.

6.5. Stimulation compatibility test through the PtNRGrids

Prior to doing the bipolar stimulation experiment on the surface of the human brain, we first conducted benchtop tests on brain models made of gelatin to investigate if any of the PtNRGrid electrode or electronics are affected by the bipolar stimulation. By adding 0.1M KCl to the gelatin model brain, the conductivity of the model brain was matched to the tissue conductivity of the pig brain. The model brain surface was wet with saline prior to bipolar stimulation as is usually done in the operation room. We stimulated with biphasic stimulating pulses (30 Hz) up to 2.55 mA with 1 ms pulse width and did not observe any damage on the PtNRGrid contacts, wires, or acquisition electronics which suggested there was no leakage current passing through the electrode itself. Potential mapping performed with the PtNRGrid (Fig. S34B) directly shows that the underlying gelatin tissue was electrically stimulated through the perfusion holes, indicating a current path formed through the gelatin channel underneath the PtNRGrid.

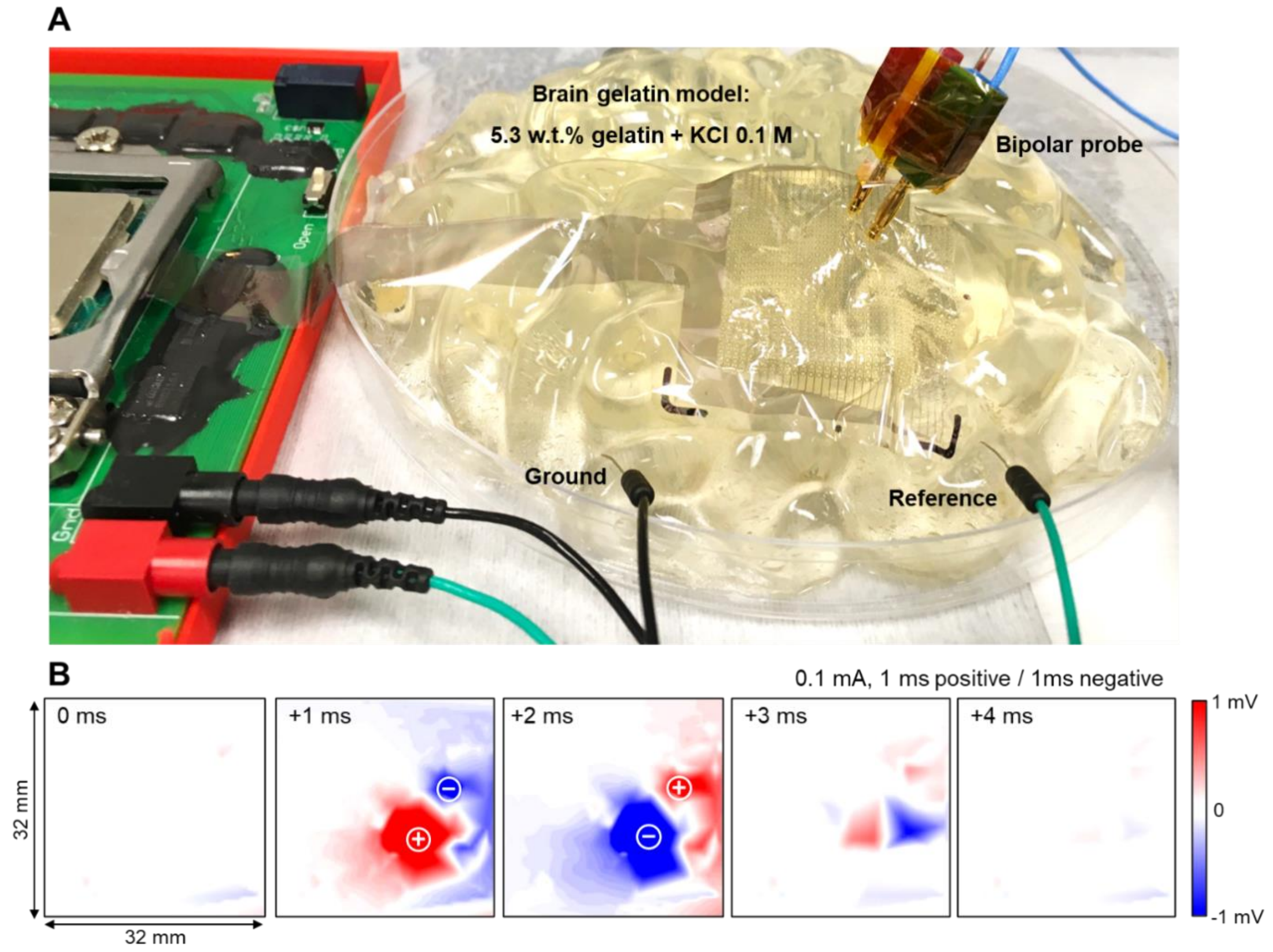

**Figure S34.** Bench top stimulation test. (A) Bipolar probe is stimulating the gelatin model brain through the perfusion holes on the PtNRGrid. (B) Potential mapping recorded with the PtNRGrid.

The effective contact area of the Ojemann bipolar stimulator probe is known to be 1.6 $mm^2$, and it is a common practice to use the current levels of 1-10 mA (A. L. Ritaccio et al., J. Clin. Neurophys. 35, 2 (2018)). When stimulating through the perfusion holes, in any of the contact points on the grid, at least 4 perfusion holes with 0.5 mm diameter were in contact to the Ojemann probe which makes up an effective contact area of >0.8 $mm^2$. Considering that the current and charge density could be doubled with the reduction in effective contact area, we limited the stimulation parameter to 0.5 - 5 mA when stimulating through the grid. The stimulation current level could go further up by increasing the density and/or size of the perfusion holes.

## 6.6. Propagating Waves From Epileptogenic Tissue

We present additional streamline analyzed epileptogenic activities at different time points during the spontaneous (Fig. S35) and stim-evoked (Fig. S36) recordings. Red streamlines are originating from all individual recording sites. In the main manuscript, we removed most of these short-range streamlines and showed the long-range propagation.

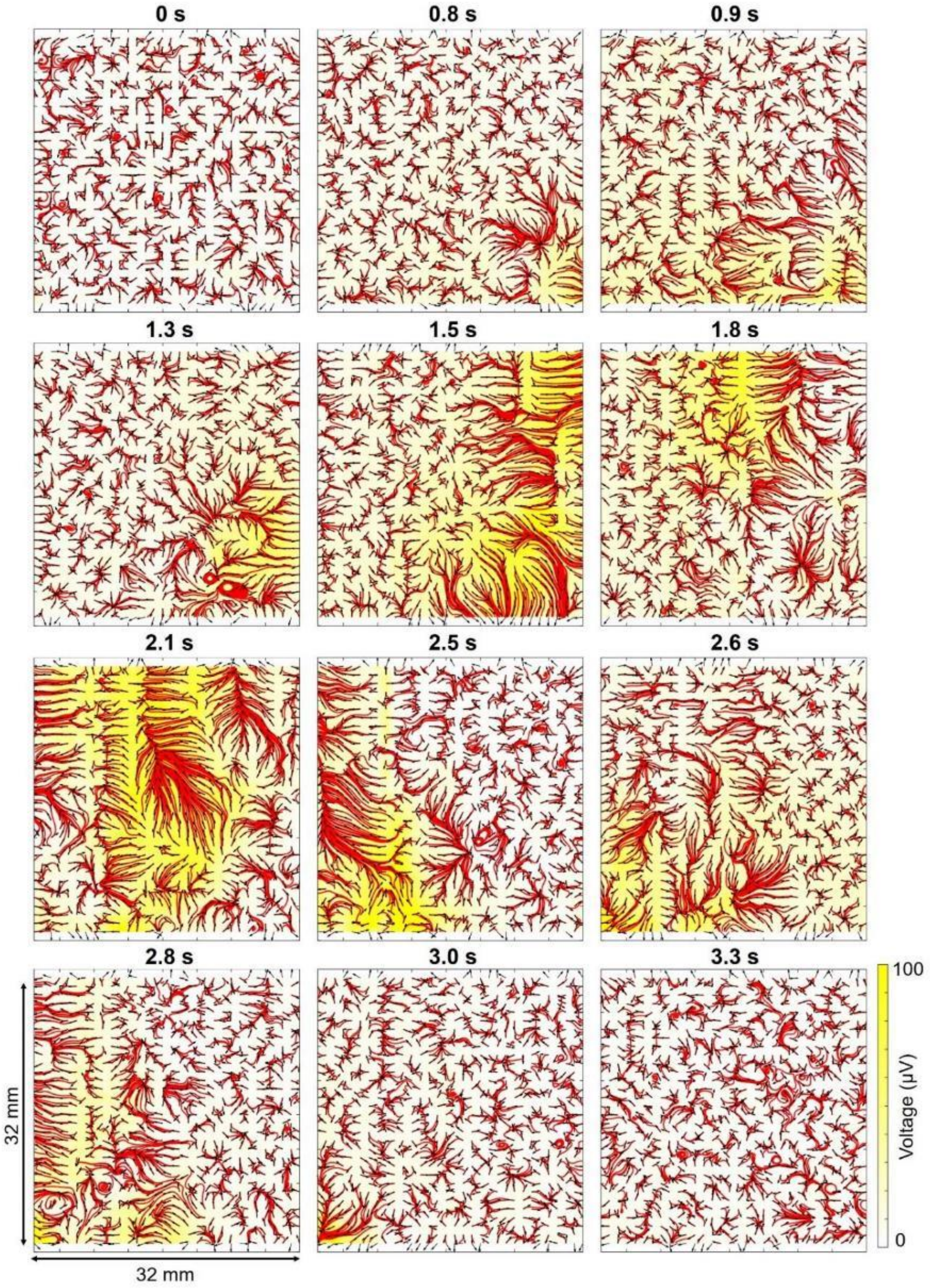

**Figure S35.** Propagating wave analysis of spontaneous epileptiform activity

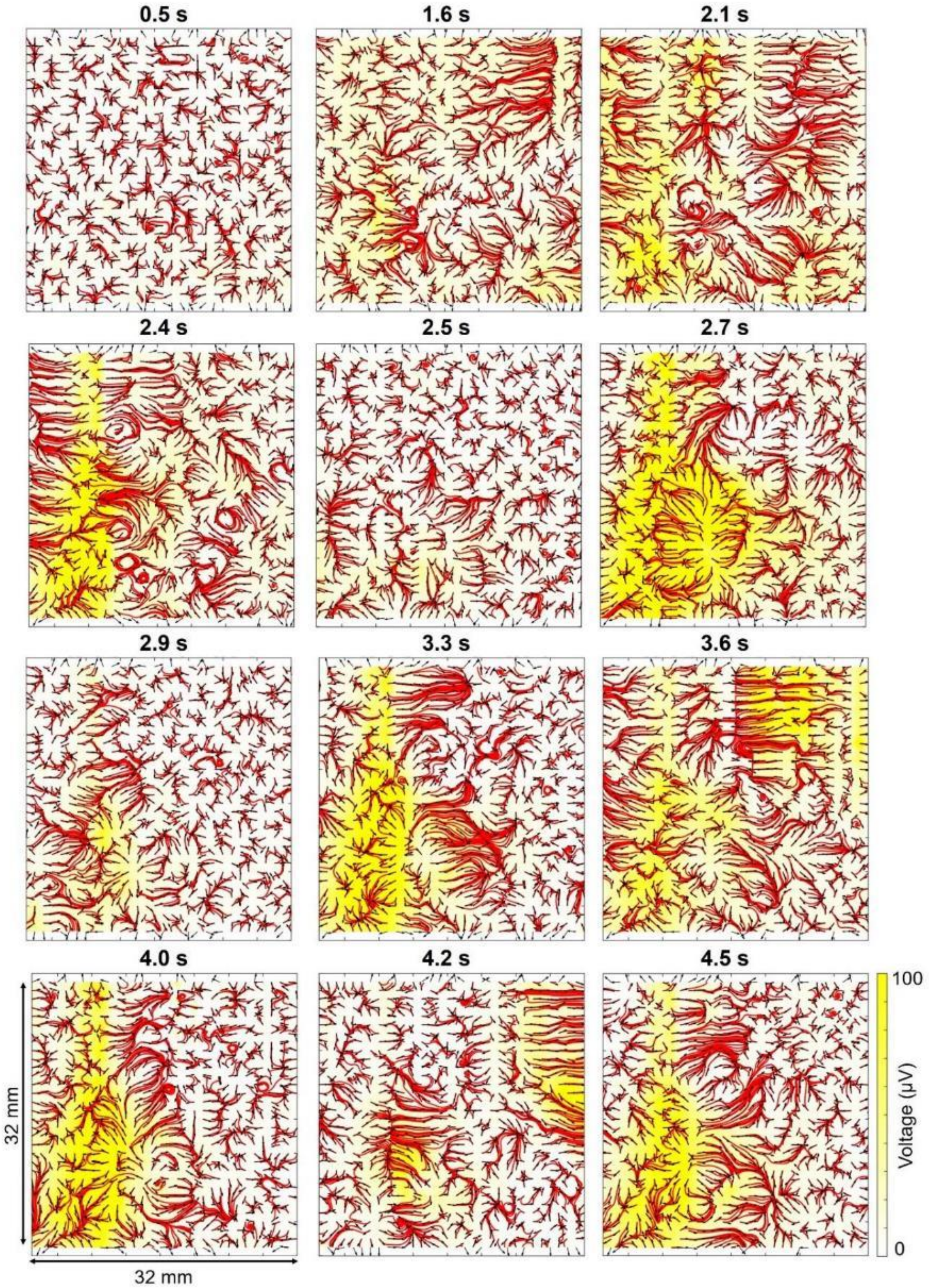

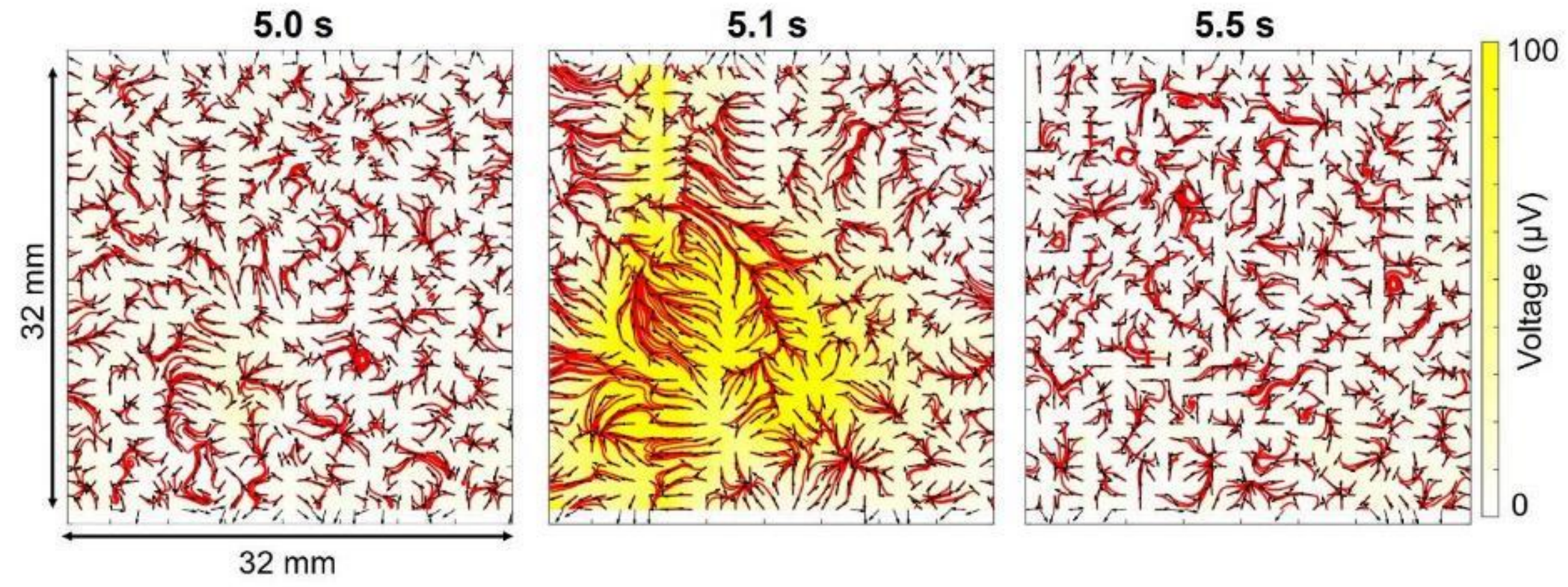

**Figure S36.** Propagating wave analysis of stim-evoked (4th stimulation) epileptiform activity

6.7. 2048 Channels Electrode

The scalable approach of PtNRGrid allowed us to scale the electrode coverage as large as 8 × 8 $cm^2$ and to multiple thousands of channels. We fabricated 2048 PtNRGrids and implanted the grid under IRB authorization at OHSU. The image of the electrode, sterilization method, and implantation method photos can be found in Fig. S37, and the 2048-channel recording system connection diagram is illustrated in Fig. S38. The craniotomy of the subject in this case was only 5 × 5 $cm^2$, so we were not able to record from the entire area of the grid. In the first implantation, including the region slid under the skull, we implanted the right half of the grid (see Figs. S37E and G). We recorded from the grid in this configuration during a motor-sensory task for 10min, and then trimmed the electrode as shown in Fig. S37D to make it narrower for ease of implantation. After trimming, the implantation was performed as shown in the lower panel of Fig. S37E without parts of the electrode being folded. The impedance mapping before trimming, after the first implantation, and after the trimming and second implantation are shown in Figs. S37F-H. The slight differences in impedance magnitudes between the left and the right side of the electrode are most likely related to the two separate reference and ground electrodes used for each side of the electrode. This difference could be reduced by splitting the reference and ground electrode connected to the two separate recording systems.

The criteria to choose the 1024 vs the 2048 versions in the in vivo human recordings is related to the size of the craniotomy. It is only possible to fit the 8 × 8 $cm^2$, 2048 version PtNRGrid for a patient who needed a larger craniotomy as deemed necessary by the clinical procedure. Most of the cases that were suitable for human electrophysiological recordings had craniotomy area ≤ 5 × 5 $cm^2$.

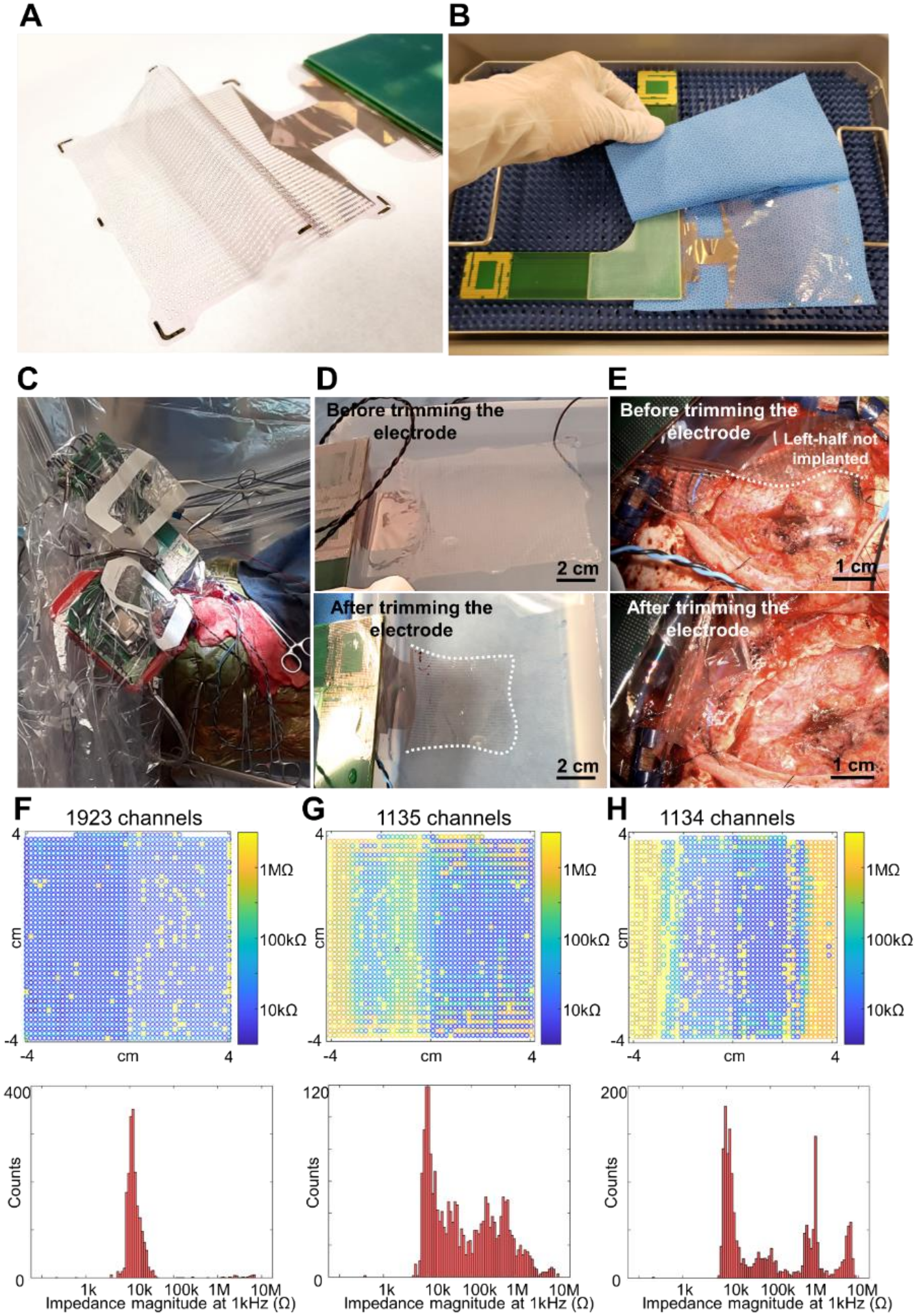

**Figure S37.** 2048 channel electrode and human recording. (A) Photo of the electrode. (B) Sterilization set up with DuraHolder pouch. Two extender boards are joined by implantable-grade autoclavable glue. (C) Intraoperative recording set up with two amplifier boards wrapped

with sterile drapes and clamped with Greenberg retractors. (D) Electrode before and after trimming. (E) Electrode implanted on the patient's brain before and after trimming. 1 kHz impedance mapping of the electrode (F) in saline after the sterilization, (G) on patient's brain with only the right half implanted, and (H) on patient's brain after trimming the electrode. The 1 kHz impedance histograms show all 2048 channels measured by the recording controller including the channels that were not placed on the cortex or trimmed out.

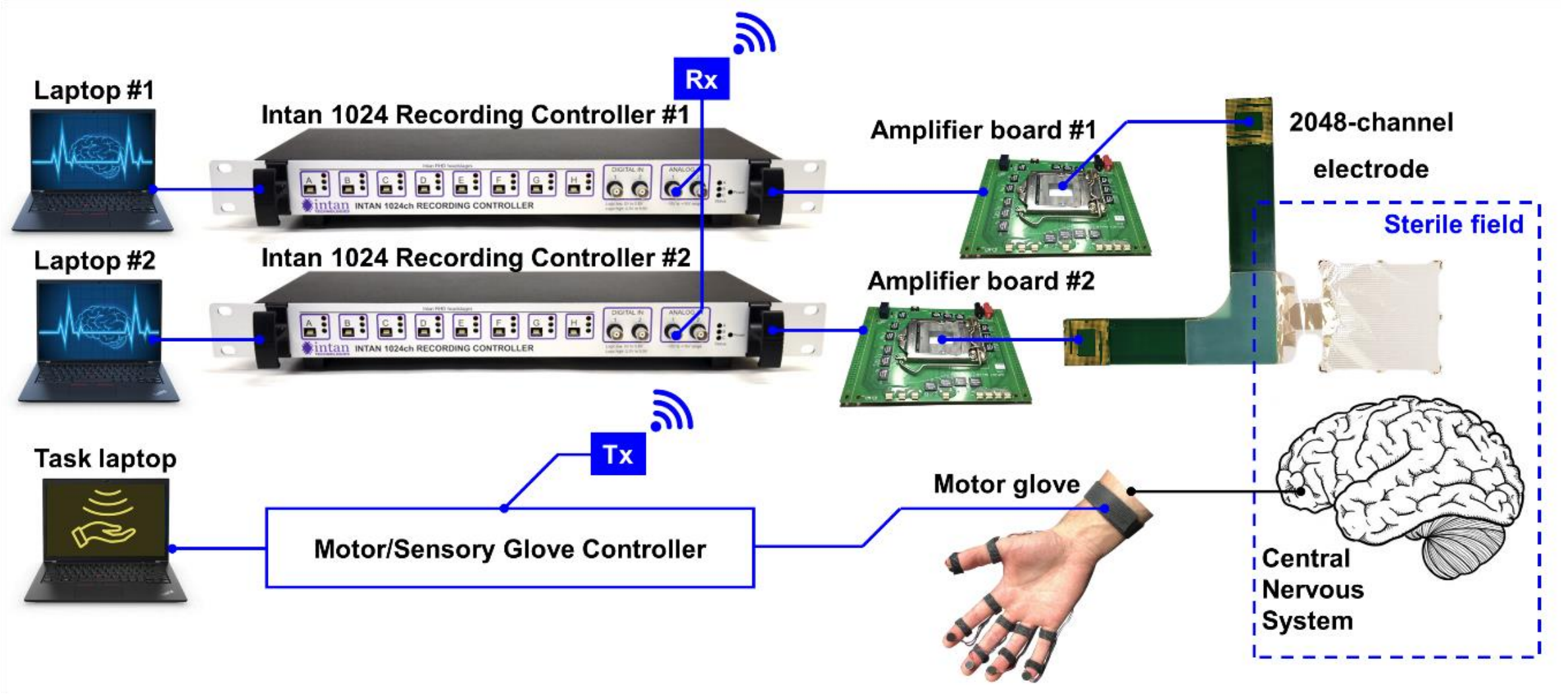

**Figure S38.** 2048-channel recording system connection diagram. Rx and Tx indicates Zigbee wireless receiver and transmitter, respectively.

6.8. Multi-Thousands Channel Electrode and Implantable Electrode

Scaling from 1024 to 2048 channels was achieved simply by adding one more set of recording system, and, as much as the surgical area allows, more recording systems could be added to further increase the channel count to several thousands. However, from a practical standpoint, the electronic acquisition units will become crowded and bulky as the channel count increases beyond 4096 channels. To further increase the channel counts while minimizing the space the recording system occupy near the surgical field, higher density connectorization system is being developed (Figure S39). The amplifier boards having 0.4 mm pitch, higher-density LGA connector (Fig. S39A) and the mock-up of implantation and connector system for 4096 channel is described in Fig. S39.

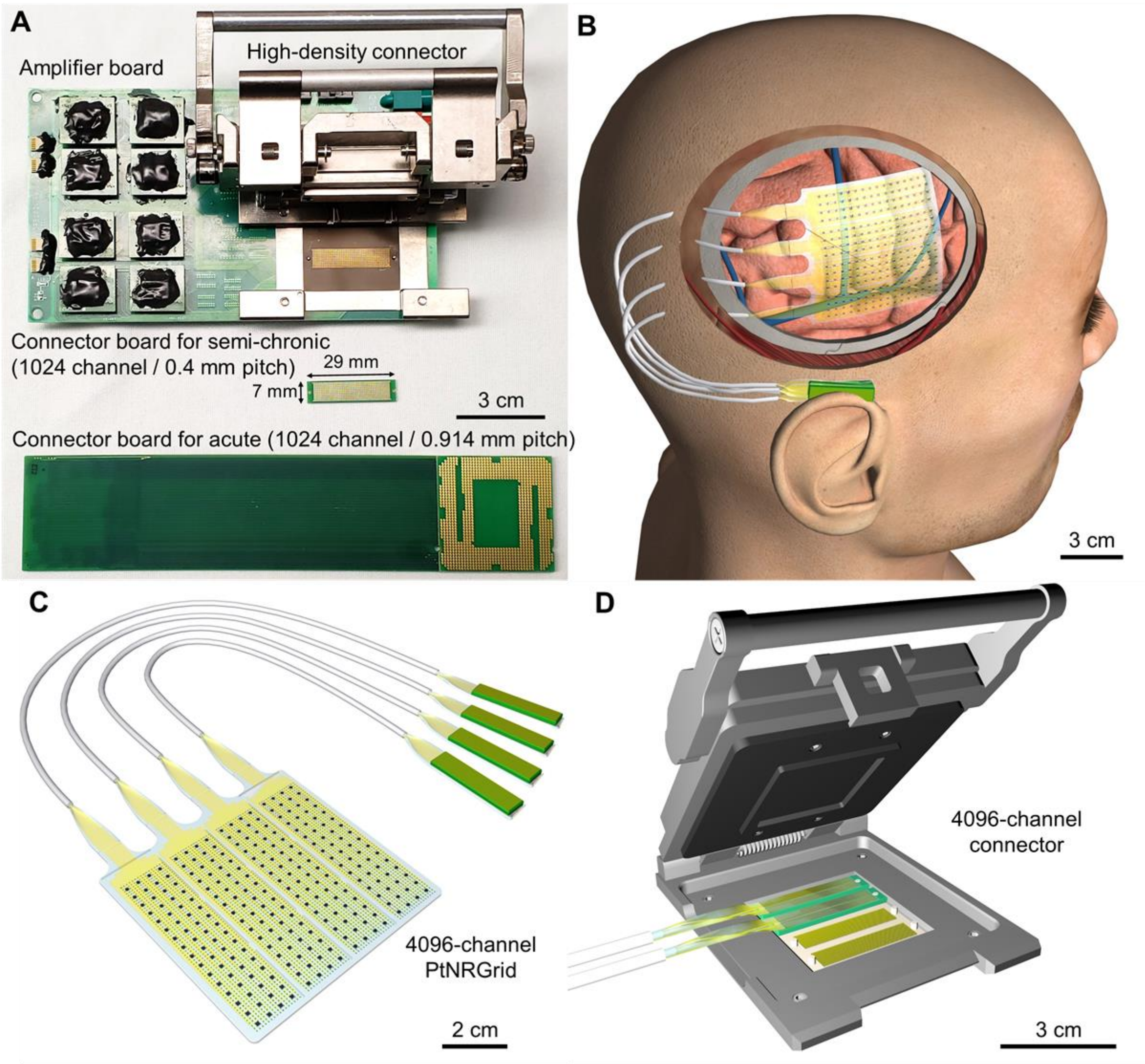

**Figure S39.** (A) High density 1024 channel connector board, (B) mock-up of implantation strategy of the 4096-channel grid in a semi-chronic setting with tunneled wires through the scalp (C) electrode and (D) connector placement in the specially designed socket.

6.9. Comparison of PtNRGrid with other recent ECoG grid technologies

**Table S2.** Comparison of PtNRGrid with other recent ECoG grid technologies published recently while this manuscript was under review.

| | | T. Kaiju *et al.*,<br>J. Neural Eng. 18<br>(2021) 036025 (*7*) | C.-H. Chiang *et al.*,<br>J. Neural Eng. 18<br>(2021) 045009 (*8*) | This work: PtNRGrid<br>Y. Tchoe *et al.* |
|---|---|---|---|---|
| Number of recording channels | Human brain | - | 244 | 1024, 2048 |
| | Non-human primate brain | 1152<br>(128 x 9 electrodes) | - | - |
| | Rodent brain | - | - | 1024 |
| Concept electrode photo without presenting recording data | | - | 1024<br>(256 x 4 electrodes) | - |
| Electrode thickness for conformity to brain curvature | | 20 ~ 60 μm | 60 μm | 6.6 μm |
| Max recording density | | 1,149 $cm^{-2}$ | 596 $cm^{-2}$ | 4,444 $cm^{-2}$ |
| Flexible material | | Parylene C | Liquid crystal polymer | Parylene C |
| Electrochemical interface material | | Pt black | PtIr | PtNR |
| Material deposition method | | Electroplating | Electroplating | Sputtering<br>& de-alloying |
| 1kHz impedance magnitude | | 26 ± 7 kΩ<br>(50 x 50 $\mu m^2$) | 4.6 ± 1.8 kΩ<br>(200 μm diameter) | 10 ± 2 kΩ<br>(30 μm diameter) |

● PtNRGrid achieved the first multithousand channel human brain recording. The human brain recordings demonstrated in C.-H. Chiang *et al.* was done with 244 channel electrodes. T. Kaiju *et al.*'s work was performed on non-human primates.

● The 1152 channel electrode demonstrated in T. Kaiju *et al.* is made by gluing nine 128 channel electrodes together in one place.

● C.-H. Chiang et al. only showed a concept photo of a 1024 channel electrode array by gluing four 256 channel electrodes without any electronics board or recordings.

● The electrodes stacked on top of each other (T. Kaiju *et al.*) may not be accepted for human use because the gaps between the layers may be hard to properly sterilize.

● Electrode thickness of T. Kaiju *et al.* and C.-H. Chiang *et al.* were 20~60 µm and 60 µm, respectively, whereas PtNRGrid was only 6.6 µm in total thickness. The thickness of the electrode is a key factor in achieving high conformality to the brain curvatures, improved recording quality and sharp delineation of cortical boundaries for neural correlates.

● Recording density of PtNRGrids was as high as 4,444 $cm^{-2}$ which is substantially higher than the other two technologies: T. Kaiju *et al.* 1,149 $cm^{-2}$, C.-H. Chiang *et al.* 596 $cm^{-2}$

● PtIr or Pt black are usually electroplated serially on top of planar contacts, one channel at a time. On the other hand, our PtNR are manufactured via a scalable process that uses co-sputtering and de-alloying to make thousands of PtNR contacts at once. The difference in this part of the fabrication process makes a substantial difference in the manufacturing time as well as the uniformity in the contact quality. Electroplating of hundreds of channels can take as much as a few hours using a manual process, whereas the typical PtNR process on 7-inch glass forms 4096 contacts at once by a few minutes of co-sputtering followed by de-alloying. Since all PtNR contacts are formed at the same time, the quality of the PtNRs is extremely uniform across thousands of channels. Additionally, the 1 kHz impedance magnitude of the PtNRs was superior to the Pt black; 30 µm diameter PtNRs had even lower 1 kHz impedance magnitude than the larger Pt black contact.

## 6.10. 1024 Channels PEDOT:PSS Electrode

The fabrication process of multi-thousand channel PtNRGrids is also compatible with poly(2,3-dihydrothieno-1,4-dioxin)-poly(styrenesulfonate) (PEDOT:PSS). Using a spin-coating and lift-off method, we successfully fabricated electrodes with PEDOT:PSS (see Fig.

S40) and recorded from patients in both OHSU and UCSD under IRB authorization. The PEDOT:PSS also effectively captured SSEPs with high SNR as demonstrated in Fig. S41.

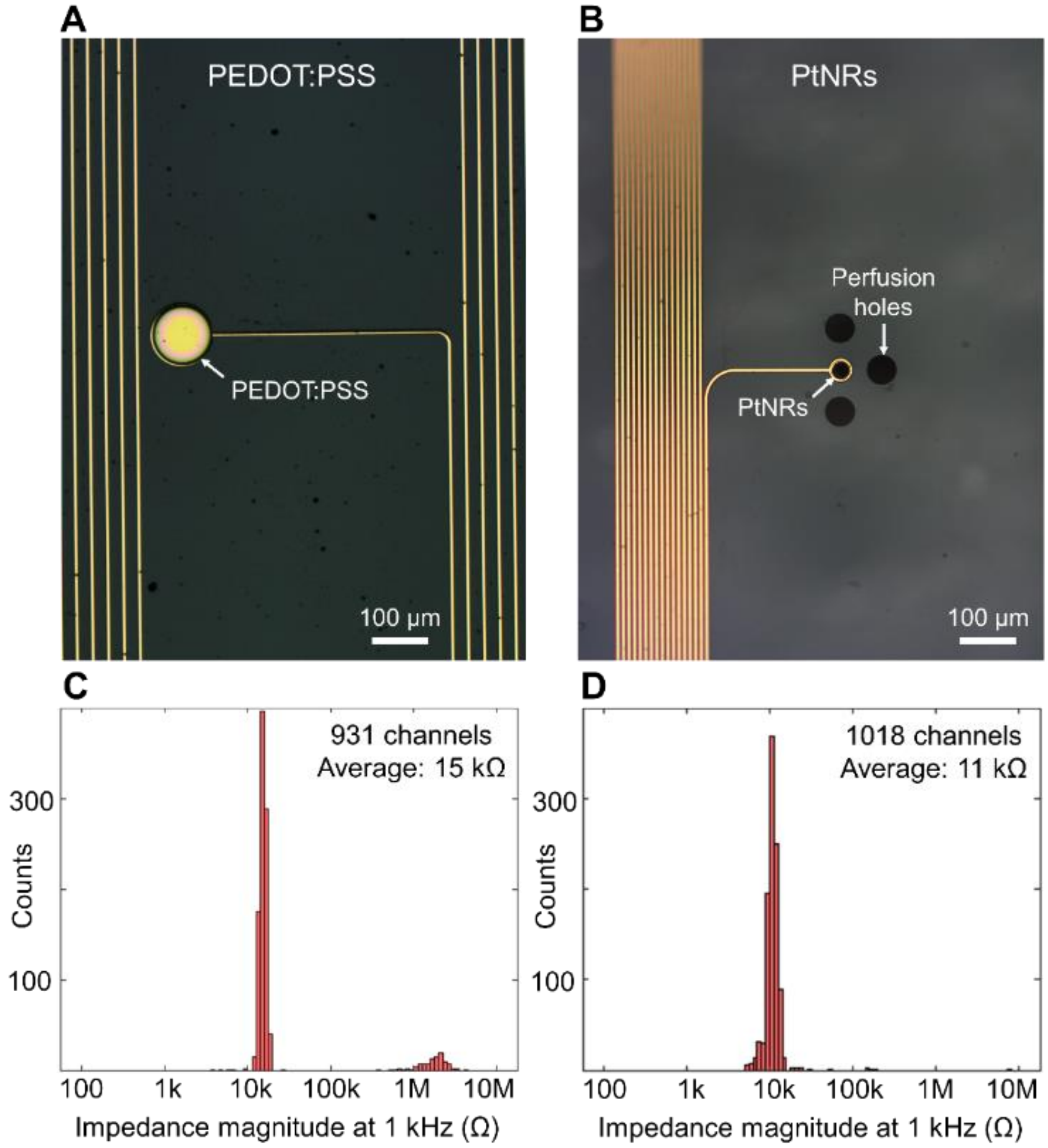

**Figure S40.** 1024 channels electrode with (A) 100μm diameter PEDOT:PSS as electrolytic interface in comparison with (B) 30 μm diameter PtNRs. Histogram of impedance magnitude at 1kHz for (C) PEDOT:PSS (100 μm diameter) and (D) PtNRs (30μm diameter).

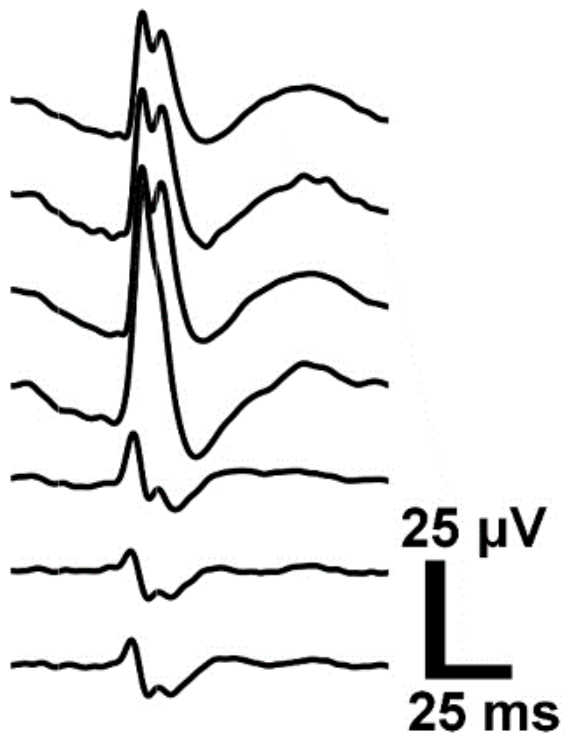

**Figure S41.** Phase reversal SSEPs recorded by 1024 channels PEDOT:PSS electrodes.

**Movie S1.** Video of the spatial mapping of HGA during the hand grabbing motion. Slowed 11 times the actual speed. https://youtu.be/nhpR0mnltQY

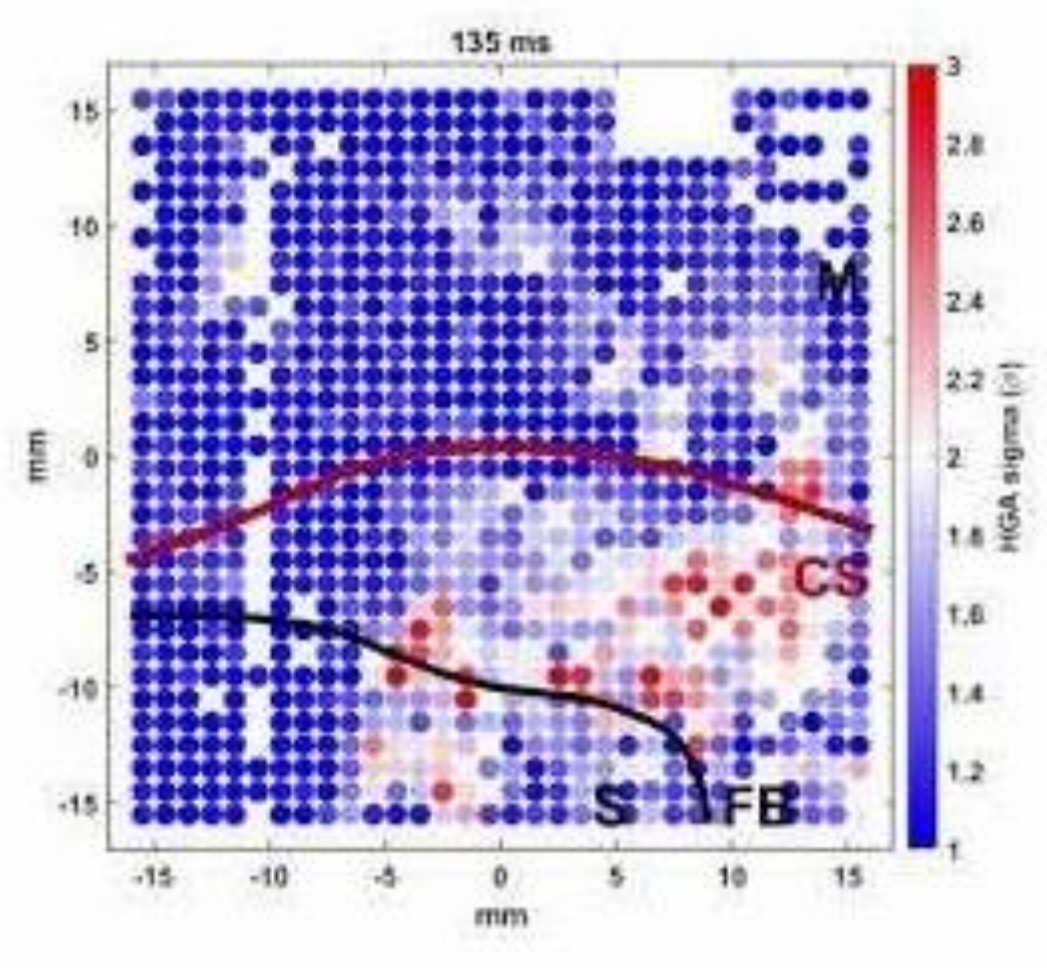

**Movie S2.** Video of the propagating beta wave prior to hand grabbing motion. Arrows indicate the vector fields of the propagating direction and the background color represents the beta wave potentials. Slowed 110 times the actual speed. https://youtu.be/zMh4iVt7SWU

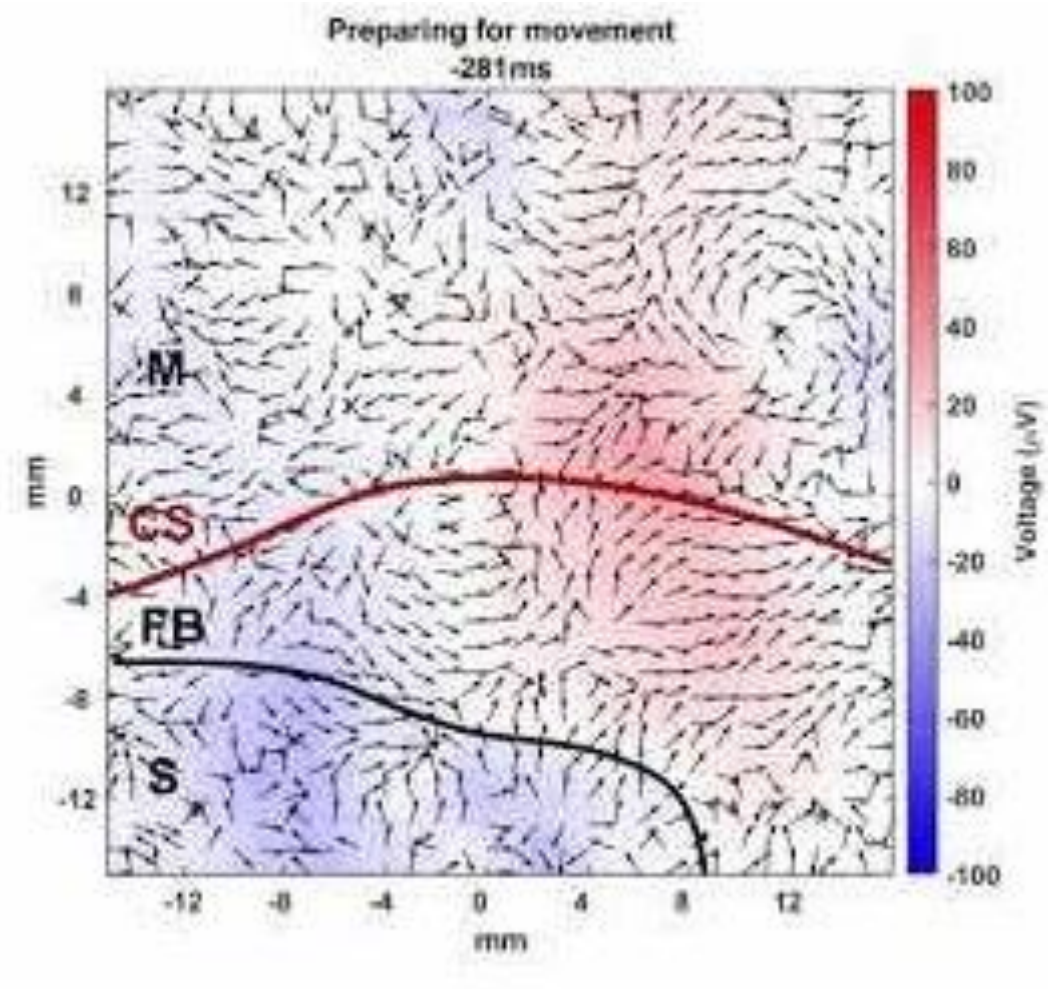

**Movie S3.** Video of the propagating beta wave after completing the hand grabbing motion. Arrows indicate the vector fields of the propagating direction and the background color represents the beta wave potentials. Slowed 110 times the actual speed. https://youtu.be/r9wr1ezsUCo

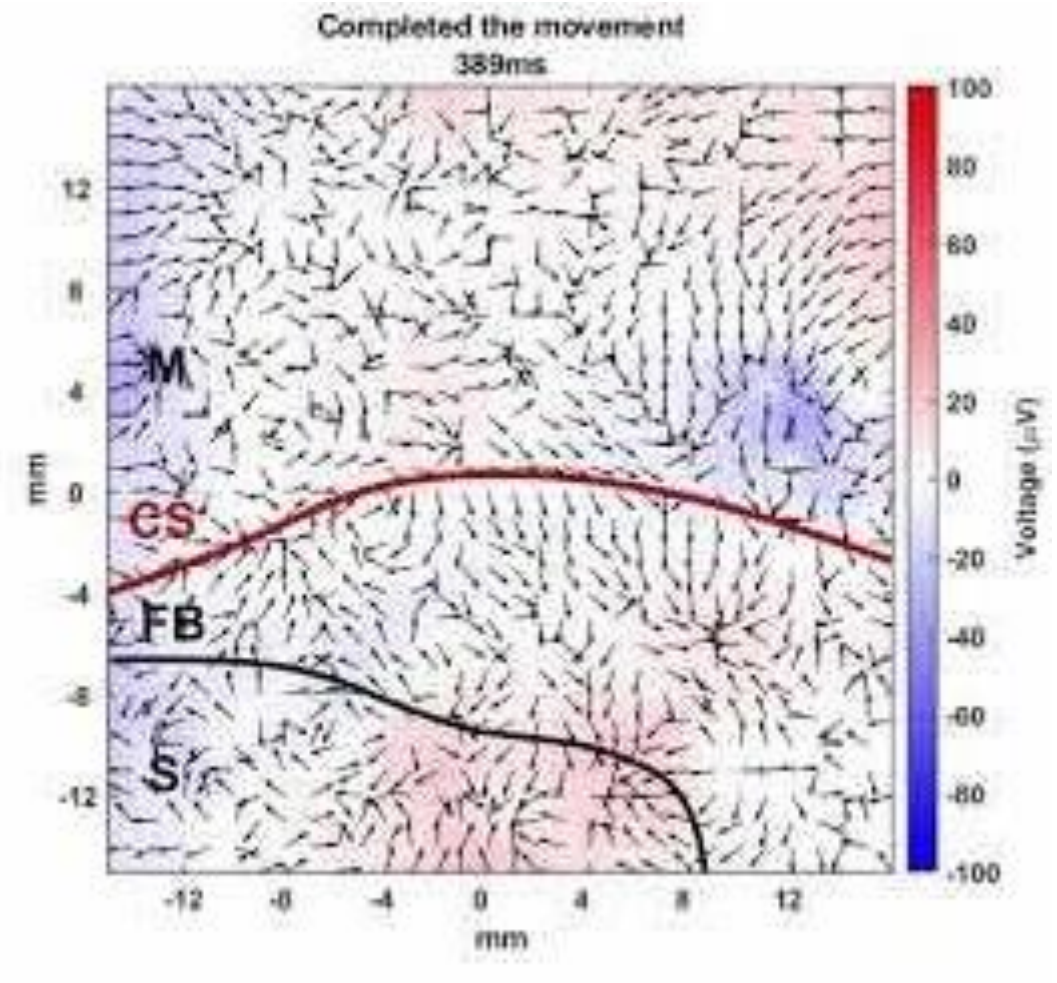

**Movie S4.** Propagating beta and high gamma activities overlayed on the brain model. The vector field and streamlines on the PtNRGrid represent the propagating beta waves, and green scattered dots represent the high gamma activity. Slowed 51 times the actual speed. https://youtu.be/Q47l1KJDv7g

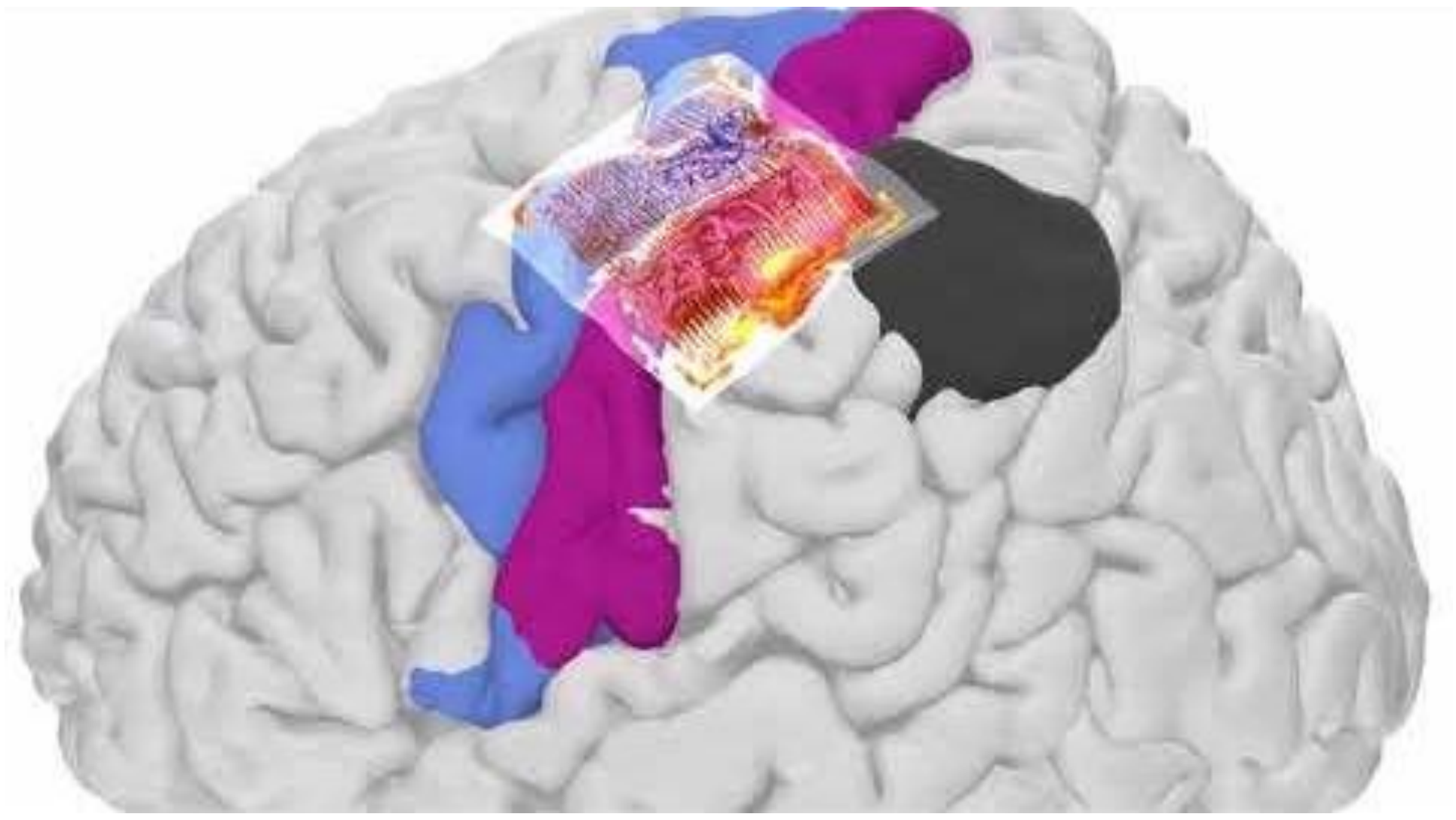

**Movie S5.** Videos of spontaneous epileptiform activity from epileptogenic tissue. Scatter map shows the 10-59 Hz brain wave amplitude. Slowed 4 times the actual speed. https://youtu.be/jFX-4lOPYYQ

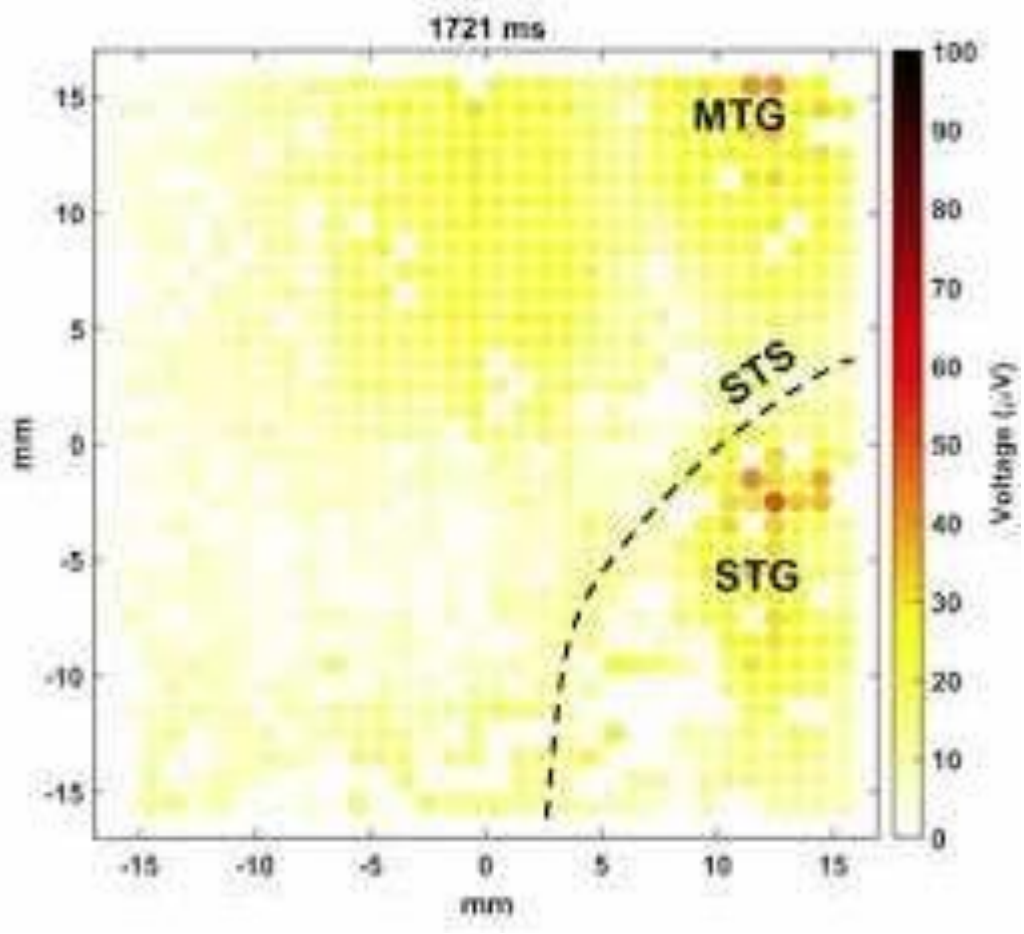

**Movie S6.** Spontaneous epileptiform discharges overlayed on the patient's brain model. Scattered dots and contours on the PtNRGrid represent the 10-59 Hz brain wave amplitude. Slowed 17 times the actual speed. https://youtu.be/YSaPGeV4aYI

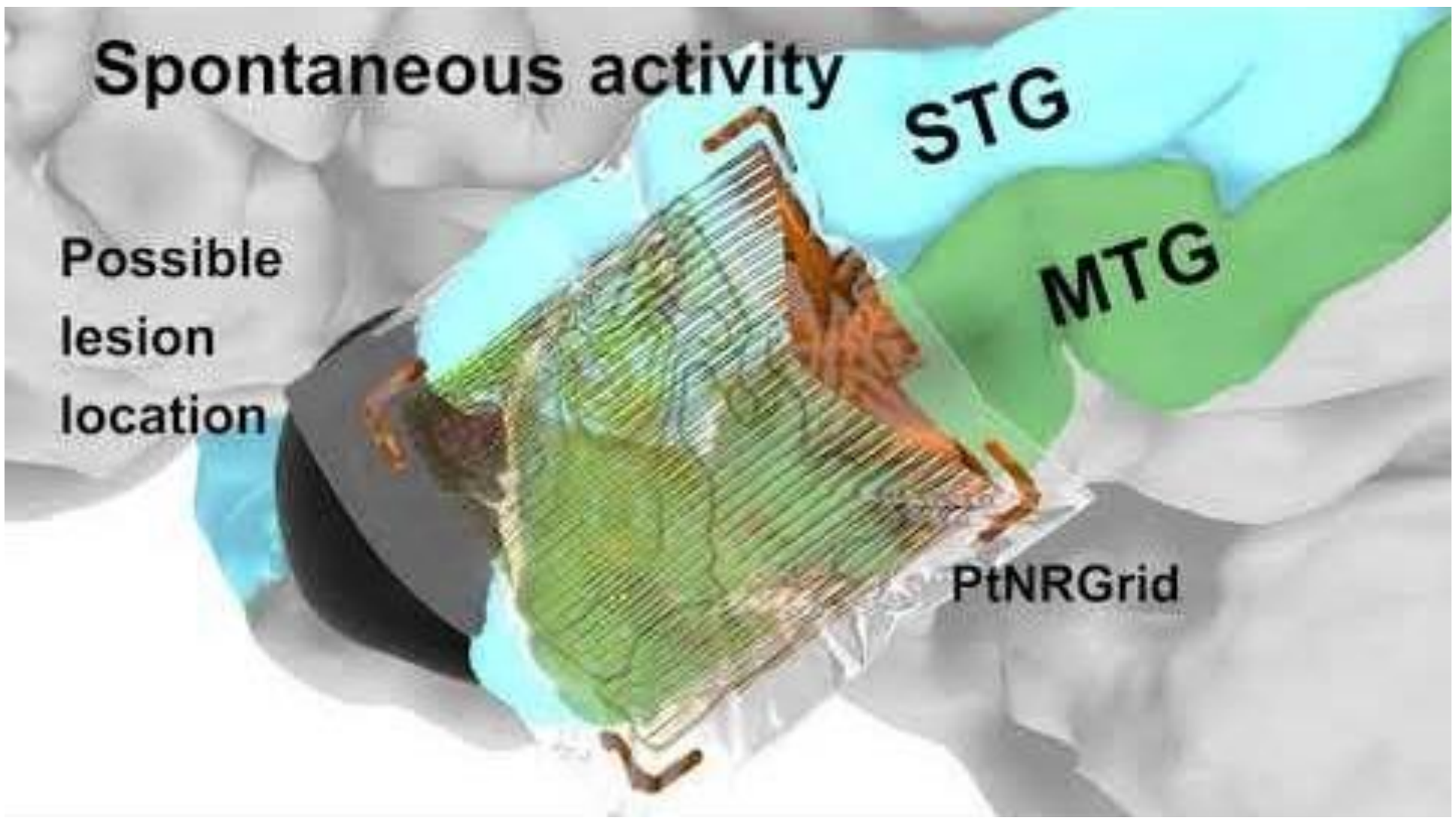

**Movie S7.** Videos of stimulation-evoked (4th stim) epileptiform activities from epileptogenic tissue. Scatter map shows the 10-59 Hz brain wave amplitude. Slowed 4 times the actual speed. https://youtu.be/wrRlPMoLb-I

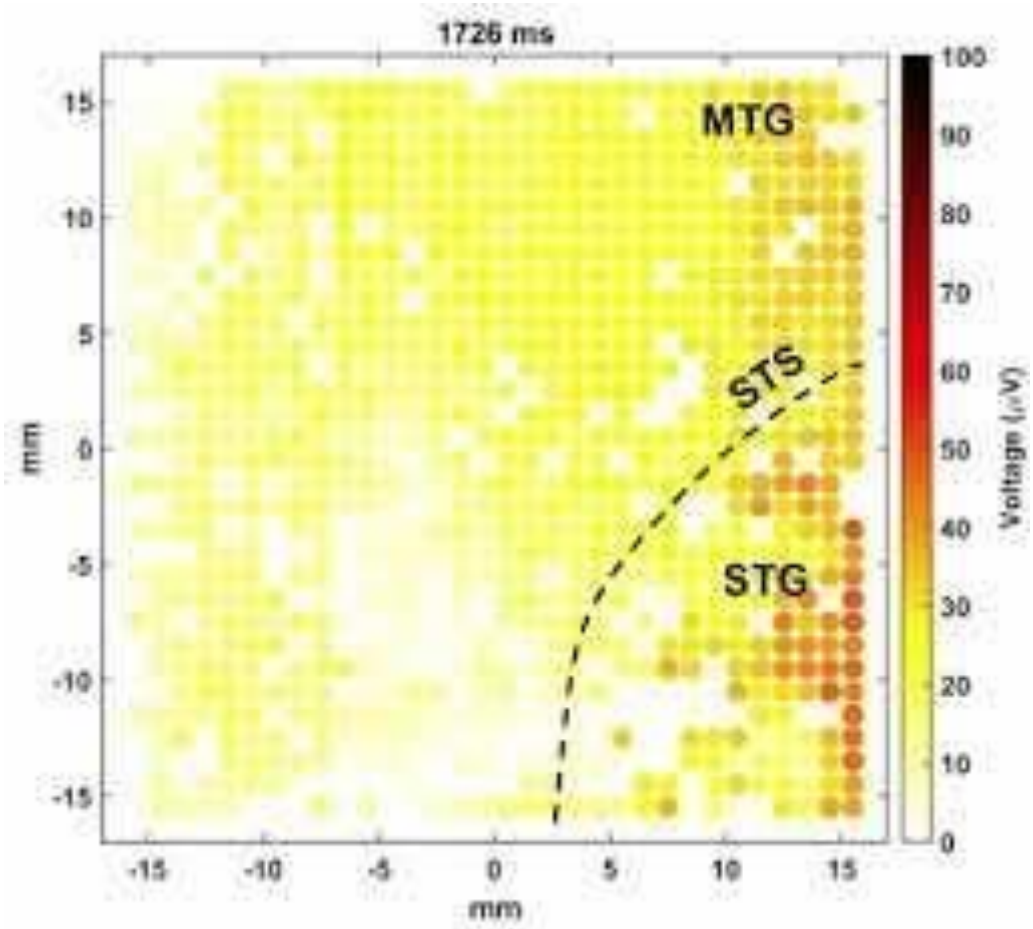

**Movie S8.** Stimulation-evoked epileptiform discharges overlaid on the patient's brain model. Scattered dots and contours on the PtNRGrid represent the 10-59 Hz brain wave amplitude. Slowed 20 times the actual speed. https://youtu.be/p7sduXUE-8U

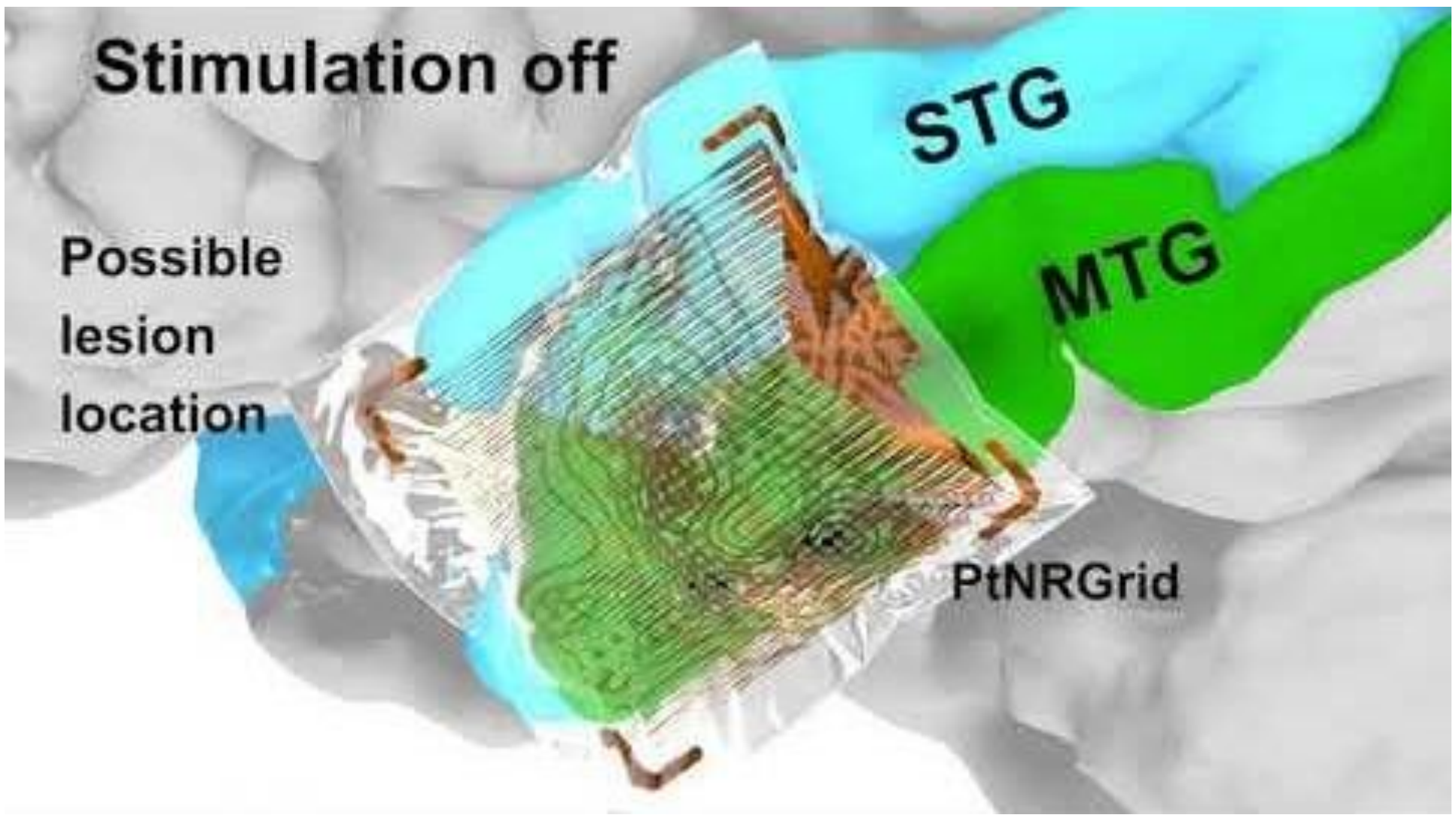